\documentclass{aa}  

\usepackage{graphicx}
\usepackage{txfonts}
\usepackage{amsmath}	
\usepackage{amssymb}	
\usepackage{enumitem}
\usepackage{array}
\usepackage{multirow}
\usepackage{lipsum}
\usepackage{siunitx} 
\usepackage{newtxtext,newtxmath}

\newcommand{\sunmass}{M\ensuremath{_\odot}}

\begin{document}



    \title{Europium enrichment and hierarchical formation of the Galactic halo}
\titlerunning{Eu enrichment in the Galactic halo}

\authorrunning{L. Cavallo et al.}

   \author{L. Cavallo\inst{\ref{unipd}}, G. Cescutti\inst{\ref{units},\ref{INAF},\ref{INFN},\ref{IFPU}}, F. Matteucci\inst{\ref{units},\ref{INAF},\ref{INFN}}}

   \institute{Dipartimento di Fisica e Astronomia, Universit\`a di Padova, Vicolo dell'Osservatorio 3, 35122 Padova, Italy\label{unipd}\\
              \email{lorenzo.cavallo@phd.unipd.it}
              \and Dipartimento di Fisica, Sezione di Astronomia, Universit\'a di Trieste, Via G. B. Tiepolo 11, 34143 Trieste, Italy\label{units}
              \and INAF, Osservatorio Astronomico di Trieste, Via Tiepolo 11, I-34143 Trieste, Italy\label{INAF}
              \and INFN, Sezione di Trieste, Via A. Valerio 2, I-34127 Trieste, Italy\label{INFN}
              \and IFPU, Institute for the Fundamental Physics of the Universe, Via Beirut, 2, I-34151 Grignano, Trieste, Italy \label{IFPU}}

   \date{Received February 11, 2022; accepted ---}


  \abstract
 {The origin of the large star-to-star variation of the [Eu/Fe] ratios observed in the extremely metal-poor (at [Fe/H]$\leq-3$) stars of the Galactic halo is still a matter of debate.}
   {In this paper, we explore this problem by putting our stochastic chemical evolution model in the hierarchical clustering framework, with the aim of explaining the observed spread in the halo.}
   {We compute the chemical enrichment of Eu occurring in the building blocks that have possibly formed the Galactic halo. In this framework, the enrichment from neutron star mergers can be influenced by the dynamics of the binary systems in the gravitational potential of the original host galaxy. In the least massive systems, the neutron stars can merge outside the host galaxy and so only a small fraction of newly produced Eu can be retained by the parent galaxy itself.}
   {In the framework of this new scenario, the accreted merging neutron stars are able to explain the presence of stars with sub-solar [Eu/Fe] ratios at [Fe/H]$\leq-3$, but only if we assume a delay time distribution for merging of the neutron stars $\propto t^{-1.5}$. We confirm the correlation between the dispersion of [Eu/Fe] at a given metallicity and the fraction of massive stars which give origin to neutron star mergers. The mixed scenario, where both neutron star mergers and magneto-rotational supernovae do produce Eu, can explain the observed spread in the Eu abundance also for a delay time distribution for mergers going either as $\propto t^{-1}$ or $\propto t^{-1.5}$.}
   {}


   \maketitle
%

\section{Introduction} \label{intro}
Presently we know that light elements and their isotopes, such as $^{1,2}$H, $^{3,4}$He and $^7$Li originated in the Big Bang. Massive stars, during their evolution and in explosive end phases, can produce elements from C to Ti, the iron-peak elements, and beyond \citep[e.g.][]{Howard1972ApJ,WOOSLEY2007269,Wanajo_2018,Curtis_2018}. However, the majority of all nuclei heavier than the iron-peak elements are produced by neutron-capture reactions and rely on high densities of free neutrons. The neutron capture processes are divided into two different classes: rapid or r-process (neutron capture timescale shorter than $\beta$ decay) and slow or s-process (in this case the neutron-capture timescale is longer than $\beta$ decay). Most neutron-capture elements are produced by both r and s-process, but for some of these heavy nuclei, the production is dominated by only one process.\\
In the last few years, many studies have tried to probe the origin of rapid neutron-capture process (r-process) elements in the universe. This challenge requires a multi-scale convolution of studies from different fields: nuclear astrophysics, stellar spectroscopy, gravitational waves, short gamma-ray bursts (SGRBs), galaxy formation theories and chemical evolution models \citep[e.g.][]{ cowan1991r,berger2014short, matteucci2014europium, cescutti2015role, Wehmeyer2015,abbott2017gw170817}.\\
Observations of heavy element abundances can provide constraints of the astrophysical site(s) of r-process nucleosynthesis.
Many observations show a large spread, up to 2 dex, of r-process elements in the metal-poor environment of the Galactic halo \citep{mcwilliam98,koch2002europium, honda2004spectroscopic, fulbright2000abundances}. On the other hand, the observed star-to-star abundance scatter of $\alpha$-elements (i.e. elements produced by massive stars like O and Mg) is smaller compared with the one of heavy elements \citep[see][]{Cayrel2004,bonifacio2012chemical}. 
The large spread of r-process elements seen in metal-poor stars indicates that the r-process nucleosynthesis in the early Universe should have been rare and prolific \citep[see][]{cescutti2015role,Hirai_2015,Wehmeyer2015, Naiman2018}.\\
In literature Eu is often indicated as a good r-process tracer for two basic reasons: i) more than 90$\%$ of Eu in the solar system has been produced by r-process \citep{cameron1982elemental, howard1986parametric, 2015MNRAS.449..506B, Prantzos2020}. ii) Europium is one of the few r-process elements that shows clean atomic lines in the visible part of the electromagnetic spectrum, and this makes Eu abundances easier to measure than other r-process elements \citep{woolf1995r}.\\
For the Eu production, have been proposed two main astrophysical sites: i) core-collapse SNe, Type II SNe during explosive nucleosynthesis \citep{cowan1991r, woosley1994r, wanajo2001r}. However, there are still many uncertainties in the physical mechanism involved in Eu production in Type II SNe \citep{Arcones2007}. On the other hand, some specific CC SNe, such as magneto-rotationally driven (MRD) supernovae \citep[][]{winteler2012magnetorotationally,Nishimura2015, 2015Natur.528..376M} and collapsars \citep[][]{Siegel_2019} have been indicated as promising sources of r-process nucleosynthesis. ii) Neutron star mergers (NSM), are able to provide a production of r-process elements rare and prolific \citep{freiburghaus1999r,Wanajo_2014,2008AstL...34..189P,1982ApL....22..143S, 2007A&A...467..395O, 2013ApJ...773...78B, 2013PhRvD..88d4026H, 2014MNRAS.443.3134P, Perego2021hgwa.bookE..13P}, as suggested by the wide scatter of [Eu/Fe] ratios at low metallicities. In particular, each event can produce a total amount of Eu from $10^{-7}$ to $10^{-5}$ $\sunmass$ as suggested by \citet{korobkin2012astrophysical}.
The detection of GW170817 by LIGO/Virgo and its electromagnetic counterpart \citep[see][]{Ciolfi2020}, known as kilonova, is the strongest proof that NSMs are able to produce $r$-process elements \citep[][]{abbott2017gw170817}. In particular, the ultraviolet, optical, and infrared emissions from GW170817 suggest a significant production of $r$-process elements \citep[][]{Cowperthwaite2017,Tanaka2017,Villar2017, Watson2019Natur.574..497W,Troja2019}. Moreover, its late-time infrared emission suggests the production of lanthanide elements (such as Eu).\\
\citet{argast2004neutron} tried to define which astrophysical site is the sole (major) producer of r-process elements. In particular, they computed the chemical evolution of Eu for the halo of our Galaxy with an inhomogeneous chemical evolution model and they concluded that NSMs cannot be the major producer of Eu due to their low rate. Later \citet{cescutti2006chemical} used a model with instantaneous mixing and found that SNe II can be the only r-process site in our Galaxy.\\
On the other hand, \citet{matteucci2014europium} showed that neutron stars (NS) can be the only production site of Eu but the time scale of coalescence (i.e. the time between the formation of the binary system of neutron stars and the merging event) cannot be longer than 1 Myr. Later on, with similar assumptions on the NSM parameters, \citet{cescutti2015role} showed that, in the framework of a stochastic chemical evolution model, NSM (alone) can reproduce the spread of [Eu/Fe] ratios observed in the halo of our Galaxy. However, they also showed that the scenario which best reproduces the observational data is the one where both NSMs and a fraction of Type II supernovae produce Eu.\\
All these studies suggested that, to reproduce both average values and spread of [Eu/Fe] as a function of [Fe/H] in our Galaxy, the coalescence time of NSM cannot be longer than 100 Myr. This assumption is in disagreement with other studies: i) population synthesis models \citep[][]{Belcynski2020,cote2019neutron, GiacobboMapeli2018b} that predicts a delay time distribution (DTD) in the form of a power-law (i.e. $\propto t^{-1.0}$ and $\propto t^{-1.5}$); ii) DTD and cosmic rate of SGRBs \citep[see][]{berger2014short, d2015short}; iii) detection of the event GW170817 occurred in an early-type galaxy \citep[][]{abbott2017gw170817}. We will take an in-deep discussion of these topics in the next section.\\
A series of recent studies has taken into account the effects of a DTD for NSM on the chemical evolution of r-process elements for both disk \citep[][]{cote2019neutron, simonetti2019new,  molero2021predicted} and halo \citep[][]{cavallo2021} environments. They concluded that when we take into account a DTD $\propto t^{-1}$ for NSMs, and we assume that these systems are the major producer of Eu, we cannot explain the average trend nor the wide spread of [Eu/Fe] in our Galaxy. In order to solve this tension, we need to invoke a second source of Eu (standard CC SNe, MRD SNe or collapsars). For the halo environment, is also possible to ease the tension between models and observations by predicting that the fraction of massive stars that can generate a binary system of neutron stars, which will eventually merge, is higher at low metallicities ([Fe/H]<$-2.5$ dex) \citep[see][]{simonetti2019new, cavallo2021}. \\
Other authors have investigated this tension and they have found different explanations. In particular, \citet{schonrich2019chemical} showed that NSM alone (with a DTD) can explain the observed abundance patterns assuming a 2-phase interstellar medium (hot and cold).\\ 
Other possible explanations can arise from the formation history of the Galactic halo. The hierarchical nature of galaxy formation in the $\Lambda$CDM cosmological paradigm, predicts that galaxies are surrounded by diffuse stellar halo components, formed by the accretion and disruption of lower mass galaxies. The early discovery of the Sagittarius stream around the Milky Way \citep[][]{Newberg2002, Majewski2003}, and later on the field of streams \citep[][]{Belokurov2006} are strongly favouring this overall picture of the formation of stellar haloes.\\
Both observations \citep[][]{Frebel2010, Ivezic2012, Aguado2021} and theoretical models \citep[][]{Cooper2010, Bullock2005, Johnston2008} point to the fact that the dwarf satellite galaxies that we observe nowadays are the remnants of a large population of ancient galaxies. Later on, these proto-galaxies merged to form the Galactic halo stellar population, although a fraction of halo stars have formed in situ.  
Recent observational surveys, such as Gaia \citep[][]{Gaia2016, Gaia2018}, are revolutionizing our understanding of the formation and evolution of the Galaxy and its stellar halo. In particular, the discovery of the Gaia-Sausage-Enceladus population of highly eccentric stars confirms a past merging event in the formation history of the MW. This population was brought in by a massive dwarf galaxy which formed the main component of the inner halo \citep[][]{Belokurov2018, Helmi2018, Fattahi2019, Vincenzo2019MNRAS.487L..47V}.\\
In the framework of the hierarchical clustering formation, \citet{Komiya2016} explored the effect of propagation of NSM eject across proto-galaxies on the r-process chemical evolution. Considering these effects, they found that NSMs alone with a DTD can reproduce the emergence of r-process elements at very low metallicity ([Fe/H] $\sim$ -3 dex).\\
In this paper, we want to resolve the tension that emerged in our last work \citet{cavallo2021}. In particular, we found that our previous model was not able to predict the presence of stars at [Fe/H]$<-3$ with [Eu/Fe]$<0$, the so-called "low metallicity tail" of abundance distribution of our data sample \citep[][]{roederer2014search}. In our previous model, we considered the Galactic halo as a unique spheroidal galaxy where all the stars are formed in situ. In this work, we change this paradigm and we consider that the halo has been formed by the subsequent accretion of smaller satellite galaxies, whose chemical evolution is computed. To do that we adopt a stochastic chemical evolution model, as proposed in \citet{cescutti2008}, that mimics an inhomogeneous mixing thanks to stochastic modelling. With this change of framework we need to take into account some "new" effects: i) the star formation efficiency varies among satellite galaxies of different mass; ii) the timescale of star formations is shorter in the least massive galaxies (due to the SNe feedback on less bounded gas) iii) the inefficient enrichment of Eu (due to NS binaries dynamic in the gravitational potential of their host galaxies) implemented following the results of \citet{bonetti2019neutron}.\\
The paper is organised as follows: in Section \ref{observational_evidence} we describe the observations; in Section \ref{CEM} we introduce the adopted chemical evolution model. In Section \ref{results} we discuss our results and finally in Section \ref{Conclusions} we draw some conclusions. 

\section{OBSERVATIONAL EVIDENCE} \label{observational_evidence}
Here we present observational evidence and their theoretical and phenomenological interpretations that can give some insights on the physics behind the possible r-process sites.
\subsection{[Eu/Fe] of metal-poor stars in the Galactic halo} \label{[Eu/Fe] halo}
The average [Eu/Fe] ratio in the Galaxy resembles the one of $\alpha$-elements, but it shows a wide spread at low metallicities (i.e. in the early evolutionary stage of the Galaxy). This wide scatter could indicate an interstellar medium not yet well mixed, that allows us to observe the chemical enrichment of single events. As discussed in Section \ref{intro}, the scatter of [Eu/Fe] is wider than the one observed for [$\alpha$/Fe] ratios, at the same metallicity. Moreover, the [$\alpha$/Fe] scatter shrinks more at lower metallicities than the one of [Eu/Fe] ([Fe/H] $\sim$ -3 and $\sim$ -2, respectively). This different behaviour seems to indicate that r-process events occur at a lower rate than supernovae. Similar conclusions have been found by \citet{cescutti2008}. In particular, they suggested that the wider spread observed in neutron-capture elements, compared to [$\alpha$/Fe] ratios, is a consequence of the difference in mass ranges between the production sites of $\alpha$ and $r-$process elements, respectively. Again, this also implies that the production of Eu, in the early Universe, must have been rare and prolific compared to the one of $\alpha$-elements.\\
To test the prediction of our model we use abundances of the halo stars contained in the JINAbase database \citep[][]{Abohalima_2018}. We excluded all upper limits, duplicates and, carbon-enhanced metal-poor (CEMP) stars. We have got left with 357 halo stars at $-4.0\leq$ [Fe/H] $\leq-0.5$ with $-0.8\leq$ [Eu/Fe] $\leq2.0$ \citep[see][]{MCW95, BUR00, FUL00, WES00, AOK02d, JOH02a, CAY04, CHR04, HON04, AOK05, BAB05, BAR05, IVA06, MAS06, PRE06, FRE07b, LAI08, HAY09, MAS10, ROE10, HOL11, HON11, ALL12, HAN12, AOK13, COH13, ISH13, LI13, MAS14, PAL14a,roederer2014search, SIQ14, SPI14, HAN15, JAC15, LI15b, LI15a}.\\
From the distribution on the [Fe/H]-[Eu/Fe] plane of the selected stars we note a group of six objects at [Fe/H]$>-2.5$ with [Eu/Fe]$>1.5$. The members of the group are: 2MASS J21054509-1836574 \citep{MCW95}, HE 2122-4707 \citep{COH13}, 2MASS J11074950+0011383 \citep{BAR05}, 2MASS J03274362-2300299 \citep{AOK02d}, 2MASS J21374577-3927223 \citep{BAB05}, and SMSS J175046.30-425506.9 \citep{JAC15}. The first five of them are reported to have [C/Fe]$>0.7$ (i.e. CEMP stars by definition) but are reported as ordinary stars in JINAbase. These stars will be plotted with open dots. The last star, SMSS J175046.30-425506.9, has [C/Fe]$=0.31$ \citep{JAC15} and is correctly reported as an ordinary star.

\subsection{Short Gamma-Ray Bursts} \label{SGRB}
The delay time between the formation of the binary system of neutron stars and the merging event plays a crucial role in the chemical evolution of Eu. Multiple observations and theoretical works have ruled out the hypothesis of short and constant delay times for NSMs. GRBs observations can provide a tool to constrain the typical time-scale of this delay.\\
Gamma-ray bursts display a bimodal duration distribution with a separation between the short and long-duration bursts at about 2 seconds. The progenitors of Long GRBs have been identified as massive stars \citep{Fruchter2006Natur.441..463F,Woosley2006ARA&A..44..507W,Wainwright2007ApJ...657..367W,Roy2021Galax...9...79R}. On the other hand, SGRBs are thought to be correlated with compact object mergers \citep{berger2014short, 1989Natur.340..126E, 2013Natur.500..547T}. This hypothesis has been recently reinforced by the observation of a SGRB \citep[see][]{2017c, Goldstein2017, Savchenko2017}, that followed the NSM event GW170817 detected by LIGO/Virgo Collaboration \citep{abbott2017gw170817}. In particular, NGC 4993, the host galaxy of GW170817, is an early-type galaxy \citep{abbott2017search, coulter2017swope} where present-day star formation is negligible.\\
In light of this, the detection pattern of SGRBs seems to provide possible constraints of delay times of neutron star mergers. Furthermore, SGRBs can connect the timescales of SNe Ia and NSMs. In fact, as for SGRBs, a fraction of SNe Ia is also observed in early-type galaxies \citep[][]{Li2011}. Using the $\sim$ 274 classified SNe Ia found in the Lick Observatory Supernova Search (LOSS), between $\sim$ $15\%$ and $\sim$ $35\%$ of SNe Ia occur in early-type galaxies \citep[][]{Leaman2011}. Similar percentages have been found, using the 103 SNe Ia with classified host galaxies from the Carnegie Supernova Project by \citet{krisciunas2017carnegie}. These percentages for SNe Ia are comparable with the ones found for SGRBs with classified host galaxies \citep[see][]{berger2014short}. This suggests that, on average, SGRBs and SNe Ia occur on similar timescales.\\
\citet{Fong_2017} found that the delay time distribution (DTD) of SGRBs has a power-law slope $\propto t^{-1}$. A similar conclusion has been found from population synthesis studies \citep[see][]{Dominik_2012,chruslinska2018double}. Moreover, a power-law with a -1 slope is similar to the one derived for SNe Ia \citep[see][]{totani2008delay, maoz2010supernova, graur2011supernovae, maoz2012type, rodney2014type}. This similarity is also consistent with the fact that the SNe Ia and SGRBs are detected in similar proportions in early-type galaxies. However, if we assume this DTD for NSMs (assuming that are the major producers of Eu) these systems cannot reproduce both the spread and the decreasing trend of [Eu/Fe] in our Galaxy \citep[][]{cote2019neutron, simonetti2019new, cavallo2021}.\\
\citet{d2015short} have derived a DTD for SGRBs with a steeper form (i.e. $\propto t^{-1.5}$). However, this slope is in disagreement with the fact that the fraction of SGRBs and SNe Ia, observed in different galaxies, look similar. This tension could be eased if the SNe Ia follow a DTD with a similar slope \citep[as suggested by][]{heringer2016type}. On the other hand, in the environment of a chemical evolution model, SNe Ia with such a slope are not able to reproduce all the [X/Fe] vs [Fe/H] trends in the Galaxy since, with a DTD $\propto t^{-1.5}$, the explosion time-scales of SNe Ia are too short \citep[see][]{matteucci2006}.

\subsection{The delay time distribution for Neutron Star Mergers}\label{DTD for NSM}
In a galaxy, the rate of NSM (i.e. the number of NS-NS merging events per time and volume) depends on the star formation history (SFH) of the galaxy and the delay time distribution of NSMs.\\
As discussed in Section \ref{SGRB}, SGRBs seem to provide a possible way to characterise the DTD of NSMs. In particular, these studies indicate a DTD that scales with the inverse of the delay time \citep[see][]{Guetta2006,Virgili2011,Davanzo2014,Wanderman,Ghirlanda2016}. \\
The DTD of NSMs can be also computed numerically with the binary population synthesis (BPS) models \citep[see][]{Tutukov1993,Dominik_2012,mennekens2014massive,giacobbo2018progenitors,Tang2020}. These models follow the evolution of individual stellar populations, where all the stars have been formed at the same time and with the same metallicity. In particular, they follow the evolution of a primordial population of neutron star binaries through the turbulent phases of mass exchanges till the merging event. Lots of these studies, predict that the delay times of NSMs should follow a power-law $\propto t^{-1}$. 
On the other hand, \citet{Greggio2021} have determined the DTD of NSMs with an alternative approach, similar to the one developed by \citet{Greggio2005} for the rate of SNe Ia. In particular, they showed that the DTD of NSMs cannot be described as a simple power law. At short delay times, the DTD should be characterised by a plateau, with a width equal to the difference between the evolutionary lifetimes of the least and most massive NS progenitors that we assume (in this case 9-50 $\sunmass$).\\
Furthermore, as already reported above, the slope of the DTD at long delay times depends on the shape of the distribution of the separations of the DNS systems at the time of the formation \citep[for more details, see][]{Greggio2021}.

\section{THE CHEMICAL EVOLUTION MODEL} \label{CEM}
The chemical evolution model adopted here is an updated version of the one used in \citet{cescutti2015role, cavallo2021}, which is based on the stochastic model developed by \citet{cescutti2008}. We review its main characteristics to improve the reader comprehension of the work. 
\subsection{Population of synthetic proto-galaxies} \label{Pop of galaxies}
In our previous model, the Galactic halo was modelled as a single massive spheroidal galaxy, in which all the stars have been formed in situ. In this work, we change this concept assuming a different formation history for the Halo. In particular, we assume that the Galactic halo has been formed by the accretion and disruption of lower mass galaxies. In the light of this, we need to generate a synthetic population of primordial satellite galaxies with given mass distribution, in agreement with the one suggested by cosmological models for the formation of MW-like haloes. In particular, \citet{Fattahi2020} have retrieved the mass distribution function of a proto-galaxy population that can form a MW-like halo, from a cosmological magneto-hydrodynamical (MHD) simulation. \\
To generate the mass of our synthetic galaxies we proceed as follows: we extract the mass randomly with the following cumulative distribution function (CDF):
\begin{equation} \label{alpha}
N(>M_{\rm gal}) = \begin{cases} 10^{-0.5\: \log(M_{\rm gal}) + 4.5} & \mbox{if } 10^5 \leq M_{\rm gal} \leq 10^9 \: \sunmass \\ 
0 & {\rm otherwise} 
\end{cases}
\end{equation}
and we continue to generate galaxies until we reach a total stellar mass equal to $M_{\rm halo}= \num{1.3e9}$ $\sunmass$ \citep[][]{Mackereth2020}. \\
The CDF that we used is a simplification of the one presented by \citet{Fattahi2020}. We make this approximation, because in this work we try to explore the overall effects of a formation history for the Galactic halo, on the results of our stochastic chemical evolution model.\\
The Galactic halo is simulated by means of N stochastic realisations. Each of them consists of a non-interacting region with the same typical volume. For typical interstellar medium (ISM) densities, a supernova remnant becomes indistinguishable from the ISM, for distances larger then $\sim$ 50 pc \citep[][]{thornton1998energy}. The size of our volumes should be large enough to include the whole supernova bubble, and at the same time sufficiently small to not lose the stochasticity (large volumes produce more homogeneous results). In light of this, we chose a typical volume of $\num{8.2e6}$ pc$^3$ with a typical radius of $\sim 125$ pc.
To each proto-galaxy it has been assigned a $N_{\rm volumes}$ number of stochastic volumes. In particular, at the $i^{\rm th}$ galaxy of mass $M_{\rm gal}^i$ have been assigned a $N_{\rm volumes}$ number of stochastic volumes proportional to its contribution to the total mass .\\\\
\noindent
In each stochastic volume, we need to compute the chemical evolution of the system. The volumes are hosted by proto-galaxies of different masses and with different star formation histories. In particular, this new environment (compared to the one used in our previous work) requires new assumptions on both star formation efficiency (SFE) and star formation timescales.\\

\subsection{Star formation efficiency} \label{Star formation efficiency}
Observational mass-metallicity relation suggest a correlation between stellar masses and mean metallicity of galaxies: $10^{\rm [Fe/H]} \propto M_{*}^{0.3}$ \citep[][]{Kirby2013}. This relation may indicate a lower SFE in lower mass galaxies \citep[][]{Calura2008A&A...484..107C, Ishimaru2015}. However, we should point out that \citet{Tremonti2004ApJ...613..898T} have suggest that the mass metallicity relation reported above can be explained by the effect of the galactic winds that remove more easily metals from low mass galaxies. In this work, we decided to implement a SFE that varies among proto-galaxies of different masses.\\
In each region, we assume the same infall law as in our previous model, based on the homogeneous model by \citet{chiappini2008new}.\\
We define the star formation rate as:
\begin{equation} \label{eqn:SFR}
    {\rm SFR}(t) = \nu \left( \frac{\sigma_{\rm gas}(t)}{\sigma_{\rm h}} \right)^{1.5}
\end{equation}
where $\sigma_{\rm gas}(t)$ is the surface density of the gas inside a volume at a certain time $t$, $\sigma_{\rm h}=80$ $pc^{-2}$ and $\nu$ is the SFE. In each volume, according to the mass of the host galaxy, we assume the same SFR law (see above) with different star formation efficiency ($\nu$). In particular, we assume a SFE $\propto \left( M_{\rm gal} \right)^{0.3}$ \citep[see][]{Komiya2016}. Then, we re-scale the SFE ($\nu_0$) that we used in \citet{cavallo2021}, obtaining the following relation:
\begin{equation} \label{eqn:SFE}
    \nu (M_{\rm gal}) = \nu_0 \left( \frac{M_{\rm gal}}{M_{\rm halo}} \right)^{0.3}
\end{equation}
where $M_{\rm gal}$ is the mass of the proto-galaxy that hosts the stochastic realisation, $M_{\rm halo}=\num{1.3e9}$ $\sunmass$, and $\nu_0 = \num{2.9e-3}$ yr$^{-1} $\\   
We also take into account an outflow of enriched gas that follows the law:
\begin{equation}
    \frac{d{\rm Gas}_{\rm out}(t)}{dt} = Wind \times {\rm SFR}(t)
\end{equation}
where the efficiency of the galactic wind ($Wind$) is set to 8 \citep{LanfranchiMatteucci2003MNRAS.345...71L}.\\
The model generates a sequence of stars where each stellar mass is selected with a random function, weighted on the initial mass function (IMF) of \citet{scalo1986stellar} in the mass range from 0.1 to 100 $\sunmass$. The sequence stops when the total mass of newborn stars reaches $M^{\rm new}_{\rm stars}$ that is the mass transformed at each time-step into stars, determined by the star formation rate. In this way, the total amount of mass transformed into new stars is the same in each region (with the same SFR) at each time step, but the total number and mass distribution of stars are different and thus the stochasticity of the model. For all the stars we also know mass and lifetime. In particular, we assume the stellar lifetimes of \citet*{maeder1989grids}.\\
When a star dies, it enriches the ISM with its newly produced elements and with the unprocessed elements present in the star since its birth. Our model considers detailed pollution from SNe core-collapse (M $> 8\sunmass$), AGB stars, NSMs, and SNe Ia, for which we follow the prescriptions for the single degenerate scenario of \citet{1986A&A...154..279M}. The iron yields for both SNe II and SNe Ia are the same as \citet{cescutti2006chemical}.\\\\
\noindent
Another important parameter, that greatly affects the chemical evolution, is the time at which the star formation in the halo stops ($t_{\rm SF}$).\\
The model uses time-steps of 1 Myr, which is shorter than any stellar lifetime considered in this model; the minimum lifetime is, in fact, 3 Myr for an 100 $\sunmass$ star, which is the maximum stellar mass considered. In our previous model, we assumed that the volumes stop their star formation all at the same time (i.e. 1 Gyr). In this work, we assume that $t_{\rm SF}$ varies among proto-galaxies of different masses.

\begin{figure*}
    \centering
    \includegraphics[width=1.\textwidth]{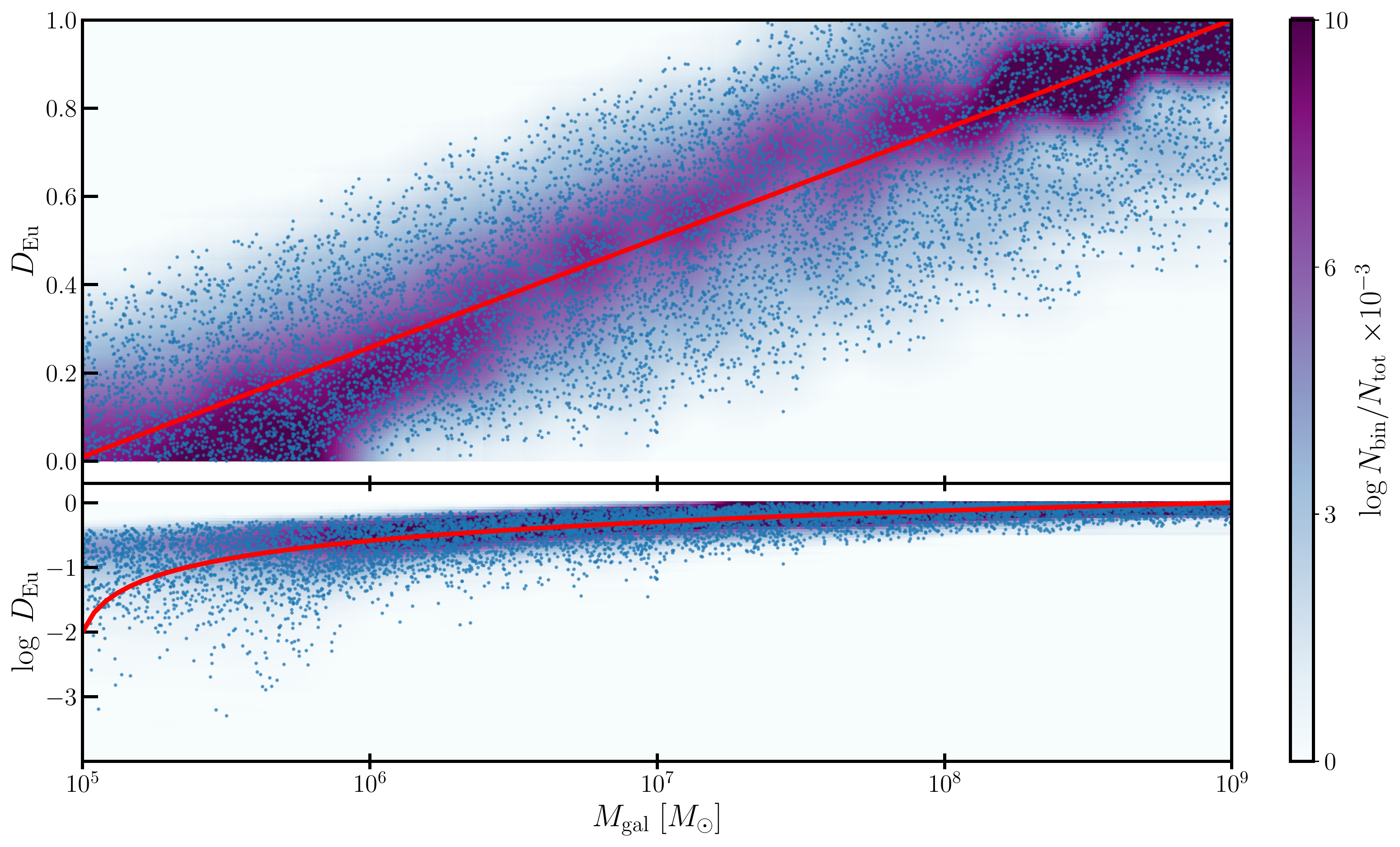}
    \caption{Density maps of $D_{\rm Eu}$ and $\log D_{\rm Eu}$ vs $M_{\rm gal}$ planes. With the red solid line is reported the linear relation between the average value of $D_{\rm Eu}$ as a function of the proto-galaxy mass. With the blue dots, we report $10^4$ random extractions of $D_{\rm Eu}$. We recall that, at given $M_{\rm gal}$ the values of $D_{\rm Eu}$ are distributed with a triangular distribution.}
    \label{fig:Eudilu}
\end{figure*}

\subsection{Timescale of star formation}
For galaxies like the Milky Way, the star formation is continuous process from $t=0$ up to $t_{\rm SF} \sim 14$ Gyr.
For smaller galaxies, the supernova feedback is expected to quench the conversion of gas into stars on shorter timescales. We decide to take into account this effect, by varying the timescale of star formation among proto-galaxies of different mass. In particular, we can expect that, in galaxies with lower masses (i.e. weaker gravitational potential), the SFR should quench (due to the supernovae feedback) in timescales shorter than in higher mass galaxies.\\
In the light of this, we develop two different empirical relations for $t_{\rm SF}$ as a function of the proto-galaxy mass. In the first, we use a bimodal approach;  the $t_{\rm SF}$ can assume two different values, based of the proto-galaxy's mass:
\begin{equation}
t_{\rm SF}(M_{\rm gal}) = \begin{cases} 250 \text{ Myr} & \mbox{if } 10^5 \leq M_{\rm gal} \leq 10^7 \: \sunmass \\ 
1000 \text{ Myr} & \mbox{if } 10^7 < M_{\rm gal} \leq 10^9 \: \sunmass\\
\end{cases}
\end{equation}
With the second one, we implement a smoother variation of $t_{\rm SF}$ over $M_{\rm gal}$ (the linear scenario): 
\begin{equation}\label{eqn:EoTlinear}
t_{\rm SF}(M_{\rm gal}) = \begin{cases} m \times \log \left( \frac{M_{\rm gal}}{M_1} \right) + q& \mbox{if } 10^5 \leq M_{\rm gal} \leq 10^9 \: \sunmass \\ 
0 & {\rm otherwise} 
\end{cases}
\end{equation}\\\\
\noindent
The chemical evolution of Eu into these lower mass galaxies is also influenced by another effect: the
inefficient enrichment of Eu due to NS binary dynamics in the gravitational potential of their host galaxy.

\subsection{Eu Nucleosynthesis} \label{Eunucleo}
For the nucleosynthesis of Eu we use, prescriptions similar to the ones used in \citet{cescutti2015role} and \citet{cavallo2021}. In particular, we use empirical values that have been chosen to reproduce the surface abundances of Eu in low-metallicity stars as well as the solar abundances of Eu \citep[see][]{cescutti2006chemical}. These values are also in agreement with the limits calculated by \citet{korobkin2012astrophysical} and also with suggestion that comes from other chemical evolution models \citep[][]{matteucci2014europium, molero2021predicted}.\\
We also consider a variable production of Eu from a single NSM event. The variation is unknown so we assume that a single NSM can produce an amount of Eu from 1$\%$ to 200$\%$ of an average value ($M_0^{\rm Eu}$). In particular, we assume that the $n^{\rm th}$ ejects a mass of Eu:
\begin{equation} \label{eqn:RandoEu}
M_{\rm NS}^{\rm Eu}(n) = M_0^{\rm Eu} \left( 0.01 +1.98 \times Rand(n) \right)    
\end{equation}
where $Rand(n)$ is an uniform random distribution in the range[0,1]. This preserves the total mass of Eu produced. Note that the variable production of Eu has a weak impact on the model stochasticity that is mainly produced by the stochastic star formation in each realisation.\\
We also test an alternative production channel of Eu: the magneto-rotationally driven(MRD) SNe. These SNe are a particular class of CC SNe. Furthermore, these events are expected to be rare and only a small fraction of CC SNe explode as MRD-SNe \citep[see][]{winteler2012magnetorotationally}. As in our previous work, we assume that the 10$\%$ of the CC-SNe explode as MRD. Moreover, this r-process channel is active only at low metallicity \citep[CC-SNe are more likely to explode as MRD SNe at low metallicity, ][]{2006A&A...460..199Y}. These assumptions are identical to the ones contained in \citep[][]{cescutti2015role}.\\\\
\noindent
The enrichment of Eu from NSMs in a low mass galaxy can be less efficient than in a higher mass galaxy due to the dynamics of NS binary systems in the gravitational potential of their host galaxy. In particular, \citet{bonetti2019neutron} have found that the motion of the binary systems due to the kick imparted by the SN explosion of the secondary star, determines a merger location potentially detached from the host galaxy, even for gravitationally bound systems. The immediate consequence is a reduction of the amount of r-process material retained by the galaxy and, consequently a decrease of [Eu/Fe] ratios. This effect is more intense for low mass proto-galaxies due to their weaker gravitational potential. Our synthetic galaxies, accordingly to their masses, should be influenced by this effect, so we decide to take it into account. 

\subsubsection{Inefficient enrichment of Eu due to NS binary dynamics} \label{sec:eudilution}
First of all, we need to introduce some useful definitions. The Eu dilution\footnote[1]{Formally the effect dilutes [Eu/Fe]. From this point when we talk about "Eu dilution" we are referring to the dilution of [Eu/Fe]} is modelled by a simple multiplicative factor ($D_{\rm Eu}$) in the range [0,1]:
\begin{equation} \label{eqn:Eudilu}
    M_{\rm dil}^{\rm Eu}(n)=M_{\rm NS}^{\rm Eu}(n) \times D_{\rm Eu}(M_{\rm gal})
\end{equation}
where $M_{\rm dil}^{\rm Eu}$ is the diluted amount of Eu from the $n^{\rm th}$ NSM and $M_{\rm NS}^{\rm Eu}(n)$ is defined in Equation \ref{eqn:RandoEu}. We point out that, in general, $D_{\rm Eu}$ can depend on the mass of the $n^{\rm th}$ NSM host galaxy.\\
From \citet{bonetti2019neutron} we extract some valuable information that allows us to calculate $D_{\rm Eu}(M_{\rm gal})$. In the lowest mass proto-galaxies (i.e. $10^5 \leq M_{\rm gal} \leq 10^7$ $\sunmass$) $\sim 30\%$ of NSMs explode beyond the galactic contours and therefore do not enrich those galaxies with newly produced Eu. On the other hand, only a fraction ($\sim 0.2$; i.e. $20\%$) of the Eu produced by the $\sim 70\%$ of NSMs is retained. For larger galaxy masses (i.e. $10^7 < M_{\rm gal} \leq 10^9$ $\sunmass$), as a consequence of the deeper gravitational potential of the galaxy, a larger fraction ($\sim 95 \%$) of NSMs explode within the host galaxy. In addition, the retained amount of Eu increases to $\sim 0.7$. For simplicity, we can summarise all these concepts with the following relations:
\begin{center}
    $\text{in galaxies with }10^5 \leq M_{\rm gal} \leq 10^7$ $\sunmass$\\
    $\overbrace{ 70\% \mbox{ of NSMs} \mbox{ enrich with the } 20 \%  \mbox{ of produced Eu}}$\\
    $30\% \mbox{ of NSMs} \mbox{ do not enrich the galaxy}   $
\end{center}
\begin{center}
    $\text{in galaxies with }10^7 < M_{\rm gal} \leq 10^9$ $\sunmass$\\
    $\overbrace{ 95\% \mbox{ of NSMs} \mbox{ enrich with the } 70 \%  \mbox{ of produced Eu}}$\\
    $5\% \: \mbox{ of NSMs} \mbox{ do not enrich the galaxy}$
\end{center}
These relations, show the average effect of dilution based on the mass of the proto-galaxy. Now, we want to compute the dilution coefficient, contained in equation \ref{eqn:Eudilu}, aworks a function of the proto-galaxy mass. For simplicity, we assumed a linear relation between $D_{\rm Eu}$ and $M_{\rm gal}$. This linear relation should pass through (or at least near) the dilution coefficients retrieved from \citet{bonetti2019neutron}. The linear function is defined as follows:
\begin{equation}\label{eqn:Dilutionlinear}
D_{\rm Eu}(M_{\rm gal}) = m \times \log \left( \frac{M_{\rm gal}}{M_1}\right) + D_1
\end{equation}
this is simply the equation of a line passing through two points, where $m=\frac{D_2-D_1}{\log{M_2/M_1}}$; $ P_1( \log(M_1), D_1 )$ and $ P_2( \log(M_2), D_2 )$.
As already introduced, these coefficients and relations describe the average fraction of Eu retained by a galaxy of a certain mass, but we can take a step further. As described in Section \ref{CEM}, our stochastic chemical evolution model follows the evolution of stars from birth to death, so we can incorporate the fraction of Eu retained by a galaxy for every single NSM event that occurs in the simulation. For example, in a galaxy with $M_{\rm gal}=10^6\: \sunmass$ the dilution fraction is (on average) $D_{\rm Eu}(M_{\rm gal}=10^6\:\sunmass)\sim0.2$; but single NSM events can enrich a larger or lower fraction of that average value. To model this possibility, for every NSM that happens in the simulation (of which we know the mass of the host galaxy) we first check if it will enrich the galaxy, with the same probabilities mentioned above, and then we extract a random value of $D_{\rm Eu}$. To this stochastic process, we associated a triangular distribution, a continuous probability distribution that depends on three parameters: the lower limit $a$, the upper limit $b$, and the mode $c$. The triangular distribution has the following general form:
\begin{equation} \label{triangular}
\begin{split}
{\rm PDF}(x) = &
\begin{cases} 
\frac{2(x-a)}{(b-a)(b-c)} & \mbox{if }a\leq x<c \\
\frac{2}{b-a} & \mbox{if } x=c\\
\frac{2(b-x)}{(b-a)(b-c)} & \mbox{if }c< x\leq b \\
\end{cases}
\end{split}
\end{equation}
with a mean value $\mu=\frac{a+b+c}{3}$. In Figure \ref{fig:Eudilu} are reported density maps and $10^4$ random extractions of $D_{\rm Eu}$ as a function of galaxy mass (blue dots). In the figure is also reported the linear relation described by equation \ref{eqn:Dilutionlinear} (red solid line). From Figure \ref{fig:Eudilu} (especially in the \textit{bottom panel}) is seen that in the least massive galaxies ($M_{\rm gal}<10^6 \: \sunmass$) we have a non negligible probability that a NSM enrich only the $1\%$ (or even lower) of the newly produced Eu. 

\begin{table*}
\centering
\begin{tabular}{c| c c c c c c c c}
  & & & & & \\
  Model Name &  DTD & $\alpha_{\rm NS}$ & $D_{\rm Eu}$ & $t_{\rm SF}$ & $M^{\rm Eu;NSM}_0$ [$\sunmass$] & $M^{\rm Eu;MRD}_0$ [$\sunmass$] \textsuperscript{$\Psi$} \\ 
  & & & & & & \\
  \hline
  \hline
  & & & & & & \\
  NSt3\textsuperscript{$\chi$} & $\propto t^{-1.5}$ & 0.02 & no & constant (1 Gyr) & $\num{4.0e-6}$ (varying as eq. \ref{eqn:RandoEu}) & no production\\
  MSFR & " & 0.005 & " & " & $\num{1.6e-5}$ (varying as eq. \ref{eqn:RandoEu})& "\\
  & & & & & \\
   \hline
  & & & & & \\
  M10 & $\propto t^{-1.5}$ & 0.005 & linear & bi-modal & $\num{1.6e-5}$ (varying as eq. \ref{eqn:RandoEu})& no production\\
  M11 & " & " & linear & linear & " & "\\
   & & & & & \\
   \hline
  & & & & & \\
  HC0 & $\propto t^{-1.5}$ & 0.005 & linear & linear & $\num{1.6e-5}$ (varying as eq. \ref{eqn:RandoEu}) & no production \\
  HC1 & " & 0.01 & " & " & $\num{8.0e-6}$ (varying as eq. \ref{eqn:RandoEu}) & " \\
  HC2 & " & 0.02 & " & " & $\num{4.0e-6}$ (varying as eq. \ref{eqn:RandoEu}) & " \\
   & & & & & \\
   \hline
  & & & & & \\
  MRD0 & $\propto t^{-1}$ & 0.005 & linear & constant (1 Gyr) & $\num{2.0e-5}$ (varying as eq. \ref{eqn:RandoEu}) & $\num{0.5e-6}$ (varying as eq. \ref{eqn:RandoEu})\\
  MRD1 & $\propto t^{-1.5}$ & " & " & " & $\num{1.4e-5}$ (varying as eq. \ref{eqn:RandoEu}) & $\num{0.3e-6}$ (varying as eq. \ref{eqn:RandoEu}) \\
  & & & & & \\
  \end{tabular}
  \caption{This table summarises the parameters of the models that we test during this work. It is organised as follows: in column 1, name of the model, in column 2, assumed DTD for coalescence time, in column 3, assumed fraction of massive stars that could lead to NSM, in column 4, regime of the law used for the Eu dilution effect, in column 5, regime of the law used for the timescale of the quenching of star formation, in column 6, assumed yield for NSM, in column 7, assumed yield for MRD SNe.\\
  $^{\chi}$ From \citet{cavallo2021}. This model will ease the comparison between old and new results.\\
  $^{\Psi}$ When we take into account the Eu production by MRD-SNe we set $\alpha_{\rm MRD}$=0.10.}
  \label{tab:Models}
\end{table*}

\subsection{Delay times of NSMs}
In this work, we used power-law DTD functions defined as follow \citep[see][]{cavallo2021}: 
\begin{equation} \label{DTDs}
\begin{split}
{\rm DTD}(t) = &
\begin{cases} 
0 & \mbox{if } t<t_{\rm min}^{\rm c}\\ 
A_x t^{-x} & \mbox{if }t_{\rm min}^{\rm c} <t<10 \mbox{ Gyr} \\
0 & \mbox{if } t>10 \mbox{ Gyr}\\
\end{cases}\\
&{\rm with} \, x=\{1,1.5\}\, and \, A_x = 1/ \int  \tau^{-x} d\tau;
\end{split}
\end{equation}
where $t_{\rm min}^{\rm c}$ is the minimum coalescence time (in this work will be always set to $1$ Myr), and $A_x$ is the normalisation constant. We choose a maximum delay time of 10 Gyr that not include all the coalescence timescales of NS-NS systems derived in \citet{tauris2017formation}. In order to include them we should choose a maximum delay time equal to $\infty$, without a significant impact on our model's results (it would have changed only the normalisation of the DTD).\\
We decide to test two different pure power-laws DTD with different slopes ($\propto t^{-1}$ and $\propto t^{-1.5}$).\\
In general, the rate NSMs at a certain time $t$ con be computed by means of the following equation:
\begin{equation}\label{eqn:rate}
    R_{\rm NSM} = \alpha_{\rm NS} k_{\alpha} \int_{t_{\rm min}}^t \psi(t-\tau)f(\tau) d\tau
\end{equation}
where $\psi(t)$ is the SFR, $f(\tau)$ is the DTD and, $t_{\rm min}$ is the minimum coalescence timescale.\\
$k_{\alpha}$ is the number of stars in the mass range of NS progenitors per solar masses. This parameter can be computed from the integration of the initial mass function (IMF) in the mass range of NS progenitors. In this work, we assumed that the progenitors of NS are in the mass range from 9 to 50 $\sunmass$ \citep[same as][]{matteucci2014europium}. {With this assumption and using a \citet{Salpeter1955ApJ...121..161S} IMF we get $k_{\alpha}=\num{6e-3}$ $\sunmass^{-1}$.}\\
On the other side, the parameter $\alpha_{\rm NS}$ is defined as the fraction of massive stars that generate a binary system of neutron stars that will eventually merge. The value of this parameter should be chosen in order to reproduce the present rate of NSM suggested by \citet{Abbott2021ApJ...913L...7A} ($\mathcal{R}=320^{+490}_{-240}$). From equation \ref{eqn:rate}, using the SFR density of \citet{MD2014} and a DTD $\propto t^{-1.5}$ we obtain that $\alpha_{\rm NS}$ between 0.0009 and 0.009. In this work, we assume $\alpha_{\rm NS}=0.005$ that is consistent with the error interval of the observed rate of NSM by \citet{Abbott2021ApJ...913L...7A}, and reproduces its median value.
\begin{figure*}
  \centering
  \includegraphics[trim={0cm 0cm 0cm 0cm}, clip, width=0.9\textwidth]{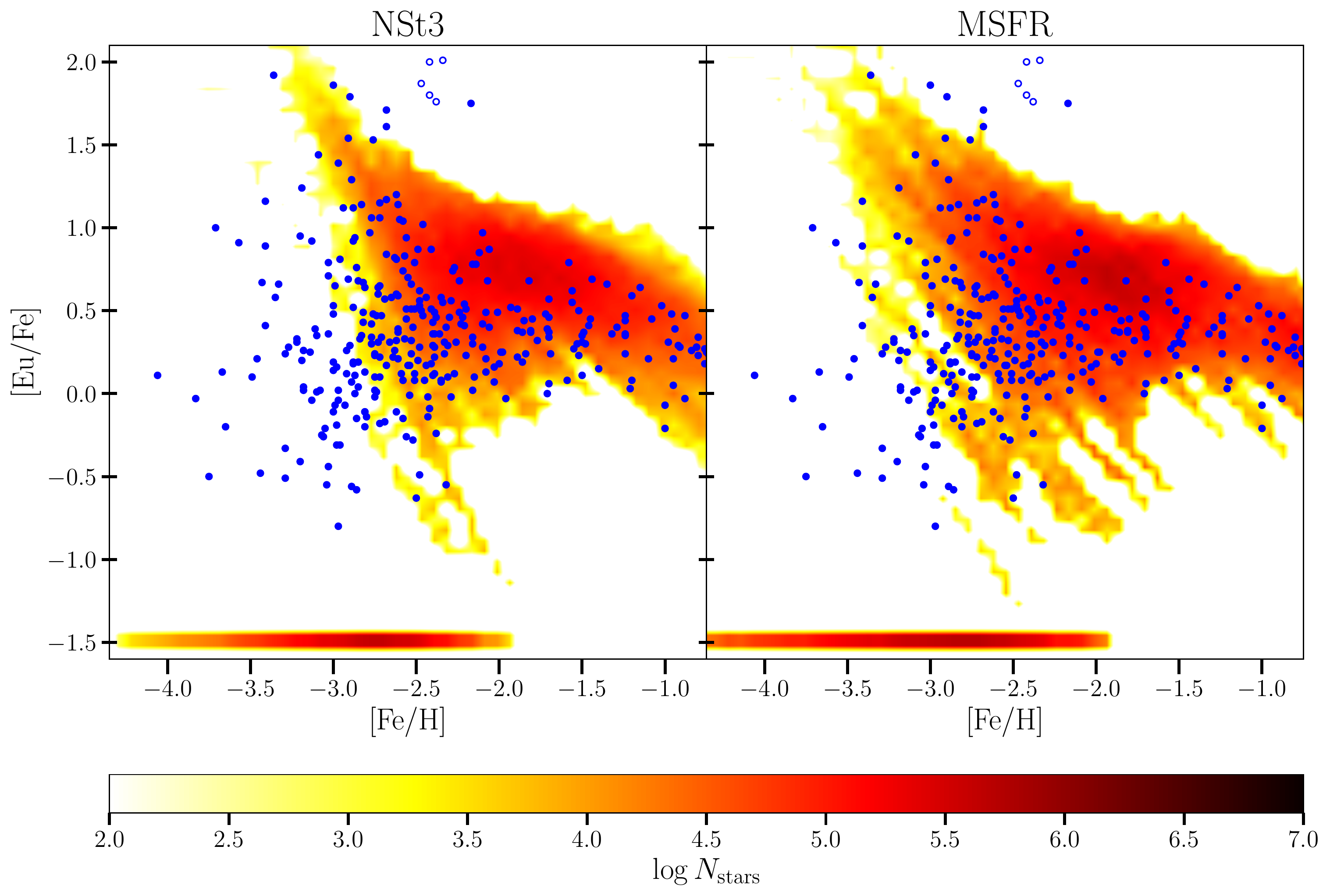}
  \caption{\textit{Left panel}: results of [Eu/Fe] vs [Fe/H] for model NSt3 contained in \citet{cavallo2021}. The density plot is the distribution of simulated long-living stars for our model (see the bar below the figure for the colour scale). The long-living stars formed without Eu (formally [Eu/Fe] = $-\infty$) are shown at [Eu/Fe] = $-1.5$ dex. The model predictions are compared to the halo stars contained in JINAbase (see Section \ref{[Eu/Fe] halo}) plotted with blue dots; we show as open dots the five CEMP stars discussed in Section \ref{[Eu/Fe] halo}. \textit{Right panel}: same as \textit{left panel} but for model MSFR. This model contains the modified SFRs assumed in our scenario. Note that we do not include the effects of dilution and quenching time of SF. }
  \label{fig:Model1}
\end{figure*}

\section{Results} \label{results}
\begin{figure*}
  \centering
  \includegraphics[trim={0cm 3.5cm 0cm 0cm}, clip, width=0.9\textwidth]{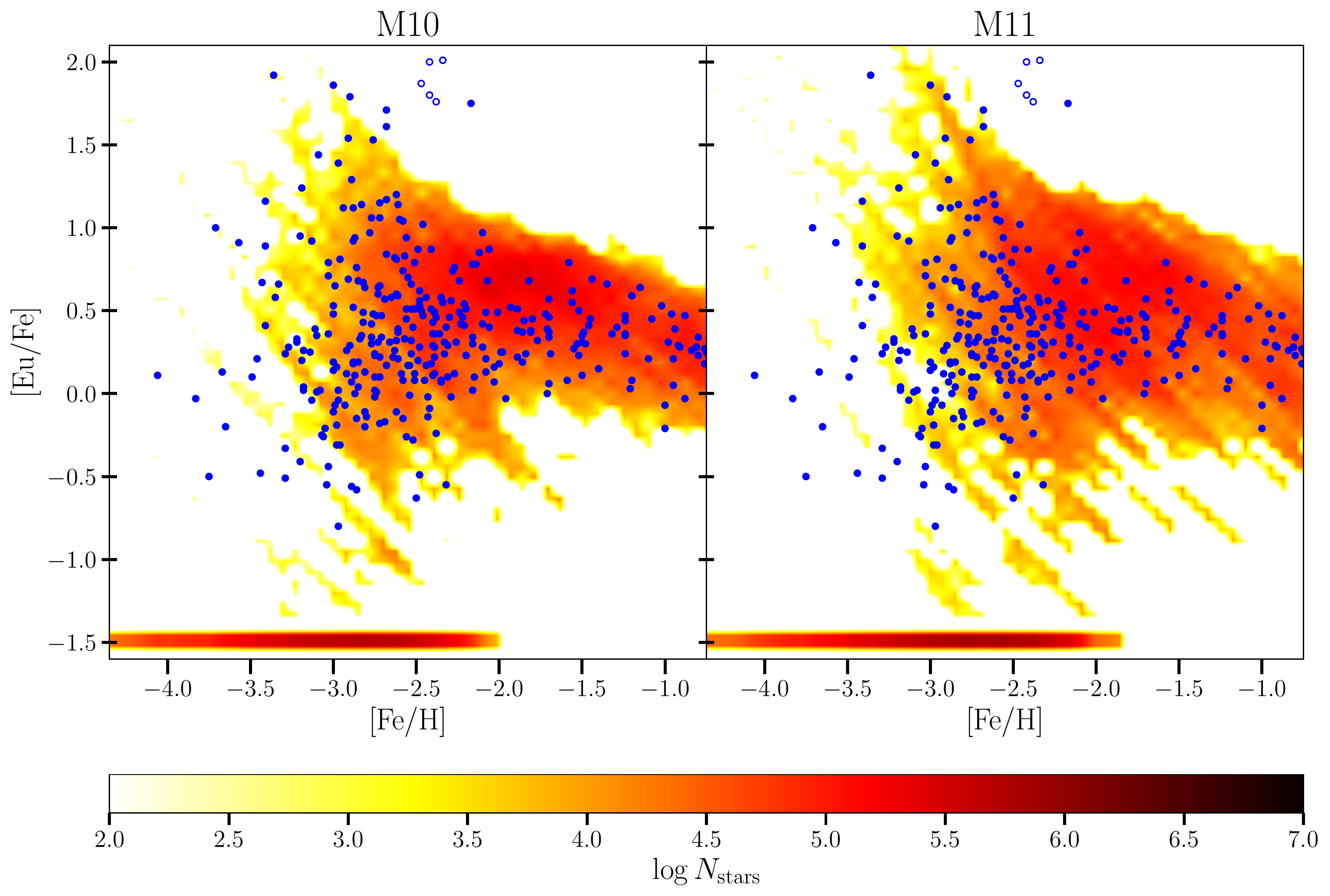}
  \caption{\textit{Left panel}: results of [Eu/Fe] vs [Fe/H] for model M10. \textit{Right panel}: same as \textit{left panel} but for model M11. In model M10 we included a $t_{SF}$ that can assume only two different values, based on the mass of the galaxy. On the other hand, model M11 assumes that the quenching time varies linearly with the galaxy mass.}
  \label{fig:Model2}
\end{figure*}

\begin{figure*}
\centering
  \includegraphics[trim={0cm 0cm 0cm 0cm}, clip, width=0.9\textwidth]{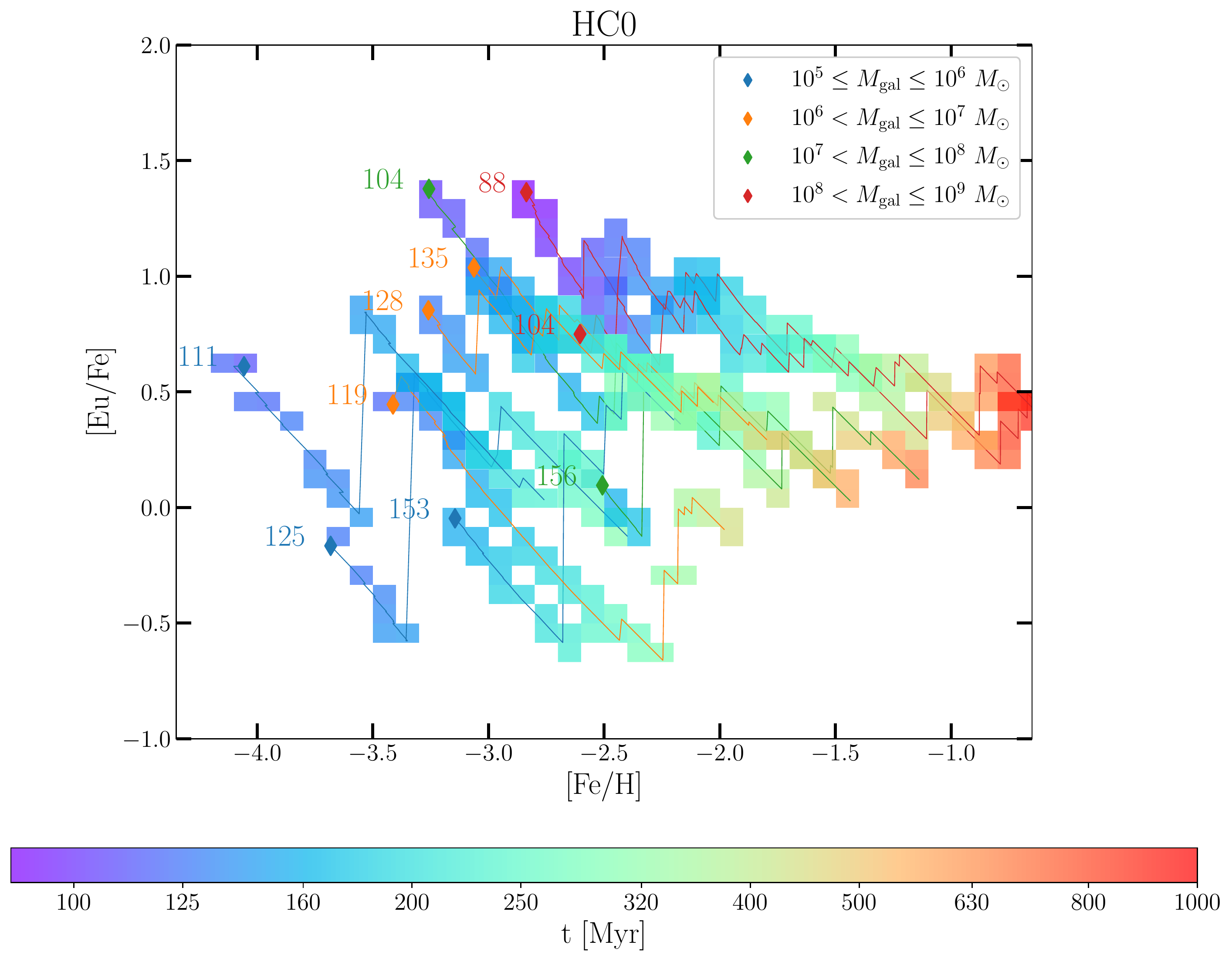}
  \caption{Results of [Eu/Fe] vs [Fe/H] for ten realisations of HC0 model. With the colour map, we show the time at which the realisation passes through a certain point in the [Eu/Fe] vs [Fe/H] plane. We also report the initial point and the time at which the first NSM exploded. With solid lines are reported the paths that each realisation follows. The path of each realisation is labelled with different colours that indicate the mass of the proto-galaxy to whom the realisation belongs.}
  \label{fig:paths} 
\end{figure*}

\begin{figure*}
\centering
  \includegraphics[trim={0cm 6cm 0cm 0cm}, clip, width=\textwidth]{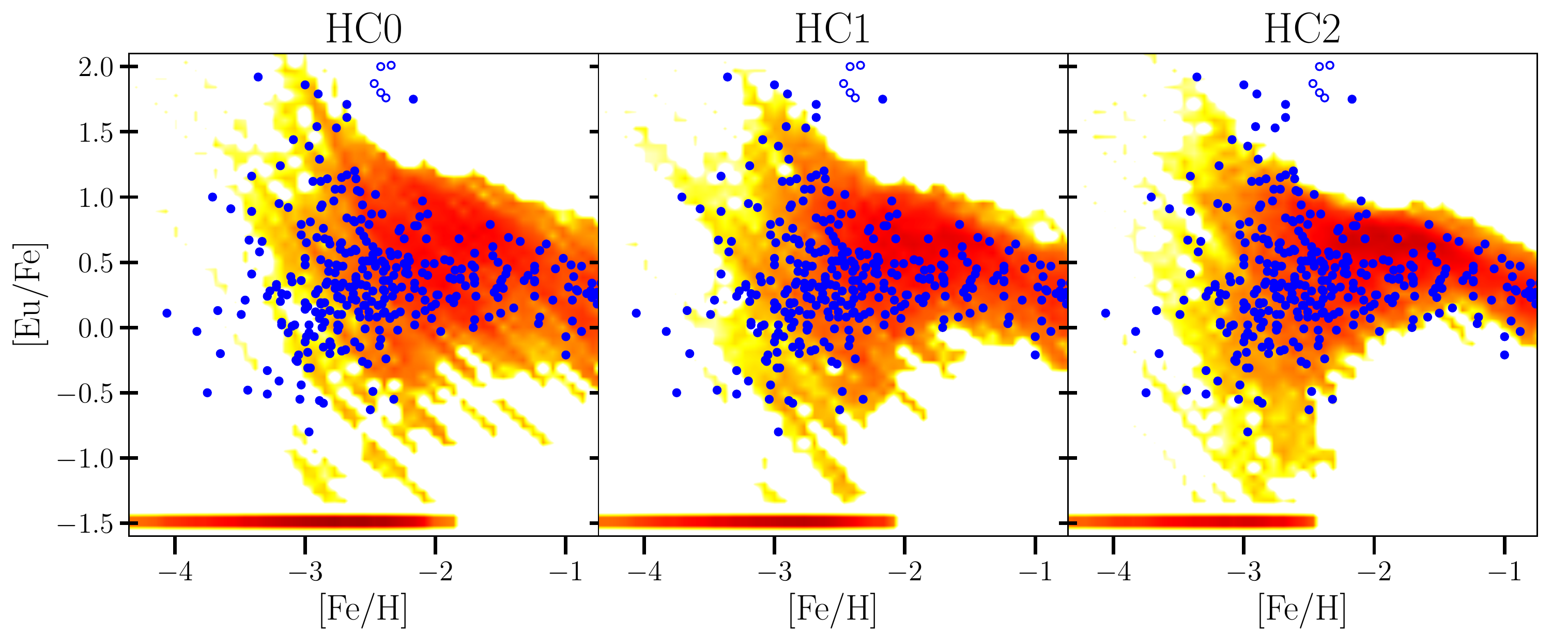}
  \caption{\textit{Left panel}: results of [Eu/Fe] vs [Fe/H] for model HC0. This model is identical to model M11. We plot it again to emphasise the consequences of the variation of $\alpha_{\rm NS}$. \textit{Central panel}: same as \textit{left panel} but for model HC1. In this case $\alpha_{\rm NS}=0.01$. To maintain constant the total amount of produced Eu, we reduce $M_0^{\rm Eu}$ to $\num{8.0e-6} \sunmass$. \textit{Right panel}: same as \textit{left panel} but for model HC2 ($\alpha_{\rm NS}=0.02$ and $M_0^{\rm Eu}=\num{4.0e-6} \sunmass$). }
  \label{fig:Model3}
\end{figure*}

\begin{figure*}
\centering
  \includegraphics[trim={0cm 3.5cm 0cm 0cm}, clip, width=0.9\textwidth]{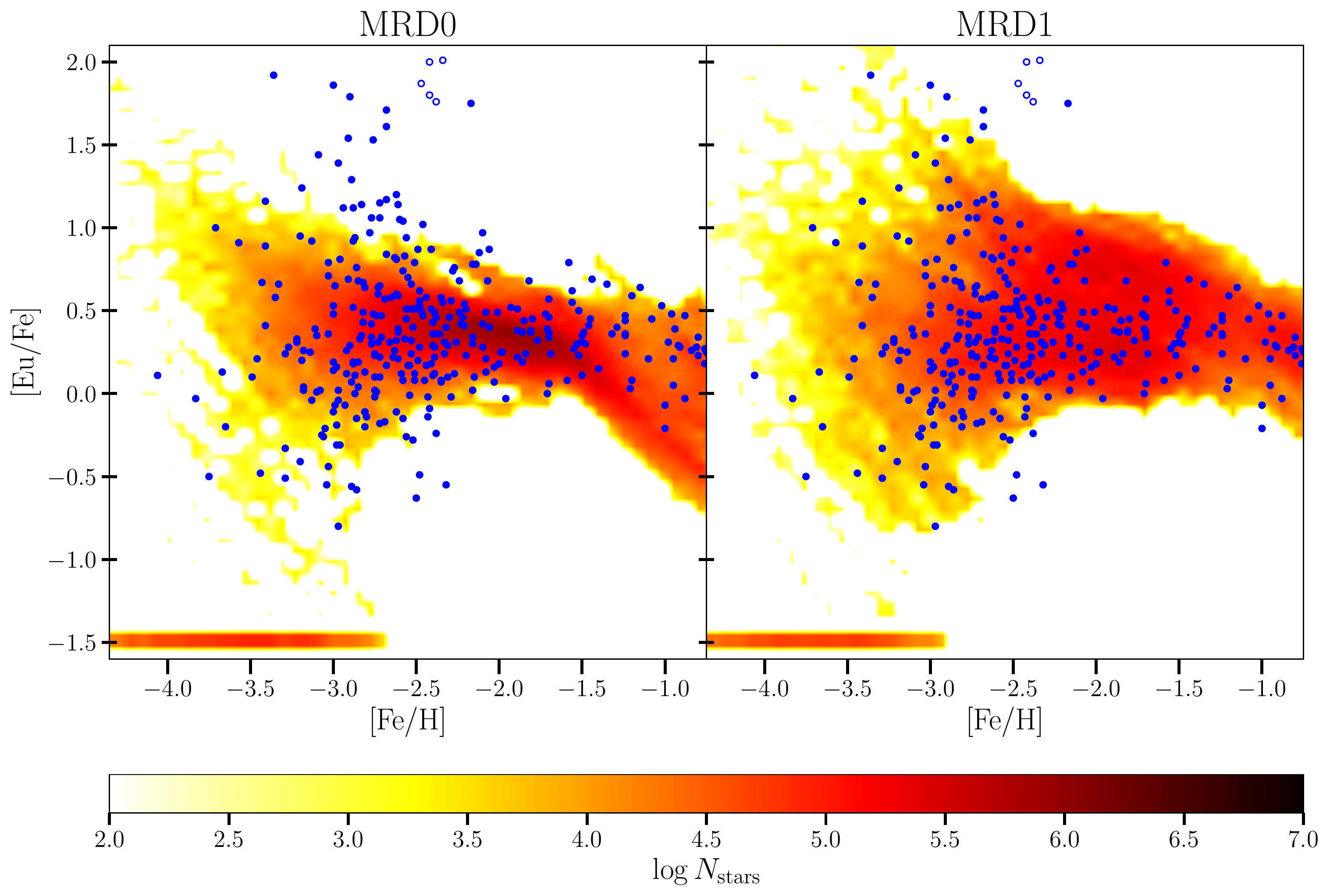}
  \caption{\textit{Left panel}: results of [Eu/Fe] vs [Fe/H] for model MRD0. This model has a DTD $\propto t^{-1}$ and Eu is produced by both NSMs and MRD SNe. MRD SNe are the 10$\%$ of CC-SNe only at Z$<10^{-3}$. \textit{Right panel}: same as before but for model MRD1. The model is identical to MRD0 apart from the assumed DTD (in this case $\propto t^{-1.5}$).}
  \label{fig:Model4}
\end{figure*}

In this section, we present the results of our models (summarised in Table \ref{tab:Models}), and we discuss the impact of our assumptions on the chemical evolution of Eu.\\ 
First of all, we present the effect of the new environment on the model result (i.e. MSFR) comparing it with the result of model NSt3 contained in \citet{cavallo2021}. After that, we include the effect of inefficient enrichment of europium and a varying star-formation quenching time on models results (i.e. M10 and M11). Afterwards, we test the impact of $\alpha_{\rm NS}$ on the predicted spread of [Eu/Fe] at intermediate metallicities (i.e. HC0, HC1, and HC2), to confirm the correlation reported in \citet{cavallo2021}. Finally, we discuss a scenario where both NSMs and MRD SNe are Eu producers (MRD0 and MRD1).\\
In Figure \ref{fig:Model1}; \ref{fig:Model2}; \ref{fig:paths}; \ref{fig:Model3}; \ref{fig:Model4} are shown the results of our models, in the [Eu/Fe]vs[Fe/H] plane. In the plots, at [Eu/Fe] = $-1.5$ dex, we also report the long-living stars formed without Eu (formally [Eu/Fe] $=-\infty$).

\subsection{Impact of the new environment} \label{resultNSM}
To test and track the effects of the new environment on the chemical evolution of Eu we used the NSt3 model, contained in \citet{cavallo2021}, as a baseline. In the first instance, we add the hierarchical clustering scenario by computing the chemical evolution of the primordial population of proto-galaxies. Each galaxy has a star formation history described by the SFR law (see equation \ref{eqn:SFR}). This model (MSFR) does not take into account the dilution of Eu or the quenching timescale of star formation. From Figure \ref{fig:Model1} it is seen that the hierarchical clustering scenario assumed here, allows the model to better reproduce the observational data. In particular, with this new scenario we can explain the presence of stars with [Eu/Fe]$\sim0$ at extremely low metallicities ([Fe/H]$\sim3$). However, this model is still not able to fit the low-metallicity tail present in the abundances distributions of halo stars.\\
As discussed in Section \ref{Eunucleo}, this new environment adds two effects that have a great effect on the chemical evolution of Eu: i) the motion of NS binary systems in the gravitational potential of the host galaxy can determine a merger location detached from the galaxy itself and so, only a fraction of the produced Eu is retained by the galaxy; ii) we expect that in low mass galaxies, the SF should quench at short timescales. We implemented an Eu dilution effect that is stochastic on single NSM events but, on average, is more intense in the least massive galaxies (details in Section \ref{sec:eudilution}). On the other hand, for the quenching time of SF and its dependence on the mass of the galaxy, we developed two empirical relations: bi-modal and linear. To test these two effects we developed a couple of models: M10 and M11. The prescriptions of these models are identical apart from the relation used for $t_{\rm SF}$: in model M10 we assume the bi-modal behaviour of $t_{\rm SF}$ and so the parameter can assume only two values, based on the mass of the galaxy; on the other side in model M11, the value $t_{\rm SF}$ scales linearly with the mass of the galaxy. Parameters assumed in models M10 and M11 are reported in Table \ref{tab:Models}. In Figure \ref{fig:Model2} are reported the results of model M10 (\textit{left panel}) and model M11 (\textit{right panel}). As seen in Figure \ref{fig:Model2}, both models show a good fit to the observed stars and, in particular, can explain the presence of stars with [Eu/Fe]$<0$ in the extremely metal-poor environment (i.e. [Fe/H]$<-3$).\\
From now on, we will refer to M11 as our best model (of the NS-only scenario). Even if M11 and M10 produce similar results, M11 assumes a $t_{\rm SF}$ vs $M_{\rm gal}$ relation more realistic than the one assumed in M10 (linear against bi-modal).\\\\
To have a better comprehension of the impact of the hierarchical clustering framework, now we will describe how our stochastic chemical evolution model works. As we introduced in Section \ref{CEM}, the stochastic model consists of $\sim200$ realisations in which we compute the chemical evolution. In this work, each realisation is associated with one galaxy in our population (note that a single galaxy can be linked to more than one stochastic realisation, depending on its mass). So, with this construction, at each time-step, we can study the chemical composition of a single realisation and, therefore, trace its time evolution. In Figure \ref{fig:paths} are plotted ten realisations of model HC0 (identical to model M11) on the [Eu/Fe] vs [Fe/H] plane. The path of each realisation through the [Eu/Fe] vs [Fe/H] plane is colour coded with a colour map that shows the time at which the realisation passes a certain point in the plane. In the figure are also reported the initial point (with coloured diamonds) and the time (in Myr) at which the first NSM has exploded. Moreover, starting points, epochs of the first NSM event, and paths of single realisations are labelled with different colours, based on the mass of the galaxy linked to them. In the following some important features of the model, that can be inferred by Figure \ref{fig:paths}, will be discussed.\\
\begin{list}{--}{} 
    \item In our model, the star formation is stochastic and so it is the formation of binary systems of NS. For this reason, we can have realisations where the first NSM explode at $\sim90$ Myr, but also realisation where this happened later at $\sim160$ Myr. Regarding the time of the first NSM explosion, we can also identify a clear trend: in the least massive galaxies, the first NSM appear later than in most massive ones. In fact looking at Figure \ref{fig:paths}, is seen that in low mass galaxies (with $10^5\leq M_{\rm gal} \leq 10^6\: \sunmass$; labelled in blue) the first NSM occurs, on average, at $\sim125$ Myr; on the other side, in the most massive galaxies (with $10^8< M_{\rm gal} \leq 10^9\: \sunmass$; labelled in red) the fist NSM explode at $\sim95$ Myr. This trend is a direct effect of the fact low mass galaxies transform less mass of gas into stars at each time-step (i.e. lower star formation efficiency; see equation \ref{eqn:SFE}) compared to the most massive ones. However, also in realisations that belong to galaxies of the same mass class, we can notice variability in the epoch of the first NSM explosion. This is caused by the fact that SF is stochastic.\\\\
    \item The trace of each realisation starts with a different value of [Eu/Fe] at different metallicities. Also, in this case, it is possible to identify a trend: galaxies of low mass (labelled in blue) enter in the plane at lower [Eu/Fe] compared to the "red" (plotted in red) ones: the dilution effect is stronger in galaxies with lower masses and therefore the ratio [Eu/Fe] is lower.\\\\
    \item The realisations follow different paths in the [Eu/Fe] vs [Fe/H] plane. Into these paths, is possible to notice some patterns, which can be understood in terms of the enrichment that takes place in that volume. For example, when a realisation moves horizontally (at constant [Eu/Fe]) towards lower metallicities, no events are enriching the ISM in iron and europium, and the gas is diluted by the in-falling gas with primordial composition. Then, when an event produces Fe, the realisation moves to higher metallicities and lower [Eu/Fe] ratios. If a NSM explodes, the realisation makes a "jump" towards higher [Eu/Fe] values. The height of these "jumps" varies among different realisations due to the combination of two facts: i) the amount of Eu that a single NSM can produce is variable (as equation \ref{eqn:RandoEu}). ii) the fraction of Eu retained by the host galaxy varies from one NSM to another (see Figure \ref{fig:Eudilu}).
\end{list}
In our scenario least massive galaxies could not produce stars with [Eu/Fe]$>1$. However, r-II stars (highly enhanced r-process elements stars with [Eu/Fe]$>0.7$) have been observed in Reticulum II, an ultra-faint dwarf galaxy \citep[][]{Ji2016Natur.531..610J, Roederer2016AJ....151...82R}. In our scenario, where the enrichment in dwarf galaxies is strongly diluted, galaxies like Reticulum II could not be enriched by NSMs but by a secondary source(s) of Eu (e.g. MRD and/or CC SNe).\\
Other works, such as \citet{Brauer2019ApJ...871..247B}, \citet{Hattori2022arXiv220704110H}, and \citet{Hirai2022arXiv220604060H}, found that (at least a fraction) of metal-poor r-II stars in the Galactic halo formed in ancient ultra-faint dwarf (UFD) galaxies (similar to Reticulum II). With our model, we can recover those conclusions by assuming a different probability distribution for $D_{Eu}$ (see Figure \ref{fig:Eudilu}). In this work, we implement the inefficient enrichment of Eu in low-mass galaxies following the prescriptions of \citet{bonetti2019neutron}.
That model predicts that the Eu enrichment in galaxies similar to Reticulum II is rare and not enough robust to explain the observations of Reticulum II. However is still possible that, with Reticulum II, we detect a rare member of a large population of poorly enriched UFD galaxies. Future observations of r-II stars in several UFD would certainly help to solve the tension with the picture that we adopt.

\subsection{Effects of $\alpha_{\rm NS}$}
As we introduced in previous sections, the shape of the DTD has a great impact on the chemical evolution of Eu. Another parameter, that is linked with the DTD by definition is the fraction of massive stars that are in a binary system with the right characteristics to lead to a NSM: $\alpha_{\rm NS}$. This parameter can be calculated, given the DTD, SFR, and the mass range of NS progenitors, imposing that the theoretical rate of NSM at the present time (see equation \ref{eqn:rate}) is consistent with the observed one. For the DTDs used in this paper, previous works suggest a wide range of $\alpha_{\rm NS}$, from $1-\num{2e-2}$ \citep[][]{giacobbo2018progenitors, simonetti2019new, Greggio2021} up to $5-\num{6e-2}$ \citep{molero2021predicted}. More recent estimates of the value of the present rate of NSM \citep[e.g.][]{Abbott2021ApJ...913L...7A} point to $\alpha_{\rm NS}<0.01$. The uncertainty of $\alpha_{\rm NS}$ values is correlated with the large error bars in the rate of NSM observed at the present day calculated and revised by \citet{abbott2017gw170817, abbott2020, Abbott2021ApJ...913L...7A}.\\
\citet{giacobbo2018progenitors}, with a binary population synthesis model, suggest that the formation of binary systems of NS is influenced by the environment metallicity. This can be parametrized by assuming a dependence of $\alpha_{\rm NS}$ on [Fe/H]. Moreover, this dependency has a weak influence on the predicted present rate of NSM, as shown by \citet{simonetti2019new}. However, no strong observational evidence has been found to support this dependency.\\
In \citet{cavallo2021}, we presented the correlation between the dispersion of the [Eu/Fe] values and the fraction of massive stars that can generate a NSM ($\alpha_{\rm NS}$). In particular, the spread shrinks when we assume higher $\alpha_{\rm NS}$ values.\\
For these reasons, we decide to test the dependence of [Eu/Fe] dispersion on $\alpha_{\rm NS}$ with three models: HC0, HC1, and HC2 (see Table \ref{tab:Models}). These models assume different values for $\alpha_{\rm NS}$ and, as a consequence, for $M_{\rm NSM}^0$: 0.005 (HC0), 0.01 (HC1), and 0.02 (HC2); results of these models are plotted in Figure \ref{fig:Model3}, where is possible to see the correlation cited above.\\
Focusing on intermediate metallicities ([Fe/H]$>-2.5$) is possible to notice the impact of $\alpha_{\rm NS}$ on the predicted spread \citep[see][]{cavallo2021}. In this metallicity range is possible to check whether or not models with a certain $\alpha_{\rm NS}$ are able to reproduce the spread observed in halo stars. Thus, our model seems to be a promising tool that allows us to estimate the $\alpha_{\rm NS}$ parameter and, as a consequence, $M_{\rm NSM}^0$. However, since the current data of Eu abundances of halo stars are affected by large uncertainties ($\sim 0.2-0.3$ dex), we cannot reject any of the tested models. Future surveys such as 4MOST \citep[][]{4MOST} and WEAVE \citep[][]{WEAVE} will produce larger and more precise data-sets so to allow us to determine the value of $\alpha_{\rm NS}$ more precisely.

\begin{figure*}
\centering
  \includegraphics[trim={0cm 0cm 0cm 0cm}, clip, width=0.9\textwidth]{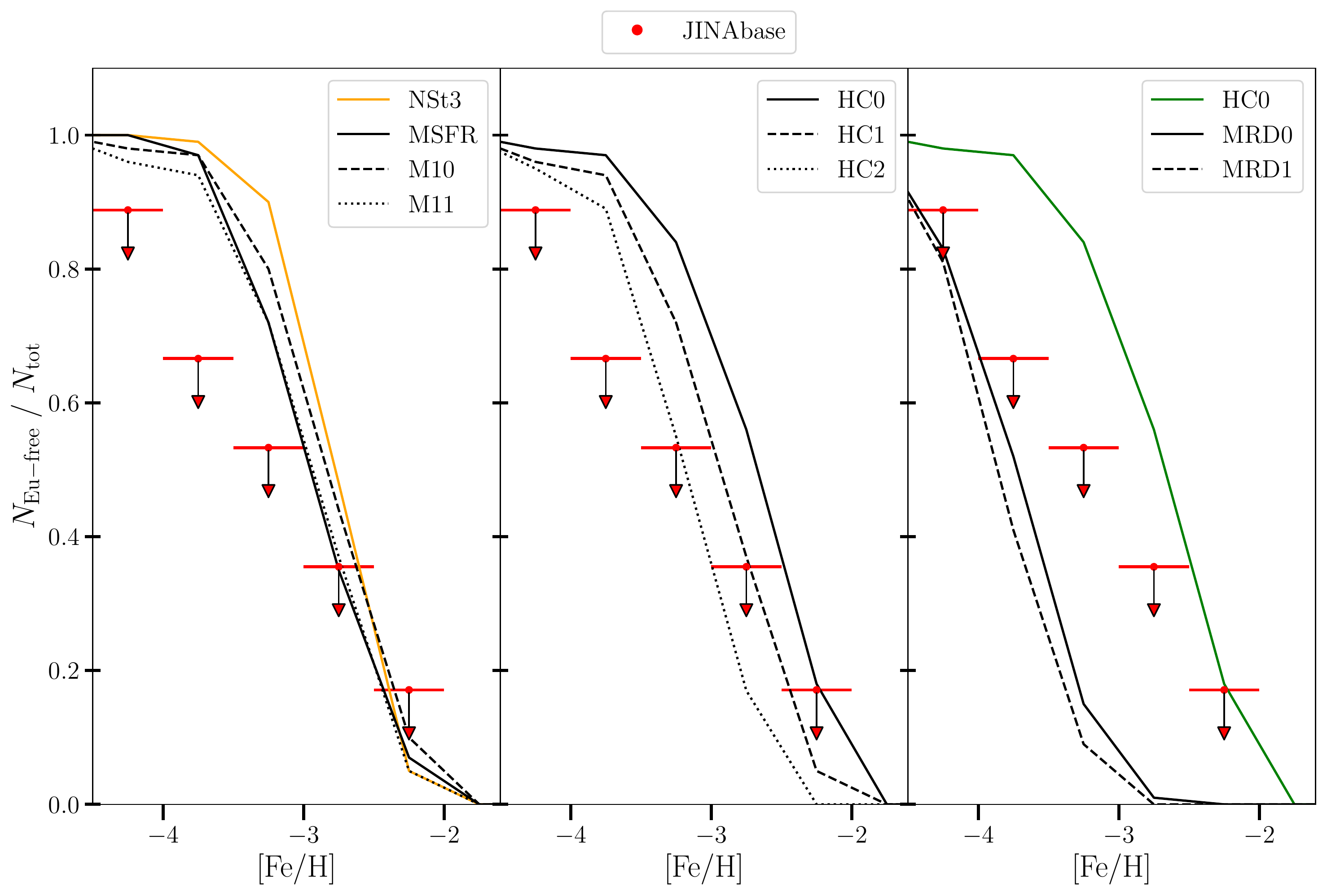}
  \caption{Comparison between the predicted and observed ratios of Eu-free stars over the total number of stars for bins of 0.5 dex in [Fe/H]. Blue and red markers are the observational proxies for the ratio; so the ratio between the number of stars for which Eu only presents an upper limit (possibly Eu-free) over the number of stars with measured Ba. Observational ratios derived from the JINAbase data-set \citet{Abohalima_2018} are plotted in red. \textit{Left panel}: Ratio of Eu-free stars for model MSFR, M10, M11 (in black), and NSt3 from \citet{cavallo2021} (in orange). \textit{Central panel}: same as \textit{left panel} but for models HC0, HC1, and HC2. \textit{Right panel}: same as \textit{left panel} but for model MRD0, MRD1 (in black), and HC0 (in green).}
  \label{fig:Eu_free}
\end{figure*}
\subsection{Eu production from MRD SNe}
As we already mentioned, lots of studies on the nucleosynthesis of r-process elements indicate NSMs as a promising r-process site \citep[][]{freiburghaus1999r,korobkin2012astrophysical, Wanajo_2014, Radice2018}. However, as we point out in \citet{cavallo2021}, in our stochastic chemical evolution model NSM with a DTD $\propto t^{-1}$ are not able to reproduce the abundances distribution of halo stars. In this section, we want to test what are the effects of adding a second source of Eu to the NSM with a DTD $\propto t^{-1}$ (and $\propto t^{-1.5}$). However, CC SNe seems to be ruled out as a major producer of Eu, as suggested by nucleosynthesis models \citep[]{Arcones2007, Arcones_2012, Wanajo_2018}. On the other hand, particular classes of CC SNe could potentially be a source of r-process elements. The major proposed candidates are collapsars \citep[][]{Siegel_2019} and MRD SNe \citep[][]{winteler2012magnetorotationally,Nishimura2015, Reichert2020}. Note that in these models $t_{\rm SF}$ is equal to $1$ Gyr for all the proto-galaxies.\\
We decide to test the scenario where both NSMs and MRD SNe (active only at Z$<10^{-3}$) produce Eu in this new framework. We develop two different models that differ only on the assumed DTD: MRD0 (DTD $\propto t^{-1}$) and MRD1 (DTD $\propto t^{-1.5}$); in Figure \ref{fig:Model4} are plotted the results of these models.\\
Looking at the \textit{right panel} of Figure \ref{fig:Model4}, is seen that with the addition of the second source of Eu (in this case MRD-SNe) the model is able to explain both the large [Eu/Fe] spread at [Fe/H]$\leq-2.5$ and the mean trend of [Eu/Fe] vs [Fe/H] at high metallicity ([Fe/H]$>-2$) without assumptions on the $t_{\rm SF}$ of the proto-galaxies.\\
In the \textit{left panel} of Figure \ref{fig:Model4} it can also be noticed that the mixed scenario is not able to reproduce the wide spread of the data when we assume DTD$\propto t^{-1}$ for the NSMs. In particular, the model fails to explain the presence of stars with [Eu/Fe]$>1.5$ and also the trend of [Eu/Fe] at higher metalicities.\\\\
\noindent
We note that all our models fail to explain the group of $\sim6$ Eu-enhanced stars at [Fe/H]$> -2.5$ present in our observational sample. As discussed in Section \ref{[Eu/Fe] halo} five of them are CEMP stars that are not the main focus of this work. On the other hand, our models fail to explain the presence of the ordinary star with [Eu/Fe]$\sim 1.7$ at [Fe/H]$\sim -2.1$. Further studies on the chemical composition of SMSS J175046.30-425506.9 could provide clues on its origin.

\subsection{Eu-free stars}
As underlined in previous versions of this model \citep{cescutti2015role,cavallo2021}, our stochastic chemical evolution model predicts the presence of Eu-free stars (i.e. [Eu/Fe]$=-\infty$). This is due to the fact that only a small fraction of stars pollutes the ISM with r-process elements (compared to the number of SNe II). So at extremely low metallicities ([Fe/H]<-3), some stars could have been born in regions where the gas is Eu-free (not polluted from r-process events). In particular, models with lower values of $\alpha_{\rm NS}$ should predict a high number of Eu-free stars. Moreover, models that assume longer delay times for NSM will produce higher fractions of Eu-free stars, at higher metallicities. In this section, we will compare the fraction of Eu-free stars predicted by our models and the observational constraints on its value.\\
The model prediction of the fraction of Eu-free stars is computed as the ratio of stars without Eu over the total number of stars in our model in a given metallicity bin. To calculate the observational ratio, we take all the halo stars for which Ba has been observed contained in JINAbase\footnote{Selected stars are contained in: \citet{RYA91,MCW95,RYA96,NOR97A,BUR00,FUL00,PRE00,WES00,PRE01,AOK02B,CAR02,COW02,JOH02a,LUC03, IVA03, SNE03, CAY04, CHR04, COH04, HON04, JOH04, AOK05, BAB05, BAR05, JON05, IVA06, JON06, MAS06, PRE06, SIV06, AOK07A, AOK07C, FRE07b, LAI07, AOK08, COH08, LAI08, ROE08, BON09, HAY09, ZHA09, BEH10, ISH10, ROE10, BEN11, CAF11B, HAN11, HON11, ALL12, AOK12, HAN12, MAS12,AOK13,COH13,CUI13,ISH13,PLA13,YON13,AOK14,MAS14,PAL14a,roederer2014search,SIQ14,SPI14,HAN15,JAC15,LI15a}.}. The number of this stars which have only an upper limit measurement of Eu over the total number of them, in a given [Fe/H] bin, is our observational proxy of Eu-free stars. Indeed, the fact that we can observe barium, but no europium could be a sign of the absence of europium in these stars; however, in most cases, it is probably an observational problem due to the weakness of europium lines compared to barium lines. Still, we can use this observational proxy as upper limits useful in providing constraints to the models.\\
In Figure \ref{fig:Eu_free} is plotted the comparison between predicted and observed ratios of Eu-free stars as a function of [Fe/H]. Looking at Figure \ref{fig:Eu_free} (\textit{left panel}), is seen that models M10 and M11 predict similar fractions of Eu-free stars at fixed [Fe/He]. Moreover, is possible to appreciate the overall reduction of predicted Eu-free stars compared to model NSt3 (orange line) presented in our last work. This reduction is caused by the change of environment, in particular, by the inefficiency in the Eu enrichment (especially in the least massive galaxies). In this case, especially at low metallicites ([Fe/H]$<-3$) the predicted fraction of Eu-free stars are above the observed ones; those are upper limits. The fact that models M10 and M11 produce too many stars formed without Eu, can be explained by the characteristic time scale of the Eu enrichment by NSMs (on average, longer than the one associated with massive stars).\\
In the \textit{central panel} of Figure \ref{fig:Eu_free} is possible to appreciate the correlation between the value of the parameter ($\alpha_{\rm NS}$) and the fraction of Eu-free stars predicted. In particular, a model with higher $\alpha_{\rm NS}$ (i.e. NSMs are less rare at all the metallicities) predicts a lower fraction of Eu-free stars. Similar to the previous case, the predictions of our models are not compatible with the chosen observational proxies, for the same reason reported above. \\
In the last panel are plotted the results of models where both NSM and MRD SNe are Eu producers and their comparison with the ones of HC0. From this panel we can observe two different things: a) models where the MRD SNe channel is active predicts a lower fraction of Eu-free stars, due to the early contribution of massive stars to the Eu enrichment (black lines vs green one); b) models with a steeper DTD (i.e. shorter delay times for NSMs) produce a lower amount of Eu-free stars, due to the shorter time between the formation of the NS binary system and the merging event. It is also important to underline that the predicted fractions of Eu-free stars are compatible or lower than the observed ones.\\
As discussed in the introduction, Eu is the perfect tracer of r-process because is mainly produced by this neutron-capture channel. So, what we define Eu free stars are stars without europium produced by an r-process. On the other side, we cannot exclude that others neutron capture channels may produce europium at this metallicity. For example, the s-process in rotating massive stars \citep[see][]{Frisch16,LC18}
could produce tiny amount of europium. From this channel, the Eu enrichment at [Fe/H]$\sim-3$ is expected to be very low with a [Eu/Fe]$<-$4 (at least two orders of magnitude less than NSMs) with an [Eu/H]$<-$7. However, given the weakness of the europium line, this pollution is likely impossible to be detected in stellar spectra at least with present spectroscopic capabilities.

\section{Conclusions} \label{Conclusions}
In this work, we use the stochastic chemical evolution model of the Galactic halo presented by \citet{cescutti2006chemical} and updated by \citet{cescutti2015role}. We aim at reproducing the large spread in the [Eu/Fe] ratio observed in Galactic halo stars and solve the tensions noted in our previous work \citep[][]{cavallo2021}. In particular, we assume that the Galactic halo has been formed by the accretion of a population of ($\sim 100$) proto-galaxies with mass from $10^5$ to $10^9$ $\sunmass$. We compute the chemical evolution of Eu in these proto-galaxies, which is influenced by two effects: i) the lower increase of the Eu abundance caused by the dynamics of NS binary systems which can easily escape from the gravitational potential of the host galaxy \citep[see][]{bonetti2019neutron}, and ii) SF timescales shorter in low mass galaxies due to the low binding energy of the gas inside the proto-galaxy. These effects depend on the mass of the proto-galaxy that we are considering. Therefore, we introduce relations that depend on the mass of the proto-galaxy for both these effects. In particular, for the dilution of Eu, we also add event-to-event stochasticity. We decide to take model NSt3 from \citep{cavallo2021} and we add the effects mentioned above. In models where NSM are the sole producers of Eu we assume that these systems follow a delay time distribution (DTD) $\propto t^{-1.5}$ with a minimum coalescence time of $1$ Myr. For NSMs Eu yields, we followed the prescriptions of \citet{matteucci2014europium} and \citet{cescutti2015role}. We also tested the scenario where both NSMs and MRD SNe (a particular class of CC SNe) produce Eu.\\
Our main conclusions can be summarised as follows:
\begin{list}{--}{} 
\item if we assume that NSMs are the sole producers of Eu, the hierarchical formation of the Galactic halo, combined with the inefficient enrichment of Eu in low mass proto-galaxies, can explain the presence of extremely metal-poor stars (i.e. [Fe/H]$\leq-3$) that present sub-solar [Eu/Fe] ratios. In this framework, the coalescence timescale of binary systems of neutron stars should follow a DTD $\propto t^{-1.5}$, each NSM event produces an average of $\num{1.6e-5}$ $\sunmass$ of Eu, $\alpha_{\rm NS}$ is fixed at $0.005$, and progenitors of NS should be in the mass range from $9$ to $50$ $\sunmass$.\\\\
\item Adopting our best model (M11/HC0) we confirm the correlation, studied in our previous work \citep[see][]{cavallo2021}, between $\alpha_{\rm NS}$ and the dispersion of [Eu/Fe] at a given metallicity, and reach similar conclusions: present literature data cannot allow us to put strong constrains of $\alpha_{\rm NS}$. However, future high-resolution spectroscopical surveys, such as 4MOST \citep[][]{4MOST} and WEAVE \citep{WEAVE}, will produce the necessary statistic to constrain at best this parameter.\\\\
\item NSMs with a DTD $\propto t^{-1}$ are not able to explain the large star-to-star scatter of [Eu/Fe] observed in the Galactic halo, also in mixed scenario in which both NSMs and MRD SNe produce Eu. However, this scenario is in agreement with observational data when we assume a DTD for NSMs $\propto t^{-1.5}$\\\\ 
\item the fractions of Eu-free stars of models in which we assume that Eu is only produced by the NSMs cannot be reconciled with observations. This is a potential issue for the concept of r-process produced exclusively by NSMs. On the other hand, when we add the second source of Eu (i.e. the MRD SNe) the models predict ratios of Eu-free stars in agreement with observations. In particular, these ratios are below or comparable to the observational values but that is not an issue. We know that a potentially large number of these observed Eu-free stars are most likely to be false Eu-free stars, and so the observational values should be considered as upper limits. Furthermore, the value of Eu-free stars cannot be clearly evaluated and no firm conclusions can be drawn.\\\\
\end{list}
Finally, we accomplished our initial goal: solve the tension noted in our previous work. In order to do that, we included our chemical evolution model in the hierarchical scenario and we also took into account the effects on chemical evolution produced by the environment of small proto-galaxies.\\
In future work, we aim at reproducing the solar abundances and present-day NSMs rate with a new version of this model able to mimic the Galactic disk environment.


\section*{Acknowledgements}
We thank an anonymous referee for his/her comments which improved the quality of this paper. \\
LC thanks Matteo Bonetti for useful discussions on the core part of this work. This work was partially supported by the European Union (ChETEC-INFRA, project no. 101008324).

%
%
\bibliographystyle{aa}
\bibliography{Bibliography} 

\begin{thebibliography}{212}
\expandafter\ifx\csname natexlab\endcsname\relax\def\natexlab#1{#1}\fi

\bibitem[{{Abbott} {et~al.}(2017{\natexlab{a}}){Abbott}, {Abbott}, {Abbott},
  {Abernathy}, {Acernese}, {Ackley}, {Adams}, {Adams}, {Addesso}, {Adhikari},
  {Adya}, {Affeldt}, {Agathos}, {Agatsuma}, {Aggarwal}, {Aguiar}, {Aiello},
  {Ain}, {Allen}, {Allocca}, {Altin}, {Anderson}, {Anderson}, {Arai}, {Araya},
  {Arceneaux}, {Areeda}, {Arnaud}, {Arun}, {Ascenzi}, {Ashton}, {Ast}, {Aston},
  {Astone}, {Aufmuth}, {Aulbert}, {Babak}, {Bacon}, {Bader}, {Baker},
  {Baldaccini}, {Ballardin}, {Ballmer}, {Barayoga}, {Barclay}, {Barish},
  {Barker}, {Barone}, {Barr}, {Barsotti}, {Barsuglia}, {Barta}, {Bartlett},
  {Bartos}, {Bassiri}, {Basti}, {Batch}, {Baune}, {Bavigadda}, {Bazzan},
  {Bejger}, {Bell}, {Berger}, {Bergmann}, {Berry}, {Bersanetti}, {Bertolini},
  {Betzwieser}, {Bhagwat}, {Bhandare}, {Bilenko}, {Billingsley}, {Birch},
  {Birney}, {Biscans}, {Bisht}, {Bitossi}, {Biwer}, {Bizouard}, {Blackburn},
  {Blair}, {Blair}, {Blair}, {Bloemen}, {Bock}, {Boer}, {Bogaert}, {Bogan},
  {Bohe}, {Bond}, {Bondu}, {Bonnand}, {Boom}, {Bork}, {Boschi}, {Bose},
  {Bouffanais}, {Bozzi}, {Bradaschia}, {Brady}, {Braginsky}, {Branchesi},
  {Brau}, {Briant}, {Brillet}, {Brinkmann}, {Brisson}, {Brockill}, {Broida},
  {Brooks}, {Brown}, {Brown}, {Brown}, {Brunett}, {Buchanan}, {Buikema},
  {Bulik}, {Bulten}, {Buonanno}, {Buskulic}, {Buy}, {Byer}, {Cabero},
  {Cadonati}, {Cagnoli}, {Cahillane}, {Calder{\'o}n Bustillo}, {Callister},
  {Calloni}, {Camp}, {Cannon}, {Cao}, {Capano}, {Capocasa}, {Carbognani},
  {Caride}, {Casanueva Diaz}, {Casentini}, {Caudill}, {Cavagli{\`a}},
  {Cavalier}, {Cavalieri}, {Cella}, {Cepeda}, {Cerboni Baiardi}, {Cerretani},
  {Cesarini}, {Chamberlin}, {Chan}, {Chao}, {Charlton}, {Chassande-Mottin},
  {Cheeseboro}, {Chen}, {Chen}, {Cheng}, {Chincarini}, {Chiummo}, {Cho}, {Cho},
  {Chow}, {Christensen}, {Chu}, {Chua}, {Chung}, {Ciani}, {Clara}, {Clark},
  {Cleva}, {Coccia}, {Cohadon}, {Colla}, {Collette}, {Cominsky}, {Constancio},
  {Conte}, {Conti}, {Cook}, {Corbitt}, {Cornish}, {Corsi}, {Cortese}, {Costa},
  {Coughlin}, {Coughlin}, {Coulon}, {Countryman}, {Couvares}, {Cowan},
  {Coward}, {Cowart}, {Coyne}, {Coyne}, {Craig}, {Creighton}, {Creighton},
  {Cripe}, {Crowder}, {Cumming}, {Cunningham}, {Cuoco}, {Dal Canton},
  {Danilishin}, {D'Antonio}, {Danzmann}, {Darman}, {Dasgupta}, {Da Silva
  Costa}, {Dattilo}, {Dave}, {Davier}, {Davies}, {Daw}, {Day}, {De}, {DeBra},
  {Debreczeni}, {Degallaix}, {De Laurentis}, {Del{\'e}glise}, {Del Pozzo},
  {Denker}, {Dent}, {Dergachev}, {De Rosa}, {DeRosa}, {DeSalvo}, {Devine},
  {Dhurandhar}, {D{\'\i}az}, {Di Fiore}, {Di Giovanni}, {Di Girolamo}, {Di
  Lieto}, {Di Pace}, {Di Palma}, {Di Virgilio}, {Dolique}, {Donovan}, {Dooley},
  {Doravari}, {Douglas}, {Downes}, {Drago}, {Drever}, {Driggers}, {Ducrot},
  {Dwyer}, {Edo}, {Edwards}, {Effler}, {Eggenstein}, {Ehrens}, {Eichholz},
  {Eikenberry}, {Engels}, {Essick}, {Etzel}, {Evans}, {Evans}, {Everett},
  {Factourovich}, {Fafone}, {Fair}, {Fan}, {Fang}, {Farinon}, {Farr}, {Farr},
  {Favata}, {Fays}, {Fehrmann}, {Fejer}, {Fenyvesi}, {Ferrante}, {Ferreira},
  {Ferrini}, {Fidecaro}, {Fiori}, {Fiorucci}, {Fisher}, {Flaminio}, {Fletcher},
  {Fournier}, {Frasca}, {Frasconi}, {Frei}, {Freise}, {Frey}, {Frey},
  {Fritschel}, {Frolov}, {Fulda}, {Fyffe}, {Gabbard}, {Gair}, {Gammaitoni},
  {Gaonkar}, {Garufi}, {Gaur}, {Gehrels}, {Gemme}, {Geng}, {Genin}, {Gennai},
  {George}, {Gergely}, {Germain}, {Ghosh}, {Ghosh}, {Ghosh}, {Giaime},
  {Giardina}, {Giazotto}, {Gill}, {Glaefke}, {Goetz}, {Goetz}, {Gondan},
  {Gonz{\'a}lez}, {Gonzalez Castro}, {Gopakumar}, {Gordon}, {Gorodetsky},
  {Gossan}, {Gosselin}, {Gouaty}, {Grado}, {Graef}, {Graff}, {Granata},
  {Grant}, {Gras}, {Gray}, {Greco}, {Green}, {Groot}, {Grote}, {Grunewald},
  {Guidi}, {Guo}, {Gupta}, {Gupta}, {Gushwa}, {Gustafson}, {Gustafson},
  {Hacker}, {Hall}, {Hall}, {Hammond}, {Haney}, {Hanke}, {Hanks}, {Hanna},
  {Hanson}, {Hardwick}, {Harms}, {Harry}, {Harry}, {Hart}, {Hartman}, {Haster},
  {Haughian}, {Heidmann}, {Heintze}, {Heitmann}, {Hello}, {Hemming}, {Hendry},
  {Heng}, {Hennig}, {Henry}, {Heptonstall}, {Heurs}, {Hild}, {Hoak}, {Hofman},
  {Holt}, {Holz}, {Hopkins}, {Hough}, {Houston}, {Howell}, {Hu}, {Huang},
  {Huerta}, {Huet}, {Hughey}, {Husa}, {Huttner}, {Huynh-Dinh}, {Indik},
  {Ingram}, {Inta}, {Isa}, {Isac}, {Isi}, {Isogai}, {Iyer}, {Izumi}, {Jacqmin},
  {Jang}, {Jani}, {Jaranowski}, {Jawahar}, {Jian}, {Jim{\'e}nez-Forteza},
  {Johnson}, {Jones}, {Jones}, {Jonker}, {Ju}, {Haris}, {Kalaghatgi},
  {Kalogera}, {Kandhasamy}, {Kang}, {Kanner}, {Kapadia}, {Karki}, {Karvinen},
  {Kasprzack}, {Katsavounidis}, {Katzman}, {Kaufer}, {Kaur}, {Kawabe},
  {K{\'e}f{\'e}lian}, {Kehl}, {Keitel}, {Kelley}, {Kells}, {Kennedy}, {Key},
  {Khalili}, {Khan}, {Khan}, {Khan}, {Khazanov}, {Kijbunchoo}, {Kim}, {Kim},
  {Kim}, {Kim}, {Kim}, {Kim}, {Kim}, {Kimbrell}, {King}, {King}, {Kissel},
  {Klein}, {Kleybolte}, {Klimenko}, {Koehlenbeck}, {Koley}, {Kondrashov},
  {Kontos}, {Korobko}, {Korth}, {Kowalska}, {Kozak}, {Kringel}, {Krishnan},
  {Kr{\'o}lak}, {Krueger}, {Kuehn}, {Kumar}, {Kumar}, {Kuo}, {Kutynia},
  {Lackey}, {Landry}, {Lange}, {Lantz}, {Lasky}, {Laxen}, {Lazzaro}, {Leaci},
  {Leavey}, {Lebigot}, {Lee}, {Lee}, {Lee}, {Lee}, {Lenon}, {Leonardi},
  {Leong}, {Leroy}, {Letendre}, {Levin}, {Lewis}, {Li}, {Libson}, {Littenberg},
  {Lockerbie}, {Lombardi}, {London}, {Lord}, {Lorenzini}, {Loriette},
  {Lormand}, {Losurdo}, {Lough}, {L{\"u}ck}, {Lundgren}, {Lynch}, {Ma},
  {Machenschalk}, {MacInnis}, {Macleod}, {Maga{\~n}a-Sandoval}, {Maga{\~n}a
  Zertuche}, {Magee}, {Majorana}, {Maksimovic}, {Malvezzi}, {Man}, {Mandic},
  {Mangano}, {Mansell}, {Manske}, {Mantovani}, {Marchesoni}, {Marion},
  {M{\'a}rka}, {M{\'a}rka}, {Markosyan}, {Maros}, {Martelli}, {Martellini},
  {Martin}, {Martynov}, {Marx}, {Mason}, {Masserot}, {Massinger}, {Masso-Reid},
  {Mastrogiovanni}, {Matichard}, {Matone}, {Mavalvala}, {Mazumder}, {McCarthy},
  {McClelland}, {McCormick}, {McGuire}, {McIntyre}, {McIver}, {McManus},
  {McRae}, {McWilliams}, {Meacher}, {Meadors}, {Meidam}, {Melatos}, {Mendell},
  {Mercer}, {Merilh}, {Merzougui}, {Meshkov}, {Messenger}, {Messick},
  {Metzdorff}, {Meyers}, {Mezzani}, {Miao}, {Michel}, {Middleton}, {Mikhailov},
  {Milano}, {Miller}, {Miller}, {Miller}, {Miller}, {Millhouse}, {Minenkov},
  {Ming}, {Mirshekari}, {Mishra}, {Mitra}, {Mitrofanov}, {Mitselmakher},
  {Mittleman}, {Moggi}, {Mohan}, {Mohapatra}, {Montani}, {Moore}, {Moore},
  {Moraru}, {Moreno}, {Morriss}, {Mossavi}, {Mours}, {Mow-Lowry}, {Mueller},
  {Muir}, {Mukherjee}, {Mukherjee}, {Mukherjee}, {Mukund}, {Mullavey}, {Munch},
  {Murphy}, {Murray}, {Mytidis}, {Nardecchia}, {Naticchioni}, {Nayak},
  {Nedkova}, {Nelemans}, {Nelson}, {Neri}, {Neunzert}, {Newton}, {Nguyen},
  {Nielsen}, {Nissanke}, {Nitz}, {Nocera}, {Nolting}, {Normandin}, {Nuttall},
  {Oberling}, {Ochsner}, {O'Dell}, {Oelker}, {Ogin}, {Oh}, {Oh}, {Ohme},
  {Oliver}, {Oppermann}, {Oram}, {O'Reilly}, {O'Shaughnessy}, {Ottaway},
  {Overmier}, {Owen}, {Pai}, {Pai}, {Palamos}, {Palashov}, {Palomba},
  {Pal-Singh}, {Pan}, {Pankow}, {Pannarale}, {Pant}, {Paoletti}, {Paoli},
  {Papa}, {Paris}, {Parker}, {Pascucci}, {Pasqualetti}, {Passaquieti},
  {Passuello}, {Patel}, {Patricelli}, {Patrick}, {Pearlstone}, {Pedraza},
  {Pedurand}, {Pekowsky}, {Pele}, {Penn}, {Perreca}, {Perri}, {Phelps},
  {Piccinni}, {Pichot}, {Piergiovanni}, {Pierro}, {Pillant}, {Pinard}, {Pinto},
  {Pitkin}, {Poe}, {Poggiani}, {Popolizio}, {Post}, {Powell}, {Prasad},
  {Predoi}, {Prestegard}, {Price}, {Prijatelj}, {Principe}, {Privitera},
  {Prix}, {Prodi}, {Prokhorov}, {Puncken}, {Punturo}, {Puppo}, {P{\"u}rrer},
  {Qi}, {Qin}, {Qiu}, {Quetschke}, {Quintero}, {Quitzow-James}, {Raab},
  {Rabeling}, {Radkins}, {Raffai}, {Raja}, {Rajan}, {Rakhmanov}, {Rapagnani},
  {Raymond}, {Razzano}, {Re}, {Read}, {Reed}, {Regimbau}, {Rei}, {Reid},
  {Reitze}, {Rew}, {Reyes}, {Ricci}, {Riles}, {Rizzo}, {Robertson}, {Robie},
  {Robinet}, {Rocchi}, {Rolland}, {Rollins}, {Roma}, {Romano}, {Romanov},
  {Romie}, {Rosi{\'n}ska}, {Rowan}, {R{\"u}diger}, {Ruggi}, {Ryan}, {Sachdev},
  {Sadecki}, {Sadeghian}, {Sakellariadou}, {Salconi}, {Saleem}, {Salemi},
  {Samajdar}, {Sammut}, {Sanchez}, {Sandberg}, {Sandeen}, {Sanders},
  {Sassolas}, {Saulson}, {Sauter}, {Savage}, {Sawadsky}, {Schale}, {Schilling},
  {Schmidt}, {Schmidt}, {Schnabel}, {Schofield}, {Sch{\"o}nbeck}, {Schreiber},
  {Schuette}, {Schutz}, {Scott}, {Scott}, {Sellers}, {Sengupta}, {Sentenac},
  {Sequino}, {Sergeev}, {Setyawati}, {Shaddock}, {Shaffer}, {Shahriar},
  {Shaltev}, {Shapiro}, {Shawhan}, {Sheperd}, {Shoemaker}, {Shoemaker},
  {Siellez}, {Siemens}, {Sieniawska}, {Sigg}, {Silva}, {Singer}, {Singer},
  {Singh}, {Singh}, {Singhal}, {Sintes}, {Slagmolen}, {Smith}, {Smith},
  {Smith}, {Son}, {Sorazu}, {Sorrentino}, {Souradeep}, {Srivastava}, {Staley},
  {Steinke}, {Steinlechner}, {Steinlechner}, {Steinmeyer}, {Stephens}, {Stone},
  {Strain}, {Straniero}, {Stratta}, {Strauss}, {Strigin}, {Sturani}, {Stuver},
  {Summerscales}, {Sun}, {Sunil}, {Sutton}, {Swinkels}, {Szczepa{\'n}czyk},
  {Tacca}, {Talukder}, {Tanner}, {T{\'a}pai}, {Tarabrin}, {Taracchini},
  {Taylor}, {Theeg}, {Thirugnanasambandam}, {Thomas}, {Thomas}, {Thomas},
  {Thorne}, {Thrane}, {Tiwari}, {Tiwari}, {Tokmakov}, {Toland}, {Tomlinson},
  {Tonelli}, {Tornasi}, {Torres}, {Torrie}, {T{\"o}yr{\"a}}, {Travasso},
  {Traylor}, {Trifir{\`o}}, {Tringali}, {Trozzo}, {Tse}, {Turconi},
  {Tuyenbayev}, {Ugolini}, {Unnikrishnan}, {Urban}, {Usman}, {Vahlbruch},
  {Vajente}, {Valdes}, {van Bakel}, {van Beuzekom}, {van den Brand}, {Van Den
  Broeck}, {Vander-Hyde}, {van der Schaaf}, {van Heijningen}, {van Veggel},
  {Vardaro}, {Vass}, {Vas{\'u}th}, {Vaulin}, {Vecchio}, {Vedovato}, {Veitch},
  {Veitch}, {Venkateswara}, {Verkindt}, {Vetrano}, {Vicer{\'e}}, {Vinciguerra},
  {Vine}, {Vinet}, {Vitale}, {Vo}, {Vocca}, {Vorvick}, {Voss}, {Vousden},
  {Vyatchanin}, {Wade}, {Wade}, {Wade}, {Walker}, {Wallace}, {Walsh}, {Wang},
  {Wang}, {Wang}, {Wang}, {Wang}, {Ward}, {Warner}, {Was}, {Weaver}, {Wei},
  {Weinert}, {Weinstein}, {Weiss}, {Wen}, {We{\ss}els}, {Westphal}, {Wette},
  {Whelan}, {Whiting}, {Williams}, {Williamson}, {Willis}, {Willke}, {Wimmer},
  {Winkler}, {Wipf}, {Wittel}, {Woan}, {Woehler}, {Worden}, {Wright}, {Wu},
  {Wu}, {Yablon}, {Yam}, {Yamamoto}, {Yancey}, {Yu}, {Yvert}, {Zadro{\.Z}ny},
  {Zangrando}, {Zanolin}, {Zendri}, {Zevin}, {Zhang}, {Zhang}, {Zhang}, {Zhao},
  {Zhou}, {Zhou}, {Zhu}, {Zucker}, {Zuraw}, {Zweizig}, {Sigurdsson}, {LIGO
  Scientific Collaboration}, \& {Virgo Collaboration}}]{abbott2017search}
{Abbott}, B.~P., {Abbott}, R., {Abbott}, T.~D., {et~al.} 2017{\natexlab{a}},
  \prd, 95, 082005

\bibitem[{{Abbott} {et~al.}(2020){Abbott}, {Abbott}, {Abbott}, {Abraham},
  {Acernese}, {Ackley}, {Adams}, {Adhikari}, {Adya}, {Affeldt}, {Agathos},
  {Agatsuma}, {Aggarwal}, {Aguiar}, {Aiello}, {Ain}, {Ajith}, {Allen},
  {Allocca}, {Aloy}, {Altin}, {Amato}, {Anand}, {Ananyeva}, {Anderson},
  {Anderson}, {Angelova}, {Antier}, {Appert}, {Arai}, {Araya}, {Areeda},
  {Ar{\`e}ne}, {Arnaud}, {Aronson}, {Arun}, {Ascenzi}, {Ashton}, {Aston},
  {Astone}, {Aubin}, {Aufmuth}, {AultONeal}, {Austin}, {Avendano},
  {Avila-Alvarez}, {Babak}, {Bacon}, {Badaracco}, {Bader}, {Bae}, {Baird},
  {Baker}, {Baldaccini}, {Ballardin}, {Ballmer}, {Bals}, {Banagiri},
  {Barayoga}, {Barbieri}, {Barclay}, {Barish}, {Barker}, {Barkett}, {Barnum},
  {Barone}, {Barr}, {Barsotti}, {Barsuglia}, {Barta}, {Bartlett}, {Bartos},
  {Bassiri}, {Basti}, {Bawaj}, {Bayley}, {Baylor}, {Bazzan}, {B{\'e}csy},
  {Bejger}, {Belahcene}, {Bell}, {Beniwal}, {Benjamin}, {Berger}, {Bergmann},
  {Bernuzzi}, {Berry}, {Bersanetti}, {Bertolini}, {Betzwieser}, {Bhandare},
  {Bidler}, {Biggs}, {Bilenko}, {Bilgili}, {Billingsley}, {Birney},
  {Birnholtz}, {Biscans}, {Bischi}, {Biscoveanu}, {Bisht}, {Bitossi},
  {Bizouard}, {Blackburn}, {Blackman}, {Blair}, {Blair}, {Blair}, {Bloemen},
  {Bobba}, {Bode}, {Boer}, {Boetzel}, {Bogaert}, {Bondu}, {Bonnand}, {Booker},
  {Boom}, {Bork}, {Boschi}, {Bose}, {Bossilkov}, {Bosveld}, {Bouffanais},
  {Bozzi}, {Bradaschia}, {Brady}, {Bramley}, {Branchesi}, {Brau}, {Breschi},
  {Briant}, {Briggs}, {Brighenti}, {Brillet}, {Brinkmann}, {Brockill},
  {Brooks}, {Brooks}, {Brown}, {Brunett}, {Buikema}, {Bulik}, {Bulten},
  {Buonanno}, {Buskulic}, {Buy}, {Byer}, {Cabero}, {Cadonati}, {Cagnoli},
  {Cahillane}, {Calder{\'o}n Bustillo}, {Callister}, {Calloni}, {Camp},
  {Campbell}, {Canepa}, {Cannon}, {Cao}, {Cao}, {Carapella}, {Carbognani},
  {Caride}, {Carney}, {Carullo}, {Casanueva Diaz}, {Casentini}, {Caudill},
  {Cavagli{\`a}}, {Cavalier}, {Cavalieri}, {Cella}, {Cerd{\'a}-Dur{\'a}n},
  {Cesarini}, {Chaibi}, {Chakravarti}, {Chamberlin}, {Chan}, {Chao},
  {Charlton}, {Chase}, {Chassande-Mottin}, {Chatterjee}, {Chaturvedi},
  {Chatziioannou}, {Cheeseboro}, {Chen}, {Chen}, {Chen}, {Cheng}, {Cheong},
  {Chia}, {Chiadini}, {Chincarini}, {Chiummo}, {Cho}, {Cho}, {Cho},
  {Christensen}, {Chu}, {Chua}, {Chung}, {Chung}, {Ciani}, {Cie{\'s}lar},
  {Ciobanu}, {Ciolfi}, {Cipriano}, {Cirone}, {Clara}, {Clark}, {Clearwater},
  {Cleva}, {Coccia}, {Cohadon}, {Cohen}, {Colleoni}, {Collette}, {Collins},
  {Colpi}, {Cominsky}, {Constancio}, {Conti}, {Cooper}, {Corban}, {Corbitt},
  {Cordero-Carri{\'o}n}, {Corezzi}, {Corley}, {Cornish}, {Corre}, {Corsi},
  {Cortese}, {Costa}, {Cotesta}, {Coughlin}, {Coughlin}, {Coulon},
  {Countryman}, {Couvares}, {Covas}, {Cowan}, {Coward}, {Cowart}, {Coyne},
  {Coyne}, {Creighton}, {Creighton}, {Cripe}, {Croquette}, {Crowder}, {Cullen},
  {Cumming}, {Cunningham}, {Cuoco}, {Dal Canton}, {D{\'a}lya}, {D'Angelo},
  {Danilishin}, {D'Antonio}, {Danzmann}, {Dasgupta}, {Da Silva Costa},
  {Datrier}, {Dattilo}, {Dave}, {Davier}, {Davis}, {Daw}, {DeBra},
  {Deenadayalan}, {Degallaix}, {De Laurentis}, {Del{\'e}glise}, {De Lillo},
  {Del Pozzo}, {DeMarchi}, {Demos}, {Dent}, {De Pietri}, {De Rosa}, {De Rossi},
  {DeSalvo}, {de Varona}, {Dhurandhar}, {D{\'\i}az}, {Dietrich}, {Di Fiore},
  {DiFronzo}, {Di Giorgio}, {Di Giovanni}, {Di Giovanni}, {Di Girolamo}, {Di
  Lieto}, {Ding}, {Di Pace}, {Di Palma}, {Di Renzo}, {Divakarla}, {Dmitriev},
  {Doctor}, {Donovan}, {Dooley}, {Doravari}, {Dorrington}, {Downes}, {Drago},
  {Driggers}, {Du}, {Ducoin}, {Dudi}, {Dupej}, {Durante}, {Dwyer}, {Easter},
  {Eddolls}, {Edo}, {Effler}, {Ehrens}, {Eichholz}, {Eikenberry}, {Eisenmann},
  {Eisenstein}, {Errico}, {Essick}, {Estelles}, {Estevez}, {Etienne}, {Etzel},
  {Evans}, {Evans}, {Fafone}, {Fairhurst}, {Fan}, {Farinon}, {Farr}, {Farr},
  {Fauchon-Jones}, {Favata}, {Fays}, {Fazio}, {Fee}, {Feicht}, {Fejer}, {Feng},
  {Fernandez-Galiana}, {Ferrante}, {Ferreira}, {Ferreira}, {Fidecaro}, {Fiori},
  {Fiorucci}, {Fishbach}, {Fisher}, {Fishner}, {Fittipaldi}, {Fitz-Axen},
  {Fiumara}, {Flaminio}, {Fletcher}, {Floden}, {Flynn}, {Fong}, {Font},
  {Forsyth}, {Fournier}, {Vivanco}, {Frasca}, {Frasconi}, {Frei}, {Freise},
  {Frey}, {Frey}, {Fritschel}, {Frolov}, {Fronz{\`e}}, {Fulda}, {Fyffe},
  {Gabbard}, {Gadre}, {Gaebel}, {Gair}, {Gamba}, {Gammaitoni}, {Gaonkar},
  {Garc{\'\i}a-Quir{\'o}s}, {Garufi}, {Gateley}, {Gaudio}, {Gaur}, {Gayathri},
  {Gemme}, {Genin}, {Gennai}, {George}, {George}, {George}, {Gergely},
  {Ghonge}, {Ghosh}, {Ghosh}, {Ghosh}, {Giacomazzo}, {Giaime}, {Giardina},
  {Gibson}, {Gill}, {Glover}, {Gniesmer}, {Godwin}, {Goetz}, {Goetz},
  {Goncharov}, {Gonz{\'a}lez}, {Castro}, {Gopakumar}, {Gossan}, {Gosselin},
  {Gouaty}, {Grace}, {Grado}, {Granata}, {Grant}, {Gras}, {Grassia}, {Gray},
  {Gray}, {Greco}, {Green}, {Green}, {Gretarsson}, {Grimaldi}, {Grimm},
  {Groot}, {Grote}, {Grunewald}, {Gruning}, {Guidi}, {Gulati}, {Guo}, {Gupta},
  {Gupta}, {Gupta}, {Gustafson}, {Gustafson}, {Haegel}, {Halim}, {Hall},
  {Hall}, {Hamilton}, {Hammond}, {Haney}, {Hanke}, {Hanks}, {Hanna}, {Hannam},
  {Hannuksela}, {Hansen}, {Hanson}, {Harder}, {Hardwick}, {Haris}, {Harms},
  {Harry}, {Harry}, {Hasskew}, {Haster}, {Haughian}, {Hayes}, {Healy},
  {Heidmann}, {Heintze}, {Heitmann}, {Hellman}, {Hello}, {Hemming}, {Hendry},
  {Heng}, {Hennig}, {Heurs}, {Hild}, {Hinderer}, {Ho}, {Hochheim}, {Hofman},
  {Holgado}, {Holland}, {Holt}, {Holz}, {Hopkins}, {Horst}, {Hough}, {Howell},
  {Hoy}, {Huang}, {H{\"u}bner}, {Huerta}, {Huet}, {Hughey}, {Hui}, {Husa},
  {Huttner}, {Huynh-Dinh}, {Idzkowski}, {Iess}, {Inchauspe}, {Ingram}, {Inta},
  {Intini}, {Irwin}, {Isa}, {Isac}, {Isi}, {Iyer}, {Jacqmin}, {Jadhav}, {Jani},
  {Janthalur}, {Jaranowski}, {Jariwala}, {Jenkins}, {Jiang}, {Johnson},
  {Johnson-McDaniel}, {Jones}, {Jones}, {Jones}, {Jones}, {Jonker}, {Ju},
  {Junker}, {Kalaghatgi}, {Kalogera}, {Kamai}, {Kandhasamy}, {Kang}, {Kanner},
  {Kapadia}, {Karki}, {Kashyap}, {Kasprzack}, {Kastaun}, {Katsanevas},
  {Katsavounidis}, {Katzman}, {Kaufer}, {Kawabe}, {Keerthana},
  {K{\'e}f{\'e}lian}, {Keitel}, {Kennedy}, {Key}, {Khalili}, {Khan}, {Khan},
  {Khazanov}, {Khetan}, {Khursheed}, {Kijbunchoo}, {Kim}, {Kim}, {Kim}, {Kim},
  {Kim}, {Kim}, {Kimball}, {King}, {Kinley-Hanlon}, {Kirchhoff}, {Kissel},
  {Kleybolte}, {Klika}, {Klimenko}, {Knowles}, {Koch}, {Koehlenbeck},
  {Koekoek}, {Koley}, {Kondrashov}, {Kontos}, {Koper}, {Korobko}, {Korth},
  {Kovalam}, {Kozak}, {Kr{\"a}mer}, {Kringel}, {Krishnendu}, {Kr{\'o}lak},
  {Krupinski}, {Kuehn}, {Kumar}, {Kumar}, {Kumar}, {Kumar}, {Kuo}, {Kutynia},
  {Kwang}, {Lackey}, {Laghi}, {Lai}, {Lam}, {Landry}, {Landry}, {Lane}, {Lang},
  {Lange}, {Lantz}, {Lanza}, {Lartaux-Vollard}, {Lasky}, {Laxen}, {Lazzarini},
  {Lazzaro}, {Leaci}, {Leavey}, {Lecoeuche}, {Lee}, {Lee}, {Lee}, {Lee}, {Lee},
  {Lee}, {Lehmann}, {Lenon}, {Leroy}, {Letendre}, {Levin}, {Li}, {Li}, {Li},
  {Li}, {Li}, {Lin}, {Linde}, {Linker}, {Littenberg}, {Liu}, {Liu},
  {Llorens-Monteagudo}, {Lo}, {London}, {Longo}, {Lorenzini}, {Loriette},
  {Lormand}, {Losurdo}, {Lough}, {Lousto}, {Lovelace}, {Lower}, {Lucaccioni},
  {L{\"u}ck}, {Lumaca}, {Lundgren}, {Lynch}, {Ma}, {Macas}, {Macfoy},
  {MacInnis}, {Macleod}, {Macquet}, {Maga{\~n}a Hernandez},
  {Maga{\~n}a-Sandoval}, {Magee}, {Majorana}, {Maksimovic}, {Malik}, {Man},
  {Mandic}, {Mangano}, {Mansell}, {Manske}, {Mantovani}, {Mapelli},
  {Marchesoni}, {Marion}, {M{\'a}rka}, {M{\'a}rka}, {Markakis}, {Markosyan},
  {Markowitz}, {Maros}, {Marquina}, {Marsat}, {Martelli}, {Martin}, {Martin},
  {Martinez}, {Martynov}, {Masalehdan}, {Mason}, {Massera}, {Masserot},
  {Massinger}, {Masso-Reid}, {Mastrogiovanni}, {Matas}, {Matichard}, {Matone},
  {Mavalvala}, {McCann}, {McCarthy}, {McClelland}, {McCormick}, {McCuller},
  {McGuire}, {McIsaac}, {McIver}, {McManus}, {McRae}, {McWilliams}, {Meacher},
  {Meadors}, {Mehmet}, {Mehta}, {Meidam}, {Mejuto Villa}, {Melatos}, {Mendell},
  {Mercer}, {Mereni}, {Merfeld}, {Merilh}, {Merzougui}, {Meshkov}, {Messenger},
  {Messick}, {Messina}, {Metzdorff}, {Meyers}, {Meylahn}, {Miani}, {Miao},
  {Michel}, {Middleton}, {Milano}, {Miller}, {Millhouse}, {Mills},
  {Milovich-Goff}, {Minazzoli}, {Minenkov}, {Mishkin}, {Mishra}, {Mistry},
  {Mitra}, {Mitrofanov}, {Mitselmakher}, {Mittleman}, {Mo}, {Moffa}, {Mogushi},
  {Mohapatra}, {Molina-Ruiz}, {Mondin}, {Montani}, {Moore}, {Moraru},
  {Morawski}, {Moreno}, {Morisaki}, {Mours}, {Mow-Lowry}, {Muciaccia},
  {Mukherjee}, {Mukherjee}, {Mukherjee}, {Mukherjee}, {Mukund}, {Mullavey},
  {Munch}, {Mu{\~n}iz}, {Muratore}, {Murray}, {Nagar}, {Nardecchia},
  {Naticchioni}, {Nayak}, {Neil}, {Neilson}, {Nelemans}, {Nelson}, {Nery},
  {Neunzert}, {Nevin}, {Ng}, {Ng}, {Nguyen}, {Nguyen}, {Nichols}, {Nichols},
  {Nissanke}, {Nocera}, {North}, {Nuttall}, {Obergaulinger}, {Oberling},
  {O'Brien}, {Oganesyan}, {Ogin}, {Oh}, {Oh}, {Ohme}, {Ohta}, {Okada},
  {Oliver}, {Oppermann}, {Oram}, {O'Reilly}, {Ormiston}, {Ortega},
  {O'Shaughnessy}, {Ossokine}, {Ottaway}, {Overmier}, {Owen}, {Pace}, {Pagano},
  {Page}, {Pagliaroli}, {Pai}, {Pai}, {Palamos}, {Palashov}, {Palomba}, {Pan},
  {Panda}, {Pang}, {Pankow}, {Pannarale}, {Pant}, {Paoletti}, {Paoli},
  {Parida}, {Parker}, {Pascucci}, {Pasqualetti}, {Passaquieti}, {Passuello},
  {Patil}, {Patricelli}, {Payne}, {Pearlstone}, {Pechsiri}, {Pedersen},
  {Pedraza}, {Pedurand}, {Pele}, {Penn}, {Perego}, {Perez}, {P{\'e}rigois},
  {Perreca}, {Petermann}, {Pfeiffer}, {Phelps}, {Phukon}, {Piccinni}, {Pichot},
  {Piergiovanni}, {Pierro}, {Pillant}, {Pinard}, {Pinto}, {Pirello}, {Pitkin},
  {Plastino}, {Poggiani}, {Pong}, {Ponrathnam}, {Popolizio}, {Porter},
  {Powell}, {Prajapati}, {Prasad}, {Prasai}, {Prasanna}, {Pratten},
  {Prestegard}, {Principe}, {Prodi}, {Prokhorov}, {Punturo}, {Puppo},
  {P{\"u}rrer}, {Qi}, {Quetschke}, {Quinonez}, {Raab}, {Raaijmakers},
  {Radkins}, {Radulesco}, {Raffai}, {Raja}, {Rajan}, {Rajbhandari},
  {Rakhmanov}, {Ramirez}, {Ramos-Buades}, {Rana}, {Rao}, {Rapagnani},
  {Raymond}, {Razzano}, {Read}, {Regimbau}, {Rei}, {Reid}, {Reitze},
  {Rettegno}, {Ricci}, {Richardson}, {Richardson}, {Ricker}, {Riemenschneider},
  {Riles}, {Rizzo}, {Robertson}, {Robinet}, {Rocchi}, {Rolland}, {Rollins},
  {Roma}, {Romanelli}, {Romano}, {Romel}, {Romie}, {Rose}, {Rose}, {Rose},
  {Rosell}, {Rosi{\'n}ska}, {Rosofsky}, {Ross}, {Rowan}, {Roy}, {R{\"u}diger},
  {Ruggi}, {Rutins}, {Ryan}, {Sachdev}, {Sadecki}, {Sakellariadou}, {Salafia},
  {Salconi}, {Saleem}, {Samajdar}, {Sammut}, {Sanchez}, {Sanchez},
  {Sanchis-Gual}, {Sanders}, {Santiago}, {Santos}, {Sarin}, {Sassolas},
  {Sathyaprakash}, {Sauter}, {Savage}, {Schale}, {Scheel}, {Scheuer},
  {Schmidt}, {Schnabel}, {Schofield}, {Sch{\"o}nbeck}, {Schreiber}, {Schulte},
  {Schutz}, {Scott}, {Scott}, {Seidel}, {Sellers}, {Sengupta}, {Sennett},
  {Sentenac}, {Sequino}, {Sergeev}, {Setyawati}, {Shaddock}, {Shaffer},
  {Shahriar}, {Shaner}, {Sharma}, {Sharma}, {Shawhan}, {Shen}, {Shink},
  {Shoemaker}, {Shoemaker}, {Shukla}, {ShyamSundar}, {Siellez}, {Sieniawska},
  {Sigg}, {Singer}, {Singh}, {Singh}, {Singhal}, {Sintes}, {Sitmukhambetov},
  {Skliris}, {Slagmolen}, {Slaven-Blair}, {Smith}, {Smith}, {Somala}, {Son},
  {Soni}, {Sorazu}, {Sorrentino}, {Souradeep}, {Sowell}, {Spencer}, {Spera},
  {Srivastava}, {Srivastava}, {Staats}, {Stachie}, {Standke}, {Steer},
  {Steinke}, {Steinlechner}, {Steinlechner}, {Steinmeyer}, {Stevenson},
  {Stocks}, {Stone}, {Stops}, {Strain}, {Stratta}, {Strigin}, {Strunk},
  {Sturani}, {Stuver}, {Sudhir}, {Summerscales}, {Sun}, {Sunil}, {Sur},
  {Suresh}, {Sutton}, {Swinkels}, {Szczepa{\'n}czyk}, {Tacca}, {Tait},
  {Talbot}, {Tanner}, {Tao}, {T{\'a}pai}, {Tapia}, {Tasson}, {Taylor},
  {Tenorio}, {Terkowski}, {Thomas}, {Thomas}, {Thondapu}, {Thorne}, {Thrane},
  {Tiwari}, {Tiwari}, {Tiwari}, {Toland}, {Tonelli}, {Tornasi},
  {Torres-Forn{\'e}}, {Torrie}, {T{\"o}yr{\"a}}, {Travasso}, {Traylor},
  {Tringali}, {Tripathee}, {Trovato}, {Trozzo}, {Tsang}, {Tse}, {Tso},
  {Tsukada}, {Tsuna}, {Tsutsui}, {Tuyenbayev}, {Ueno}, {Ugolini},
  {Unnikrishnan}, {Urban}, {Usman}, {Vahlbruch}, {Vajente}, {Valdes},
  {Valentini}, {van Bakel}, {van Beuzekom}, {van den Brand}, {Van Den Broeck},
  {Vander-Hyde}, {van der Schaaf}, {VanHeijningen}, {van Veggel}, {Vardaro},
  {Varma}, {Vass}, {Vas{\'u}th}, {Vecchio}, {Vedovato}, {Veitch}, {Veitch},
  {Venkateswara}, {Venugopalan}, {Verkindt}, {Vetrano}, {Vicer{\'e}}, {Viets},
  {Vinciguerra}, {Vine}, {Vinet}, {Vitale}, {Vo}, {Vocca}, {Vorvick},
  {Vyatchanin}, {Wade}, {Wade}, {Wade}, {Walet}, {Walker}, {Wallace}, {Walsh},
  {Wang}, {Wang}, {Wang}, {Wang}, {Ward}, {Warden}, {Warner}, {Was}, {Watchi},
  {Weaver}, {Wei}, {Weinert}, {Weinstein}, {Weiss}, {Wellmann}, {Wen},
  {Wessel}, {We{\ss}els}, {Westhouse}, {Wette}, {Whelan}, {White}, {Whiting},
  {Whittle}, {Wilken}, {Williams}, {Williamson}, {Willis}, {Willke}, {Winkler},
  {Wipf}, {Wittel}, {Woan}, {Woehler}, {Wofford}, {Wright}, {Wu}, {Wysocki},
  {Xiao}, {Xu}, {Yamamoto}, {Yancey}, {Yang}, {Yang}, {Yang}, {Yap}, {Yazback},
  {Yeeles}, {Yu}, {Yu}, {Yuen}, {Zadro{\.z}ny}, {Zadro{\.z}ny}, {Zanolin},
  {Zelenova}, {Zendri}, {Zevin}, {Zhang}, {Zhang}, {Zhang}, {Zhao}, {Zhao},
  {Zhou}, {Zhou}, {Zhu}, {Zimmerman}, {Zucker}, \& {Zweizig}}]{abbott2020}
{Abbott}, B.~P., {Abbott}, R., {Abbott}, T.~D., {et~al.} 2020, \apjl, 892, L3

\bibitem[{{Abbott} {et~al.}(2017{\natexlab{b}}){Abbott}, {Abbott}, {Abbott},
  {Acernese}, {Ackley}, {Adams}, {Adams}, {Addesso}, {Adhikari}, {Adya},
  {Affeldt}, {Afrough}, {Agarwal}, {Agathos}, {Agatsuma}, {Aggarwal}, {Aguiar},
  {Aiello}, {Ain}, {Ajith}, {Allen}, {Allen}, {Allocca}, {Aloy}, {Altin},
  {Amato}, {Ananyeva}, {Anderson}, {Anderson}, {Angelova}, {Antier}, {Appert},
  {Arai}, {Araya}, {Areeda}, {Arnaud}, {Arun}, {Ascenzi}, {Ashton}, {Ast},
  {Aston}, {Astone}, {Atallah}, {Aufmuth}, {Aulbert}, {AultONeal}, {Austin},
  {Avila-Alvarez}, {Babak}, {Bacon}, {Bader}, {Bae}, {Baker}, {Baldaccini},
  {Ballardin}, {Ballmer}, {Banagiri}, {Barayoga}, {Barclay}, {Barish},
  {Barker}, {Barkett}, {Barone}, {Barr}, {Barsotti}, {Barsuglia}, {Barta},
  {Bartlett}, {Bartos}, {Bassiri}, {Basti}, {Batch}, {Bawaj}, {Bayley},
  {Bazzan}, {B{\'e}csy}, {Beer}, {Bejger}, {Belahcene}, {Bell}, {Berger},
  {Bergmann}, {Bero}, {Berry}, {Bersanetti}, {Bertolini}, {Betzwieser},
  {Bhagwat}, {Bhandare}, {Bilenko}, {Billingsley}, {Billman}, {Birch},
  {Birney}, {Birnholtz}, {Biscans}, {Biscoveanu}, {Bisht}, {Bitossi}, {Biwer},
  {Bizouard}, {Blackburn}, {Blackman}, {Blair}, {Blair}, {Blair}, {Bloemen},
  {Bock}, {Bode}, {Boer}, {Bogaert}, {Bohe}, {Bondu}, {Bonilla}, {Bonnand},
  {Boom}, {Bork}, {Boschi}, {Bose}, {Bossie}, {Bouffanais}, {Bozzi},
  {Bradaschia}, {Brady}, {Branchesi}, {Brau}, {Briant}, {Brillet}, {Brinkmann},
  {Brisson}, {Brockill}, {Broida}, {Brooks}, {Brown}, {Brown}, {Brunett},
  {Buchanan}, {Buikema}, {Bulik}, {Bulten}, {Buonanno}, {Buskulic}, {Buy},
  {Byer}, {Cabero}, {Cadonati}, {Cagnoli}, {Cahillane}, {Calder{\'o}n
  Bustillo}, {Callister}, {Calloni}, {Camp}, {Canepa}, {Canizares}, {Cannon},
  {Cao}, {Cao}, {Capano}, {Capocasa}, {Carbognani}, {Caride}, {Carney},
  {Casanueva Diaz}, {Casentini}, {Caudill}, {Cavagli{\`a}}, {Cavalier},
  {Cavalieri}, {Cella}, {Cepeda}, {Cerd{\'a}-Dur{\'a}n}, {Cerretani},
  {Cesarini}, {Chamberlin}, {Chan}, {Chao}, {Charlton}, {Chase},
  {Chassande-Mottin}, {Chatterjee}, {Chatziioannou}, {Cheeseboro}, {Chen},
  {Chen}, {Chen}, {Cheng}, {Chia}, {Chincarini}, {Chiummo}, {Chmiel}, {Cho},
  {Cho}, {Chow}, {Christensen}, {Chu}, {Chua}, {Chua}, {Chung}, {Chung},
  {Ciani}, {Ciolfi}, {Cirelli}, {Cirone}, {Clara}, {Clark}, {Clearwater},
  {Cleva}, {Cocchieri}, {Coccia}, {Cohadon}, {Cohen}, {Colla}, {Collette},
  {Cominsky}, {Constancio}, {Conti}, {Cooper}, {Corban}, {Corbitt},
  {Cordero-Carri{\'o}n}, {Corley}, {Cornish}, {Corsi}, {Cortese}, {Costa},
  {Coughlin}, {Coughlin}, {Coulon}, {Countryman}, {Couvares}, {Covas}, {Cowan},
  {Coward}, {Cowart}, {Coyne}, {Coyne}, {Creighton}, {Creighton}, {Cripe},
  {Crowder}, {Cullen}, {Cumming}, {Cunningham}, {Cuoco}, {Dal Canton},
  {D{\'a}lya}, {Danilishin}, {D'Antonio}, {Danzmann}, {Dasgupta}, {Da Silva
  Costa}, {Dattilo}, {Dave}, {Davier}, {Davis}, {Daw}, {Day}, {De}, {DeBra},
  {Degallaix}, {De Laurentis}, {Del{\'e}glise}, {Del Pozzo}, {Demos}, {Denker},
  {Dent}, {De Pietri}, {Dergachev}, {De Rosa}, {DeRosa}, {De Rossi}, {DeSalvo},
  {de Varona}, {Devenson}, {Dhurandhar}, {D{\'\i}az}, {Di Fiore}, {Di
  Giovanni}, {Di Girolamo}, {Di Lieto}, {Di Pace}, {Di Palma}, {Di Renzo},
  {Doctor}, {Dolique}, {Donovan}, {Dooley}, {Doravari}, {Dorrington},
  {Douglas}, {Dovale {\'A}lvarez}, {Downes}, {Drago}, {Dreissigacker},
  {Driggers}, {Du}, {Ducrot}, {Dupej}, {Dwyer}, {Edo}, {Edwards}, {Effler},
  {Eggenstein}, {Ehrens}, {Eichholz}, {Eikenberry}, {Eisenstein}, {Essick},
  {Estevez}, {Etienne}, {Etzel}, {Evans}, {Evans}, {Factourovich}, {Fafone},
  {Fair}, {Fairhurst}, {Fan}, {Farinon}, {Farr}, {Farr}, {Fauchon-Jones},
  {Favata}, {Fays}, {Fee}, {Fehrmann}, {Feicht}, {Fejer}, {Fernandez-Galiana},
  {Ferrante}, {Ferreira}, {Ferrini}, {Fidecaro}, {Finstad}, {Fiori},
  {Fiorucci}, {Fishbach}, {Fisher}, {Fitz-Axen}, {Flaminio}, {Fletcher},
  {Fong}, {Font}, {Forsyth}, {Forsyth}, {Fournier}, {Frasca}, {Frasconi},
  {Frei}, {Freise}, {Frey}, {Frey}, {Fries}, {Fritschel}, {Frolov}, {Fulda},
  {Fyffe}, {Gabbard}, {Gadre}, {Gaebel}, {Gair}, {Gammaitoni}, {Ganija},
  {Gaonkar}, {Garcia-Quiros}, {Garufi}, {Gateley}, {Gaudio}, {Gaur},
  {Gayathri}, {Gehrels}, {Gemme}, {Genin}, {Gennai}, {George}, {George},
  {Gergely}, {Germain}, {Ghonge}, {Ghosh}, {Ghosh}, {Ghosh}, {Giaime},
  {Giardina}, {Giazotto}, {Gill}, {Glover}, {Goetz}, {Goetz}, {Gomes},
  {Goncharov}, {Gonz{\'a}lez}, {Gonzalez Castro}, {Gopakumar}, {Gorodetsky},
  {Gossan}, {Gosselin}, {Gouaty}, {Grado}, {Graef}, {Granata}, {Grant}, {Gras},
  {Gray}, {Greco}, {Green}, {Gretarsson}, {Groot}, {Grote}, {Grunewald},
  {Gruning}, {Guidi}, {Guo}, {Gupta}, {Gupta}, {Gushwa}, {Gustafson},
  {Gustafson}, {Halim}, {Hall}, {Hall}, {Hamilton}, {Hammond}, {Haney},
  {Hanke}, {Hanks}, {Hanna}, {Hannam}, {Hannuksela}, {Hanson}, {Hardwick},
  {Harms}, {Harry}, {Harry}, {Hart}, {Haster}, {Haughian}, {Healy}, {Heidmann},
  {Heintze}, {Heitmann}, {Hello}, {Hemming}, {Hendry}, {Heng}, {Hennig},
  {Heptonstall}, {Heurs}, {Hild}, {Hinderer}, {Hoak}, {Hofman}, {Holt}, {Holz},
  {Hopkins}, {Horst}, {Hough}, {Houston}, {Howell}, {Hreibi}, {Hu}, {Huerta},
  {Huet}, {Hughey}, {Husa}, {Huttner}, {Huynh-Dinh}, {Indik}, {Inta}, {Intini},
  {Isa}, {Isac}, {Isi}, {Iyer}, {Izumi}, {Jacqmin}, {Jani}, {Jaranowski},
  {Jawahar}, {Jim{\'e}nez-Forteza}, {Johnson}, {Johnson-McDaniel}, {Jones},
  {Jones}, {Jonker}, {Ju}, {Junker}, {Kalaghatgi}, {Kalogera}, {Kamai},
  {Kandhasamy}, {Kang}, {Kanner}, {Kapadia}, {Karki}, {Karvinen}, {Kasprzack},
  {Kastaun}, {Katolik}, {Katsavounidis}, {Katzman}, {Kaufer}, {Kawabe},
  {K{\'e}f{\'e}lian}, {Keitel}, {Kemball}, {Kennedy}, {Kent}, {Key}, {Khalili},
  {Khan}, {Khan}, {Khan}, {Khazanov}, {Kijbunchoo}, {Kim}, {Kim}, {Kim}, {Kim},
  {Kim}, {Kim}, {Kimbrell}, {King}, {King}, {Kinley-Hanlon}, {Kirchhoff},
  {Kissel}, {Kleybolte}, {Klimenko}, {Knowles}, {Koch}, {Koehlenbeck}, {Koley},
  {Kondrashov}, {Kontos}, {Korobko}, {Korth}, {Kowalska}, {Kozak},
  {Kr{\"a}mer}, {Kringel}, {Krishnan}, {Kr{\'o}lak}, {Kuehn}, {Kumar}, {Kumar},
  {Kumar}, {Kuo}, {Kutynia}, {Kwang}, {Lackey}, {Lai}, {Landry}, {Lang},
  {Lange}, {Lantz}, {Lanza}, {Lartaux-Vollard}, {Lasky}, {Laxen}, {Lazzarini},
  {Lazzaro}, {Leaci}, {Leavey}, {Lee}, {Lee}, {Lee}, {Lee}, {Lee}, {Lehmann},
  {Lenon}, {Leonardi}, {Leroy}, {Letendre}, {Levin}, {Li}, {Linker},
  {Littenberg}, {Liu}, {Lo}, {Lockerbie}, {London}, {Lord}, {Lorenzini},
  {Loriette}, {Lormand}, {Losurdo}, {Lough}, {Lousto}, {Lovelace}, {L{\"u}ck},
  {Lumaca}, {Lundgren}, {Lynch}, {Ma}, {Macas}, {Macfoy}, {Machenschalk},
  {MacInnis}, {Macleod}, {Maga{\~n}a Hernandez}, {Maga{\~n}a-Sandoval},
  {Maga{\~n}a Zertuche}, {Magee}, {Majorana}, {Maksimovic}, {Man}, {Mandic},
  {Mangano}, {Mansell}, {Manske}, {Mantovani}, {Marchesoni}, {Marion},
  {M{\'a}rka}, {M{\'a}rka}, {Markakis}, {Markosyan}, {Markowitz}, {Maros},
  {Marquina}, {Martelli}, {Martellini}, {Martin}, {Martin}, {Martynov},
  {Mason}, {Massera}, {Masserot}, {Massinger}, {Masso-Reid}, {Mastrogiovanni},
  {Matas}, {Matichard}, {Matone}, {Mavalvala}, {Mazumder}, {McCarthy},
  {McClelland}, {McCormick}, {McCuller}, {McGuire}, {McIntyre}, {McIver},
  {McManus}, {McNeill}, {McRae}, {McWilliams}, {Meacher}, {Meadors}, {Mehmet},
  {Meidam}, {Mejuto-Villa}, {Melatos}, {Mendell}, {Mercer}, {Merilh},
  {Merzougui}, {Meshkov}, {Messenger}, {Messick}, {Metzdorff}, {Meyers},
  {Miao}, {Michel}, {Middleton}, {Mikhailov}, {Milano}, {Miller}, {Miller},
  {Miller}, {Millhouse}, {Milovich-Goff}, {Minazzoli}, {Minenkov}, {Ming},
  {Mishra}, {Mitra}, {Mitrofanov}, {Mitselmakher}, {Mittleman}, {Moffa},
  {Moggi}, {Mogushi}, {Mohan}, {Mohapatra}, {Montani}, {Moore}, {Moraru},
  {Moreno}, {Morriss}, {Mours}, {Mow-Lowry}, {Mueller}, {Muir}, {Mukherjee},
  {Mukherjee}, {Mukherjee}, {Mukund}, {Mullavey}, {Munch}, {Mu{\~n}iz},
  {Muratore}, {Murray}, {Napier}, {Nardecchia}, {Naticchioni}, {Nayak},
  {Neilson}, {Nelemans}, {Nelson}, {Nery}, {Neunzert}, {Nevin}, {Newport},
  {Newton}, {Ng}, {Nguyen}, {Nichols}, {Nielsen}, {Nissanke}, {Nitz}, {Noack},
  {Nocera}, {Nolting}, {North}, {Nuttall}, {Oberling}, {O'Dea}, {Ogin}, {Oh},
  {Oh}, {Ohme}, {Okada}, {Oliver}, {Oppermann}, {Oram}, {O'Reilly}, {Ormiston},
  {Ortega}, {O'Shaughnessy}, {Ossokine}, {Ottaway}, {Overmier}, {Owen}, {Pace},
  {Page}, {Page}, {Pai}, {Pai}, {Palamos}, {Palashov}, {Palomba}, {Pal-Singh},
  {Pan}, {Pan}, {Pang}, {Pang}, {Pankow}, {Pannarale}, {Pant}, {Paoletti},
  {Paoli}, {Papa}, {Parida}, {Parker}, {Pascucci}, {Pasqualetti},
  {Passaquieti}, {Passuello}, {Patil}, {Patricelli}, {Pearlstone}, {Pedraza},
  {Pedurand}, {Pekowsky}, {Pele}, {Penn}, {Perez}, {Perreca}, {Perri},
  {Pfeiffer}, {Phelps}, {Piccinni}, {Pichot}, {Piergiovanni}, {Pierro},
  {Pillant}, {Pinard}, {Pinto}, {Pirello}, {Pitkin}, {Poe}, {Poggiani},
  {Popolizio}, {Porter}, {Post}, {Powell}, {Prasad}, {Pratt}, {Pratten},
  {Predoi}, {Prestegard}, {Prijatelj}, {Principe}, {Privitera}, {Prodi},
  {Prokhorov}, {Puncken}, {Punturo}, {Puppo}, {P{\"u}rrer}, {Qi}, {Quetschke},
  {Quintero}, {Quitzow-James}, {Raab}, {Rabeling}, {Radkins}, {Raffai}, {Raja},
  {Rajan}, {Rajbhandari}, {Rakhmanov}, {Ramirez}, {Ramos-Buades}, {Rapagnani},
  {Raymond}, {Razzano}, {Read}, {Regimbau}, {Rei}, {Reid}, {Reitze}, {Ren},
  {Reyes}, {Ricci}, {Ricker}, {Rieger}, {Riles}, {Rizzo}, {Robertson}, {Robie},
  {Robinet}, {Rocchi}, {Rolland}, {Rollins}, {Roma}, {Romano}, {Romel},
  {Romie}, {Rosi{\'n}ska}, {Ross}, {Rowan}, {R{\"u}diger}, {Ruggi}, {Rutins},
  {Ryan}, {Sachdev}, {Sadecki}, {Sadeghian}, {Sakellariadou}, {Salconi},
  {Saleem}, {Salemi}, {Samajdar}, {Sammut}, {Sampson}, {Sanchez}, {Sanchez},
  {Sanchis-Gual}, {Sandberg}, {Sanders}, {Sassolas}, {Sathyaprakash},
  {Saulson}, {Sauter}, {Savage}, {Sawadsky}, {Schale}, {Scheel}, {Scheuer},
  {Schmidt}, {Schmidt}, {Schnabel}, {Schofield}, {Sch{\"o}nbeck}, {Schreiber},
  {Schuette}, {Schulte}, {Schutz}, {Schwalbe}, {Scott}, {Scott}, {Seidel},
  {Sellers}, {Sengupta}, {Sentenac}, {Sequino}, {Sergeev}, {Shaddock},
  {Shaffer}, {Shah}, {Shahriar}, {Shaner}, {Shao}, {Shapiro}, {Shawhan},
  {Sheperd}, {Shoemaker}, {Shoemaker}, {Siellez}, {Siemens}, {Sieniawska},
  {Sigg}, {Silva}, {Singer}, {Singh}, {Singhal}, {Sintes}, {Slagmolen},
  {Smith}, {Smith}, {Smith}, {Somala}, {Son}, {Sonnenberg}, {Sorazu},
  {Sorrentino}, {Souradeep}, {Spencer}, {Srivastava}, {Staats}, {Staley},
  {Steinke}, {Steinlechner}, {Steinlechner}, {Steinmeyer}, {Stevenson},
  {Stone}, {Stops}, {Strain}, {Stratta}, {Strigin}, {Strunk}, {Sturani},
  {Stuver}, {Summerscales}, {Sun}, {Sunil}, {Suresh}, {Sutton}, {Swinkels},
  {Szczepa{\'n}czyk}, {Tacca}, {Tait}, {Talbot}, {Talukder}, {Tanner},
  {T{\'a}pai}, {Taracchini}, {Tasson}, {Taylor}, {Taylor}, {Tewari}, {Theeg},
  {Thies}, {Thomas}, {Thomas}, {Thomas}, {Thorne}, {Thorne}, {Thrane},
  {Tiwari}, {Tiwari}, {Tokmakov}, {Toland}, {Tonelli}, {Tornasi},
  {Torres-Forn{\'e}}, {Torrie}, {T{\"o}yr{\"a}}, {Travasso}, {Traylor},
  {Trinastic}, {Tringali}, {Trozzo}, {Tsang}, {Tse}, {Tso}, {Tsukada}, {Tsuna},
  {Tuyenbayev}, {Ueno}, {Ugolini}, {Unnikrishnan}, {Urban}, {Usman},
  {Vahlbruch}, {Vajente}, {Valdes}, {van Bakel}, {van Beuzekom}, {van den
  Brand}, {Van Den Broeck}, {Vander-Hyde}, {van der Schaaf}, {van Heijningen},
  {van Veggel}, {Vardaro}, {Varma}, {Vass}, {Vas{\'u}th}, {Vecchio},
  {Vedovato}, {Veitch}, {Veitch}, {Venkateswara}, {Venugopalan}, {Verkindt},
  {Vetrano}, {Vicer{\'e}}, {Viets}, {Vinciguerra}, {Vine}, {Vinet}, {Vitale},
  {Vo}, {Vocca}, {Vorvick}, {Vyatchanin}, {Wade}, {Wade}, {Wade}, {Walet},
  {Walker}, {Wallace}, {Walsh}, {Wang}, {Wang}, {Wang}, {Wang}, {Wang}, {Ward},
  {Warner}, {Was}, {Watchi}, {Weaver}, {Wei}, {Weinert}, {Weinstein}, {Weiss},
  {Wen}, {Wessel}, {We{\ss}els}, {Westerweck}, {Westphal}, {Wette}, {Whelan},
  {Whitcomb}, {Whiting}, {Whittle}, {Wilken}, {Williams}, {Williams},
  {Williamson}, {Willis}, {Willke}, {Wimmer}, {Winkler}, {Wipf}, {Wittel},
  {Woan}, {Woehler}, {Wofford}, {Wong}, {Worden}, {Wright}, {Wu}, {Wysocki},
  {Xiao}, {Yamamoto}, {Yancey}, {Yang}, {Yap}, {Yazback}, {Yu}, {Yu}, {Yvert},
  {Zadro{\.z}ny}, {Zanolin}, {Zelenova}, {Zendri}, {Zevin}, {Zhang}, {Zhang},
  {Zhang}, {Zhang}, {Zhao}, {Zhou}, {Zhou}, {Zhu}, {Zhu}, {Zimmerman},
  {Zucker}, {Zweizig}, {(LIGO Scientific Collaboration}, {Virgo Collaboration},
  {Burns}, {Veres}, {Kocevski}, {Racusin}, {Goldstein}, {Connaughton},
  {Briggs}, {Blackburn}, {Hamburg}, {Hui}, {von Kienlin}, {McEnery}, {Preece},
  {Wilson-Hodge}, {Bissaldi}, {Cleveland}, {Gibby}, {Giles}, {Kippen},
  {McBreen}, {Meegan}, {Paciesas}, {Poolakkil}, {Roberts}, {Stanbro},
  {Gamma-ray Burst Monitor}, {Savchenko}, {Ferrigno}, {Kuulkers}, {Bazzano},
  {Bozzo}, {Brandt}, {Chenevez}, {Courvoisier}, {Diehl}, {Domingo}, {Hanlon},
  {Jourdain}, {Laurent}, {Lebrun}, {Lutovinov}, {Mereghetti}, {Natalucci},
  {Rodi}, {Roques}, {Sunyaev}, {Ubertini}, \& {(INTEGRAL}}]{2017c}
{Abbott}, B.~P., {Abbott}, R., {Abbott}, T.~D., {et~al.} 2017{\natexlab{b}},
  \apjl, 848, L13

\bibitem[{{Abbott} {et~al.}(2017{\natexlab{c}}){Abbott}, {Abbott}, {Abbott},
  {Acernese}, {Ackley}, {Adams}, {Adams}, {Addesso}, {Adhikari}, {Adya},
  {Affeldt}, {Afrough}, {Agarwal}, {Agathos}, {Agatsuma}, {Aggarwal}, {Aguiar},
  {Aiello}, {Ain}, {Ajith}, {Allen}, {Allen}, {Allocca}, {Altin}, {Amato},
  {Ananyeva}, {Anderson}, {Anderson}, {Angelova}, {Antier}, {Appert}, {Arai},
  {Araya}, {Areeda}, {Arnaud}, {Arun}, {Ascenzi}, {Ashton}, {Ast}, {Aston},
  {Astone}, {Atallah}, {Aufmuth}, {Aulbert}, {AultONeal}, {Austin},
  {Avila-Alvarez}, {Babak}, {Bacon}, {Bader}, {Bae}, {Bailes}, {Baker},
  {Baldaccini}, {Ballardin}, {Ballmer}, {Banagiri}, {Barayoga}, {Barclay},
  {Barish}, {Barker}, {Barkett}, {Barone}, {Barr}, {Barsotti}, {Barsuglia},
  {Barta}, {Barthelmy}, {Bartlett}, {Bartos}, {Bassiri}, {Basti}, {Batch},
  {Bawaj}, {Bayley}, {Bazzan}, {B{\'e}csy}, {Beer}, {Bejger}, {Belahcene},
  {Bell}, {Berger}, {Bergmann}, {Bernuzzi}, {Bero}, {Berry}, {Bersanetti},
  {Bertolini}, {Betzwieser}, {Bhagwat}, {Bhandare}, {Bilenko}, {Billingsley},
  {Billman}, {Birch}, {Birney}, {Birnholtz}, {Biscans}, {Biscoveanu}, {Bisht},
  {Bitossi}, {Biwer}, {Bizouard}, {Blackburn}, {Blackman}, {Blair}, {Blair},
  {Blair}, {Bloemen}, {Bock}, {Bode}, {Boer}, {Bogaert}, {Bohe}, {Bondu},
  {Bonilla}, {Bonnand}, {Boom}, {Bork}, {Boschi}, {Bose}, {Bossie},
  {Bouffanais}, {Bozzi}, {Bradaschia}, {Brady}, {Branchesi}, {Brau}, {Briant},
  {Brillet}, {Brinkmann}, {Brisson}, {Brockill}, {Broida}, {Brooks}, {Brown},
  {Brown}, {Brunett}, {Buchanan}, {Buikema}, {Bulik}, {Bulten}, {Buonanno},
  {Buskulic}, {Buy}, {Byer}, {Cabero}, {Cadonati}, {Cagnoli}, {Cahillane},
  {Calder{\'o}n Bustillo}, {Callister}, {Calloni}, {Camp}, {Canepa},
  {Canizares}, {Cannon}, {Cao}, {Cao}, {Capano}, {Capocasa}, {Carbognani},
  {Caride}, {Carney}, {Carullo}, {Casanueva Diaz}, {Casentini}, {Caudill},
  {Cavagli{\`a}}, {Cavalier}, {Cavalieri}, {Cella}, {Cepeda},
  {Cerd{\'a}-Dur{\'a}n}, {Cerretani}, {Cesarini}, {Chamberlin}, {Chan}, {Chao},
  {Charlton}, {Chase}, {Chassande-Mottin}, {Chatterjee}, {Chatziioannou},
  {Cheeseboro}, {Chen}, {Chen}, {Chen}, {Cheng}, {Chia}, {Chincarini},
  {Chiummo}, {Chmiel}, {Cho}, {Cho}, {Chow}, {Christensen}, {Chu}, {Chua},
  {Chua}, {Chung}, {Chung}, {Ciani}, {Ciolfi}, {Cirelli}, {Cirone}, {Clara},
  {Clark}, {Clearwater}, {Cleva}, {Cocchieri}, {Coccia}, {Cohadon}, {Cohen},
  {Colla}, {Collette}, {Cominsky}, {Constancio}, {Conti}, {Cooper}, {Corban},
  {Corbitt}, {Cordero-Carri{\'o}n}, {Corley}, {Cornish}, {Corsi}, {Cortese},
  {Costa}, {Coughlin}, {Coughlin}, {Coulon}, {Countryman}, {Couvares}, {Covas},
  {Cowan}, {Coward}, {Cowart}, {Coyne}, {Coyne}, {Creighton}, {Creighton},
  {Cripe}, {Crowder}, {Cullen}, {Cumming}, {Cunningham}, {Cuoco}, {Dal Canton},
  {D{\'a}lya}, {Danilishin}, {D'Antonio}, {Danzmann}, {Dasgupta}, {Da Silva
  Costa}, {Dattilo}, {Dave}, {Davier}, {Davis}, {Daw}, {Day}, {De}, {DeBra},
  {Degallaix}, {De Laurentis}, {Del{\'e}glise}, {Del Pozzo}, {Demos}, {Denker},
  {Dent}, {De Pietri}, {Dergachev}, {De Rosa}, {DeRosa}, {De Rossi}, {DeSalvo},
  {de Varona}, {Devenson}, {Dhurandhar}, {D{\'\i}az}, {Dietrich}, {Di Fiore},
  {Di Giovanni}, {Di Girolamo}, {Di Lieto}, {Di Pace}, {Di Palma}, {Di Renzo},
  {Doctor}, {Dolique}, {Donovan}, {Dooley}, {Doravari}, {Dorrington},
  {Douglas}, {Dovale {\'A}lvarez}, {Downes}, {Drago}, {Dreissigacker},
  {Driggers}, {Du}, {Ducrot}, {Dudi}, {Dupej}, {Dwyer}, {Edo}, {Edwards},
  {Effler}, {Eggenstein}, {Ehrens}, {Eichholz}, {Eikenberry}, {Eisenstein},
  {Essick}, {Estevez}, {Etienne}, {Etzel}, {Evans}, {Evans}, {Factourovich},
  {Fafone}, {Fair}, {Fairhurst}, {Fan}, {Farinon}, {Farr}, {Farr},
  {Fauchon-Jones}, {Favata}, {Fays}, {Fee}, {Fehrmann}, {Feicht}, {Fejer},
  {Fernandez-Galiana}, {Ferrante}, {Ferreira}, {Ferrini}, {Fidecaro},
  {Finstad}, {Fiori}, {Fiorucci}, {Fishbach}, {Fisher}, {Fitz-Axen},
  {Flaminio}, {Fletcher}, {Fong}, {Font}, {Forsyth}, {Forsyth}, {Fournier},
  {Frasca}, {Frasconi}, {Frei}, {Freise}, {Frey}, {Frey}, {Fries}, {Fritschel},
  {Frolov}, {Fulda}, {Fyffe}, {Gabbard}, {Gadre}, {Gaebel}, {Gair},
  {Gammaitoni}, {Ganija}, {Gaonkar}, {Garcia-Quiros}, {Garufi}, {Gateley},
  {Gaudio}, {Gaur}, {Gayathri}, {Gehrels}, {Gemme}, {Genin}, {Gennai},
  {George}, {George}, {Gergely}, {Germain}, {Ghonge}, {Ghosh}, {Ghosh},
  {Ghosh}, {Giaime}, {Giardina}, {Giazotto}, {Gill}, {Glover}, {Goetz},
  {Goetz}, {Gomes}, {Goncharov}, {Gonz{\'a}lez}, {Gonzalez Castro},
  {Gopakumar}, {Gorodetsky}, {Gossan}, {Gosselin}, {Gouaty}, {Grado}, {Graef},
  {Granata}, {Grant}, {Gras}, {Gray}, {Greco}, {Green}, {Gretarsson}, {Groot},
  {Grote}, {Grunewald}, {Gruning}, {Guidi}, {Guo}, {Gupta}, {Gupta}, {Gushwa},
  {Gustafson}, {Gustafson}, {Halim}, {Hall}, {Hall}, {Hamilton}, {Hammond},
  {Haney}, {Hanke}, {Hanks}, {Hanna}, {Hannam}, {Hannuksela}, {Hanson},
  {Hardwick}, {Harms}, {Harry}, {Harry}, {Hart}, {Haster}, {Haughian}, {Healy},
  {Heidmann}, {Heintze}, {Heitmann}, {Hello}, {Hemming}, {Hendry}, {Heng},
  {Hennig}, {Heptonstall}, {Heurs}, {Hild}, {Hinderer}, {Ho}, {Hoak}, {Hofman},
  {Holt}, {Holz}, {Hopkins}, {Horst}, {Hough}, {Houston}, {Howell}, {Hreibi},
  {Hu}, {Huerta}, {Huet}, {Hughey}, {Husa}, {Huttner}, {Huynh-Dinh}, {Indik},
  {Inta}, {Intini}, {Isa}, {Isac}, {Isi}, {Iyer}, {Izumi}, {Jacqmin}, {Jani},
  {Jaranowski}, {Jawahar}, {Jim{\'e}nez-Forteza}, {Johnson},
  {Johnson-McDaniel}, {Jones}, {Jones}, {Jonker}, {Ju}, {Junker}, {Kalaghatgi},
  {Kalogera}, {Kamai}, {Kandhasamy}, {Kang}, {Kanner}, {Kapadia}, {Karki},
  {Karvinen}, {Kasprzack}, {Kastaun}, {Katolik}, {Katsavounidis}, {Katzman},
  {Kaufer}, {Kawabe}, {K{\'e}f{\'e}lian}, {Keitel}, {Kemball}, {Kennedy},
  {Kent}, {Key}, {Khalili}, {Khan}, {Khan}, {Khan}, {Khazanov}, {Kijbunchoo},
  {Kim}, {Kim}, {Kim}, {Kim}, {Kim}, {Kim}, {Kimbrell}, {King}, {King},
  {Kinley-Hanlon}, {Kirchhoff}, {Kissel}, {Kleybolte}, {Klimenko}, {Knowles},
  {Koch}, {Koehlenbeck}, {Koley}, {Kondrashov}, {Kontos}, {Korobko}, {Korth},
  {Kowalska}, {Kozak}, {Kr{\"a}mer}, {Kringel}, {Krishnan}, {Kr{\'o}lak},
  {Kuehn}, {Kumar}, {Kumar}, {Kumar}, {Kuo}, {Kutynia}, {Kwang}, {Lackey},
  {Lai}, {Landry}, {Lang}, {Lange}, {Lantz}, {Lanza}, {Larson},
  {Lartaux-Vollard}, {Lasky}, {Laxen}, {Lazzarini}, {Lazzaro}, {Leaci},
  {Leavey}, {Lee}, {Lee}, {Lee}, {Lee}, {Lee}, {Lehmann}, {Lenon}, {Leon},
  {Leonardi}, {Leroy}, {Letendre}, {Levin}, {Li}, {Linker}, {Littenberg},
  {Liu}, {Liu}, {Lo}, {Lockerbie}, {London}, {Lord}, {Lorenzini}, {Loriette},
  {Lormand}, {Losurdo}, {Lough}, {Lousto}, {Lovelace}, {L{\"u}ck}, {Lumaca},
  {Lundgren}, {Lynch}, {Ma}, {Macas}, {Macfoy}, {Machenschalk}, {MacInnis},
  {Macleod}, {Maga{\~n}a Hernandez}, {Maga{\~n}a-Sandoval}, {Maga{\~n}a
  Zertuche}, {Magee}, {Majorana}, {Maksimovic}, {Man}, {Mandic}, {Mangano},
  {Mansell}, {Manske}, {Mantovani}, {Marchesoni}, {Marion}, {M{\'a}rka},
  {M{\'a}rka}, {Markakis}, {Markosyan}, {Markowitz}, {Maros}, {Marquina},
  {Marsh}, {Martelli}, {Martellini}, {Martin}, {Martin}, {Martynov}, {Marx},
  {Mason}, {Massera}, {Masserot}, {Massinger}, {Masso-Reid}, {Mastrogiovanni},
  {Matas}, {Matichard}, {Matone}, {Mavalvala}, {Mazumder}, {McCarthy},
  {McClelland}, {McCormick}, {McCuller}, {McGuire}, {McIntyre}, {McIver},
  {McManus}, {McNeill}, {McRae}, {McWilliams}, {Meacher}, {Meadors}, {Mehmet},
  {Meidam}, {Mejuto-Villa}, {Melatos}, {Mendell}, {Mercer}, {Merilh},
  {Merzougui}, {Meshkov}, {Messenger}, {Messick}, {Metzdorff}, {Meyers},
  {Miao}, {Michel}, {Middleton}, {Mikhailov}, {Milano}, {Miller}, {Miller},
  {Miller}, {Millhouse}, {Milovich-Goff}, {Minazzoli}, {Minenkov}, {Ming},
  {Mishra}, {Mitra}, {Mitrofanov}, {Mitselmakher}, {Mittleman}, {Moffa},
  {Moggi}, {Mogushi}, {Mohan}, {Mohapatra}, {Molina}, {Montani}, {Moore},
  {Moraru}, {Moreno}, {Morisaki}, {Morriss}, {Mours}, {Mow-Lowry}, {Mueller},
  {Muir}, {Mukherjee}, {Mukherjee}, {Mukherjee}, {Mukund}, {Mullavey}, {Munch},
  {Mu{\~n}iz}, {Muratore}, {Murray}, {Nagar}, {Napier}, {Nardecchia},
  {Naticchioni}, {Nayak}, {Neilson}, {Nelemans}, {Nelson}, {Nery}, {Neunzert},
  {Nevin}, {Newport}, {Newton}, {Ng}, {Nguyen}, {Nguyen}, {Nichols}, {Nielsen},
  {Nissanke}, {Nitz}, {Noack}, {Nocera}, {Nolting}, {North}, {Nuttall},
  {Oberling}, {O'Dea}, {Ogin}, {Oh}, {Oh}, {Ohme}, {Okada}, {Oliver},
  {Oppermann}, {Oram}, {O'Reilly}, {Ormiston}, {Ortega}, {O'Shaughnessy},
  {Ossokine}, {Ottaway}, {Overmier}, {Owen}, {Pace}, {Page}, {Page}, {Pai},
  {Pai}, {Palamos}, {Palashov}, {Palomba}, {Pal-Singh}, {Pan}, {Pan}, {Pang},
  {Pang}, {Pankow}, {Pannarale}, {Pant}, {Paoletti}, {Paoli}, {Papa}, {Parida},
  {Parker}, {Pascucci}, {Pasqualetti}, {Passaquieti}, {Passuello}, {Patil},
  {Patricelli}, {Pearlstone}, {Pedraza}, {Pedurand}, {Pekowsky}, {Pele},
  {Penn}, {Perez}, {Perreca}, {Perri}, {Pfeiffer}, {Phelps}, {Piccinni},
  {Pichot}, {Piergiovanni}, {Pierro}, {Pillant}, {Pinard}, {Pinto}, {Pirello},
  {Pitkin}, {Poe}, {Poggiani}, {Popolizio}, {Porter}, {Post}, {Powell},
  {Prasad}, {Pratt}, {Pratten}, {Predoi}, {Prestegard}, {Prijatelj},
  {Principe}, {Privitera}, {Prix}, {Prodi}, {Prokhorov}, {Puncken}, {Punturo},
  {Puppo}, {P{\"u}rrer}, {Qi}, {Quetschke}, {Quintero}, {Quitzow-James},
  {Raab}, {Rabeling}, {Radkins}, {Raffai}, {Raja}, {Rajan}, {Rajbhandari},
  {Rakhmanov}, {Ramirez}, {Ramos-Buades}, {Rapagnani}, {Raymond}, {Razzano},
  {Read}, {Regimbau}, {Rei}, {Reid}, {Reitze}, {Ren}, {Reyes}, {Ricci},
  {Ricker}, {Rieger}, {Riles}, {Rizzo}, {Robertson}, {Robie}, {Robinet},
  {Rocchi}, {Rolland}, {Rollins}, {Roma}, {Romano}, {Romano}, {Romel}, {Romie},
  {Rosi{\'n}ska}, {Ross}, {Rowan}, {R{\"u}diger}, {Ruggi}, {Rutins}, {Ryan},
  {Sachdev}, {Sadecki}, {Sadeghian}, {Sakellariadou}, {Salconi}, {Saleem},
  {Salemi}, {Samajdar}, {Sammut}, {Sampson}, {Sanchez}, {Sanchez},
  {Sanchis-Gual}, {Sandberg}, {Sanders}, {Sassolas}, {Sathyaprakash},
  {Saulson}, {Sauter}, {Savage}, {Sawadsky}, {Schale}, {Scheel}, {Scheuer},
  {Schmidt}, {Schmidt}, {Schnabel}, {Schofield}, {Sch{\"o}nbeck}, {Schreiber},
  {Schuette}, {Schulte}, {Schutz}, {Schwalbe}, {Scott}, {Scott}, {Seidel},
  {Sellers}, {Sengupta}, {Sentenac}, {Sequino}, {Sergeev}, {Shaddock},
  {Shaffer}, {Shah}, {Shahriar}, {Shaner}, {Shao}, {Shapiro}, {Shawhan},
  {Sheperd}, {Shoemaker}, {Shoemaker}, {Siellez}, {Siemens}, {Sieniawska},
  {Sigg}, {Silva}, {Singer}, {Singh}, {Singhal}, {Sintes}, {Slagmolen},
  {Smith}, {Smith}, {Smith}, {Somala}, {Son}, {Sonnenberg}, {Sorazu},
  {Sorrentino}, {Souradeep}, {Spencer}, {Srivastava}, {Staats}, {Staley},
  {Steinke}, {Steinlechner}, {Steinlechner}, {Steinmeyer}, {Stevenson},
  {Stone}, {Stops}, {Strain}, {Stratta}, {Strigin}, {Strunk}, {Sturani},
  {Stuver}, {Summerscales}, {Sun}, {Sunil}, {Suresh}, {Sutton}, {Swinkels},
  {Szczepa{\'n}czyk}, {Tacca}, {Tait}, {Talbot}, {Talukder}, {Tanner},
  {T{\'a}pai}, {Taracchini}, {Tasson}, {Taylor}, {Taylor}, {Tewari}, {Theeg},
  {Thies}, {Thomas}, {Thomas}, {Thomas}, {Thorne}, {Thorne}, {Thrane},
  {Tiwari}, {Tiwari}, {Tokmakov}, {Toland}, {Tonelli}, {Tornasi},
  {Torres-Forn{\'e}}, {Torrie}, {T{\"o}yr{\"a}}, {Travasso}, {Traylor},
  {Trinastic}, {Tringali}, {Trozzo}, {Tsang}, {Tse}, {Tso}, {Tsukada}, {Tsuna},
  {Tuyenbayev}, {Ueno}, {Ugolini}, {Unnikrishnan}, {Urban}, {Usman},
  {Vahlbruch}, {Vajente}, {Valdes}, {Vallisneri}, {van Bakel}, {van Beuzekom},
  {van den Brand}, {Van Den Broeck}, {Vander-Hyde}, {van der Schaaf}, {van
  Heijningen}, {van Veggel}, {Vardaro}, {Varma}, {Vass}, {Vas{\'u}th},
  {Vecchio}, {Vedovato}, {Veitch}, {Veitch}, {Venkateswara}, {Venugopalan},
  {Verkindt}, {Vetrano}, {Vicer{\'e}}, {Viets}, {Vinciguerra}, {Vine}, {Vinet},
  {Vitale}, {Vo}, {Vocca}, {Vorvick}, {Vyatchanin}, {Wade}, {Wade}, {Wade},
  {Walet}, {Walker}, {Wallace}, {Walsh}, {Wang}, {Wang}, {Wang}, {Wang},
  {Wang}, {Ward}, {Warner}, {Was}, {Watchi}, {Weaver}, {Wei}, {Weinert},
  {Weinstein}, {Weiss}, {Wen}, {Wessel}, {We{\ss}els}, {Westerweck},
  {Westphal}, {Wette}, {Whelan}, {Whitcomb}, {Whiting}, {Whittle}, {Wilken},
  {Williams}, {Williams}, {Williamson}, {Willis}, {Willke}, {Wimmer},
  {Winkler}, {Wipf}, {Wittel}, {Woan}, {Woehler}, {Wofford}, {Wong}, {Worden},
  {Wright}, {Wu}, {Wysocki}, {Xiao}, {Yamamoto}, {Yancey}, {Yang}, {Yap},
  {Yazback}, {Yu}, {Yu}, {Yvert}, {Zadro{\.Z}ny}, {Zanolin}, {Zelenova},
  {Zendri}, {Zevin}, {Zhang}, {Zhang}, {Zhang}, {Zhang}, {Zhao}, {Zhou},
  {Zhou}, {Zhu}, {Zhu}, {Zimmerman}, {Zucker}, {Zweizig}, {LIGO Scientific
  Collaboration}, \& {Virgo Collaboration}}]{abbott2017gw170817}
{Abbott}, B.~P., {Abbott}, R., {Abbott}, T.~D., {et~al.} 2017{\natexlab{c}},
  \prl, 119, 161101

\bibitem[{{Abbott} {et~al.}(2021){Abbott}, {Abbott}, {Abraham}, {Acernese},
  {Ackley}, {Adams}, {Adams}, {Adhikari}, {Adya}, {Affeldt}, {Agathos},
  {Agatsuma}, {Aggarwal}, {Aguiar}, {Aiello}, {Ain}, {Ajith}, {Allen},
  {Allocca}, {Altin}, {Amato}, {Anand}, {Ananyeva}, {Anderson}, {Anderson},
  {Angelova}, {Ansoldi}, {Antelis}, {Antier}, {Appert}, {Arai}, {Araya},
  {Areeda}, {Ar{\`e}ne}, {Arnaud}, {Aronson}, {Arun}, {Asali}, {Ascenzi},
  {Ashton}, {Aston}, {Astone}, {Aubin}, {Aufmuth}, {AultONeal}, {Austin},
  {Avendano}, {Babak}, {Badaracco}, {Bader}, {Bae}, {Baer}, {Bagnasco},
  {Baird}, {Ball}, {Ballardin}, {Ballmer}, {Bals}, {Balsamo}, {Baltus},
  {Banagiri}, {Bankar}, {Bankar}, {Barayoga}, {Barbieri}, {Barish}, {Barker},
  {Barneo}, {Barnum}, {Barone}, {Barr}, {Barsotti}, {Barsuglia}, {Barta},
  {Bartlett}, {Bartos}, {Bassiri}, {Basti}, {Bawaj}, {Bayley}, {Bazzan},
  {Becher}, {B{\'e}csy}, {Bedakihale}, {Bejger}, {Belahcene}, {Beniwal},
  {Benjamin}, {Bennett}, {Bentley}, {Bergamin}, {Berger}, {Bergmann},
  {Bernuzzi}, {Berry}, {Bersanetti}, {Bertolini}, {Betzwieser}, {Bhandare},
  {Bhandari}, {Bhattacharjee}, {Bidler}, {Bilenko}, {Billingsley}, {Birney},
  {Birnholtz}, {Biscans}, {Bischi}, {Biscoveanu}, {Bisht}, {Bitossi},
  {Bizouard}, {Blackburn}, {Blackman}, {Blair}, {Blair}, {Blair}, {Blanch},
  {Bobba}, {Bode}, {Boer}, {Boetzel}, {Bogaert}, {Boldrini}, {Bondu},
  {Bonilla}, {Bonnand}, {Booker}, {Boom}, {Bork}, {Boschi}, {Bose},
  {Bossilkov}, {Boudart}, {Bouffanais}, {Bozzi}, {Bradaschia}, {Brady},
  {Bramley}, {Branchesi}, {Brau}, {Breschi}, {Briant}, {Briggs}, {Brighenti},
  {Brillet}, {Brinkmann}, {Brockill}, {Brooks}, {Brooks}, {Brown}, {Brunett},
  {Bruno}, {Bruntz}, {Buikema}, {Bulik}, {Bulten}, {Buonanno}, {Buscicchio},
  {Buskulic}, {Byer}, {Cabero}, {Cadonati}, {Caesar}, {Cagnoli}, {Cahillane},
  {Calder{\'o}n Bustillo}, {Callaghan}, {Callister}, {Calloni}, {Camp},
  {Canepa}, {Cannon}, {Cao}, {Cao}, {Carapella}, {Carbognani}, {Carney},
  {Carpinelli}, {Carullo}, {Carver}, {Casanueva Diaz}, {Casentini}, {Caudill},
  {Cavagli{\`a}}, {Cavalier}, {Cavalieri}, {Cella}, {Cerd{\'a}-Dur{\'a}n},
  {Cesarini}, {Chaibi}, {Chakravarti}, {Chan}, {Chan}, {Chandra}, {Chanial},
  {Chao}, {Charlton}, {Chase}, {Chassande-Mottin}, {Chatterjee},
  {Chattopadhyay}, {Chaturvedi}, {Chatziioannou}, {Chen}, {Chen}, {Chen},
  {Chen}, {Cheng}, {Cheong}, {Chia}, {Chiadini}, {Chierici}, {Chincarini},
  {Chiummo}, {Cho}, {Cho}, {Cho}, {Choate}, {Christensen}, {Chu}, {Chua},
  {Chung}, {Chung}, {Ciani}, {Ciecielag}, {Cie{\'s}lar}, {Cifaldi}, {Ciobanu},
  {Ciolfi}, {Cipriano}, {Cirone}, {Clara}, {Clark}, {Clark}, {Clarke},
  {Clearwater}, {Clesse}, {Cleva}, {Coccia}, {Cohadon}, {Cohen}, {Colleoni},
  {Collette}, {Collins}, {Colpi}, {Constancio}, {Conti}, {Cooper}, {Corban},
  {Corbitt}, {Cordero-Carri{\'o}n}, {Corezzi}, {Corley}, {Cornish}, {Corre},
  {Corsi}, {Cortese}, {Costa}, {Cotesta}, {Coughlin}, {Coughlin}, {Coulon},
  {Countryman}, {Couvares}, {Covas}, {Coward}, {Cowart}, {Coyne}, {Coyne},
  {Creighton}, {Creighton}, {Croquette}, {Crowder}, {Cudell}, {Cullen},
  {Cumming}, {Cummings}, {Cunningham}, {Cuoco}, {Curylo}, {Dal Canton},
  {D{\'a}lya}, {Dana}, {DaneshgaranBajastani}, {D'Angelo}, {Danilishin},
  {D'Antonio}, {Danzmann}, {Darsow-Fromm}, {Dasgupta}, {Datrier}, {Dattilo},
  {Dave}, {Davier}, {Davies}, {Davis}, {Daw}, {Dean}, {DeBra}, {Deenadayalan},
  {Degallaix}, {De Laurentis}, {Del{\'e}glise}, {Del Favero}, {De Lillo}, {De
  Lillo}, {Del Pozzo}, {DeMarchi}, {De Matteis}, {D'Emilio}, {Demos}, {Denker},
  {Dent}, {Depasse}, {De Pietri}, {De Rosa}, {De Rossi}, {DeSalvo}, {de
  Varona}, {Dhurandhar}, {D{\'\i}az}, {Diaz-Ortiz}, {Didio}, {Dietrich}, {Di
  Fiore}, {DiFronzo}, {Di Giorgio}, {Di Giovanni}, {Di Giovanni}, {Di
  Girolamo}, {Di Lieto}, {Ding}, {Di Pace}, {Di Palma}, {Di Renzo},
  {Divakarla}, {Dmitriev}, {Doctor}, {D'Onofrio}, {Donovan}, {Dooley},
  {Doravari}, {Dorrington}, {Downes}, {Drago}, {Driggers}, {Du}, {Ducoin},
  {Dupej}, {Durante}, {D'Urso}, {Duverne}, {Dwyer}, {Easter}, {Eddolls},
  {Edelman}, {Edo}, {Edy}, {Effler}, {Eichholz}, {Eikenberry}, {Eisenmann},
  {Eisenstein}, {Ejlli}, {Errico}, {Essick}, {Estell{\'e}s}, {Estevez},
  {Etienne}, {Etzel}, {Evans}, {Evans}, {Ewing}, {Fafone}, {Fair}, {Fairhurst},
  {Fan}, {Farah}, {Farinon}, {Farr}, {Farr}, {Fauchon-Jones}, {Favata}, {Fays},
  {Fazio}, {Feicht}, {Fejer}, {Feng}, {Fenyvesi}, {Ferguson},
  {Fernandez-Galiana}, {Ferrante}, {Ferreira}, {Fidecaro}, {Figura}, {Fiori},
  {Fiorucci}, {Fishbach}, {Fisher}, {Fishner}, {Fittipaldi}, {Fitz-Axen},
  {Fiumara}, {Flaminio}, {Floden}, {Flynn}, {Fong}, {Font}, {Forsyth},
  {Fournier}, {Frasca}, {Frasconi}, {Frei}, {Freise}, {Frey}, {Frey},
  {Fritschel}, {Frolov}, {Fronz{\'e}}, {Fulda}, {Fyffe}, {Gabbard}, {Gadre},
  {Gaebel}, {Gair}, {Gais}, {Galaudage}, {Gamba}, {Ganapathy}, {Ganguly},
  {Gaonkar}, {Garaventa}, {Garc{\'\i}a-Quir{\'o}s}, {Garufi}, {Gateley},
  {Gaudio}, {Gayathri}, {Gemme}, {Gennai}, {George}, {George}, {Gergely},
  {Ghonge}, {Ghosh}, {Ghosh}, {Ghosh}, {Giacomazzo}, {Giacoppo}, {Giaime},
  {Giardina}, {Gibson}, {Gier}, {Gill}, {Giri}, {Glanzer}, {Gleckl}, {Godwin},
  {Goetz}, {Goetz}, {Gohlke}, {Goncharov}, {Gonz{\'a}lez}, {Gopakumar},
  {Gossan}, {Gosselin}, {Gouaty}, {Grace}, {Grado}, {Granata}, {Granata},
  {Grant}, {Gras}, {Grassia}, {Gray}, {Gray}, {Greco}, {Green}, {Green},
  {Gretarsson}, {Griggs}, {Grignani}, {Grimaldi}, {Grimes}, {Grimm}, {Grote},
  {Grunewald}, {Gruning}, {Guerrero}, {Guidi}, {Guimaraes}, {Guix{\'e}},
  {Gulati}, {Guo}, {Gupta}, {Gupta}, {Gupta}, {Gustafson}, {Gustafson},
  {Guzman}, {Haegel}, {Halim}, {Hall}, {Hamilton}, {Hammond}, {Haney}, {Hanke},
  {Hanks}, {Hanna}, {Hannuksela}, {Hannuksela}, {Hansen}, {Hansen}, {Hanson},
  {Harder}, {Hardwick}, {Haris}, {Harms}, {Harry}, {Harry}, {Hartwig},
  {Hasskew}, {Haster}, {Haughian}, {Hayes}, {Healy}, {Heidmann}, {Heintze},
  {Heinze}, {Heinzel}, {Heitmann}, {Hellman}, {Hello}, {Helmling-Cornell},
  {Hemming}, {Hendry}, {Heng}, {Hennes}, {Hennig}, {Hennig}, {Hernandez
  Vivanco}, {Heurs}, {Hild}, {Hill}, {Hines}, {Hochheim}, {Hofgard}, {Hofman},
  {Hohmann}, {Holgado}, {Holland}, {Hollows}, {Holmes}, {Holt}, {Holz},
  {Hopkins}, {Horst}, {Hough}, {Howell}, {Hoy}, {Hoyland}, {Huang},
  {H{\"u}bner}, {Huddart}, {Huerta}, {Hughey}, {Hui}, {Husa}, {Huttner},
  {Hutzler}, {Huxford}, {Huynh-Dinh}, {Idzkowski}, {Iess}, {Imperato},
  {Inchauspe}, {Ingram}, {Intini}, {Isi}, {Iyer}, {JaberianHamedan}, {Jacqmin},
  {Jadhav}, {Jadhav}, {James}, {Jani}, {Janssens}, {Janthalur}, {Jaranowski},
  {Jariwala}, {Jaume}, {Jenkins}, {Jeunon}, {Jiang}, {Johns}, {Jones}, {Jones},
  {Jones}, {Jones}, {Jones}, {Jonker}, {Ju}, {Junker}, {Kalaghatgi},
  {Kalogera}, {Kamai}, {Kandhasamy}, {Kang}, {Kanner}, {Kapadia}, {Kapasi},
  {Karathanasis}, {Karki}, {Kashyap}, {Kasprzack}, {Kastaun}, {Katsanevas},
  {Katsavounidis}, {Katzman}, {Kawabe}, {K{\'e}f{\'e}lian}, {Keitel}, {Key},
  {Khadka}, {Khalili}, {Khan}, {Khan}, {Khazanov}, {Khetan}, {Khursheed},
  {Kijbunchoo}, {Kim}, {Kim}, {Kim}, {Kim}, {Kim}, {Kim}, {Kimball}, {King},
  {Kinley-Hanlon}, {Kirchhoff}, {Kissel}, {Kleybolte}, {Klimenko}, {Knowles},
  {Knyazev}, {Koch}, {Koehlenbeck}, {Koekoek}, {Koley}, {Kolstein}, {Komori},
  {Kondrashov}, {Kontos}, {Koper}, {Korobko}, {Korth}, {Kovalam}, {Kozak},
  {Kr{\"a}mer}, {Kringel}, {Krishnendu}, {Kr{\'o}lak}, {Kuehn}, {Kumar},
  {Kumar}, {Kumar}, {Kumar}, {Kuns}, {Kwang}, {Lackey}, {Laghi}, {Lalande},
  {Lam}, {Lamberts}, {Landry}, {Lane}, {Lang}, {Lange}, {Lantz}, {Lanza}, {La
  Rosa}, {Lartaux-Vollard}, {Lasky}, {Laxen}, {Lazzarini}, {Lazzaro}, {Leaci},
  {Leavey}, {Lecoeuche}, {Lee}, {Lee}, {Lee}, {Lee}, {Lehmann}, {Leon},
  {Leroy}, {Letendre}, {Levin}, {Li}, {Li}, {Li}, {Li}, {Li}, {Linde},
  {Linker}, {Linley}, {Littenberg}, {Liu}, {Liu}, {Llorens-Monteagudo}, {Lo},
  {Lockwood}, {London}, {Longo}, {Lorenzini}, {Loriette}, {Lormand}, {Losurdo},
  {Lough}, {Lousto}, {Lovelace}, {L{\"u}ck}, {Lumaca}, {Lundgren}, {Ma},
  {Macas}, {MacInnis}, {Macleod}, {MacMillan}, {Macquet}, {Maga{\~n}a
  Hernandez}, {Maga{\~n}a-Sandoval}, {Magazz{\`u}}, {Magee}, {Majorana},
  {Maksimovic}, {Maliakal}, {Malik}, {Man}, {Mandic}, {Mangano}, {Mansell},
  {Manske}, {Mantovani}, {Mapelli}, {Marchesoni}, {Marion}, {M{\'a}rka},
  {M{\'a}rka}, {Markakis}, {Markosyan}, {Markowitz}, {Maros}, {Marquina},
  {Marsat}, {Martelli}, {Martin}, {Martin}, {Martinez}, {Martinez}, {Martynov},
  {Masalehdan}, {Mason}, {Massera}, {Masserot}, {Massinger}, {Masso-Reid},
  {Mastrogiovanni}, {Matas}, {Mateu-Lucena}, {Matichard}, {Matiushechkina},
  {Mavalvala}, {Maynard}, {McCann}, {McCarthy}, {McClelland}, {McCormick},
  {McCuller}, {McGuire}, {McIsaac}, {McIver}, {McManus}, {McRae}, {McWilliams},
  {Meacher}, {Meadors}, {Mehmet}, {Mehta}, {Melatos}, {Melchor}, {Mendell},
  {Menendez-Vazquez}, {Mercer}, {Mereni}, {Merfeld}, {Merilh}, {Merritt},
  {Merzougui}, {Meshkov}, {Messenger}, {Messick}, {Metzdorff}, {Meyers},
  {Meylahn}, {Mhaske}, {Miani}, {Miao}, {Michaloliakos}, {Michel}, {Middleton},
  {Milano}, {Miller}, {Miller}, {Millhouse}, {Mills}, {Milotti},
  {Milovich-Goff}, {Minazzoli}, {Minenkov}, {Mir}, {Mishkin}, {Mishra},
  {Mistry}, {Mitra}, {Mitrofanov}, {Mitselmakher}, {Mittleman}, {Mo},
  {Mogushi}, {Mohapatra}, {Mohite}, {Molina}, {Molina-Ruiz}, {Mondin},
  {Montani}, {Moore}, {Moraru}, {Morawski}, {Moreno}, {Morisaki}, {Mours},
  {Mow-Lowry}, {Mozzon}, {Muciaccia}, {Mukherjee}, {Mukherjee}, {Mukherjee},
  {Mukherjee}, {Mukund}, {Mullavey}, {Munch}, {Mu{\~n}iz}, {Murray}, {Nadji},
  {Nagar}, {Nardecchia}, {Naticchioni}, {Nayak}, {Neil}, {Neilson}, {Nelemans},
  {Nelson}, {Nery}, {Neunzert}, {Ng}, {Ng}, {Nguyen}, {Nguyen}, {Nguyen},
  {Nichols}, {Nissanke}, {Nocera}, {Noh}, {North}, {Nothard}, {Nuttall},
  {Oberling}, {O'Brien}, {O'Dell}, {Oganesyan}, {Ogin}, {Oh}, {Oh}, {Ohme},
  {Ohta}, {Okada}, {Olivetto}, {Oppermann}, {Oram}, {O'Reilly}, {Ormiston},
  {Ormsby}, {Ortega}, {O'Shaughnessy}, {Ossokine}, {Osthelder}, {Ottaway},
  {Overmier}, {Owen}, {Pace}, {Pagano}, {Page}, {Pagliaroli}, {Pai}, {Pai},
  {Palamos}, {Palashov}, {Palomba}, {Pan}, {Panda}, {Pang}, {Pankow},
  {Pannarale}, {Pant}, {Paoletti}, {Paoli}, {Paolone}, {Parker}, {Pascucci},
  {Pasqualetti}, {Passaquieti}, {Passuello}, {Patel}, {Patricelli}, {Payne},
  {Pechsiri}, {Pedraza}, {Pegoraro}, {Pele}, {Penn}, {Perego}, {Perez},
  {P{\'e}rigois}, {Perreca}, {Perri{\`e}s}, {Petermann}, {Petterson},
  {Pfeiffer}, {Pham}, {Phukon}, {Piccinni}, {Pichot}, {Piendibene},
  {Piergiovanni}, {Pierini}, {Pierro}, {Pillant}, {Pilo}, {Pinard}, {Pinto},
  {Piotrzkowski}, {Pirello}, {Pitkin}, {Placidi}, {Plastino}, {Pluchar},
  {Poggiani}, {Polini}, {Pong}, {Ponrathnam}, {Popolizio}, {Porter},
  {Poverman}, {Powell}, {Pracchia}, {Prajapati}, {Prasai}, {Prasanna},
  {Pratten}, {Prestegard}, {Principe}, {Prodi}, {Prokhorov}, {Prosposito},
  {Puecher}, {Punturo}, {Puosi}, {Puppo}, {P{\"u}rrer}, {Qi}, {Quetschke},
  {Quinonez}, {Quitzow-James}, {Raab}, {Raaijmakers}, {Radkins}, {Radulesco},
  {Raffai}, {Rafferty}, {Rail}, {Raja}, {Rajan}, {Rajbhandari}, {Rakhmanov},
  {Ramirez}, {Ramirez}, {Ramos-Buades}, {Rana}, {Rao}, {Rapagnani}, {Rapol},
  {Ratto}, {Raymond}, {Razzano}, {Read}, {Regimbau}, {Rei}, {Reid}, {Reitze},
  {Rettegno}, {Ricci}, {Richardson}, {Richardson}, {Richardson}, {Ricker},
  {Riemenschneider}, {Riles}, {Rizzo}, {Robertson}, {Robinet}, {Rocchi},
  {Rocha}, {Rodriguez}, {Rodriguez-Soto}, {Rolland}, {Rollins}, {Roma},
  {Romanelli}, {Romano}, {Romel}, {Romero}, {Romero-Shaw}, {Romie}, {Ronchini},
  {Rose}, {Rose}, {Rose}, {Rosell}, {Rosi{\'n}ska}, {Rosofsky}, {Ross},
  {Rowan}, {Rowlinson}, {Roy}, {Roy}, {Ruggi}, {Ryan}, {Sachdev}, {Sadecki},
  {Sakellariadou}, {Salafia}, {Salconi}, {Saleem}, {Samajdar}, {Sanchez},
  {Sanchez}, {Sanchez}, {Sanchis-Gual}, {Sanders}, {Santiago}, {Santos},
  {Saravanan}, {Sarin}, {Sassolas}, {Sathyaprakash}, {Sauter}, {Savage},
  {Savant}, {Sawant}, {Sayah}, {Schaetzl}, {Schale}, {Scheel}, {Scheuer},
  {Schindler-Tyka}, {Schmidt}, {Schnabel}, {Schofield}, {Sch{\"o}nbeck},
  {Schreiber}, {Schulte}, {Schutz}, {Schwarm}, {Schwartz}, {Scott}, {Scott},
  {Seglar-Arroyo}, {Seidel}, {Sellers}, {Sengupta}, {Sennett}, {Sentenac},
  {Sequino}, {Sergeev}, {Setyawati}, {Shaffer}, {Shahriar}, {Sharifi},
  {Sharma}, {Sharma}, {Shawhan}, {Shen}, {Shikauchi}, {Shink}, {Shoemaker},
  {Shoemaker}, {Shukla}, {ShyamSundar}, {Sieniawska}, {Sigg}, {Singer},
  {Singh}, {Singh}, {Singha}, {Singhal}, {Sintes}, {Sipala}, {Skliris},
  {Slagmolen}, {Slaven-Blair}, {Smetana}, {Smith}, {Smith}, {Somala}, {Son},
  {Soni}, {Sorazu}, {Sordini}, {Sorrentino}, {Sorrentino}, {Soulard},
  {Souradeep}, {Sowell}, {Spencer}, {Spera}, {Srivastava}, {Srivastava},
  {Staats}, {Stachie}, {Steer}, {Steinke}, {Steinlechner}, {Steinlechner},
  {Steinmeyer}, {Stevenson}, {Stolle-McAllister}, {Stops}, {Stover}, {Strain},
  {Stratta}, {Strunk}, {Sturani}, {Stuver}, {S{\"u}dbeck}, {Sudhagar},
  {Sudhir}, {Suh}, {Summerscales}, {Sun}, {Sun}, {Sunil}, {Sur}, {Suresh},
  {Sutton}, {Swinkels}, {Szczepa{\'n}czyk}, {Tacca}, {Tait}, {Talbot},
  {Tanasijczuk}, {Tanner}, {Tao}, {Tapia}, {Tapia San Martin}, {Tasson},
  {Taylor}, {Tenorio}, {Terkowski}, {Thirugnanasambandam}, {Thomas}, {Thomas},
  {Thomas}, {Thompson}, {Thondapu}, {Thorne}, {Thrane}, {Tiwari}, {Tiwari},
  {Tiwari}, {Toland}, {Tolley}, {Tonelli}, {Tornasi}, {Torres-Forn{\'e}},
  {Torrie}, {Tosta e Melo}, {T{\"o}yr{\"a}}, {Tran}, {Trapananti}, {Travasso},
  {Traylor}, {Tringali}, {Tripathee}, {Trovato}, {Trudeau}, {Tsai}, {Tsang},
  {Tse}, {Tso}, {Tsukada}, {Tsuna}, {Tsutsui}, {Turconi}, {Ubhi}, {Udall},
  {Ueno}, {Ugolini}, {Unnikrishnan}, {Urban}, {Usman}, {Utina}, {Vahlbruch},
  {Vajente}, {Vajpeyi}, {Valdes}, {Valentini}, {Valsan}, {van Bakel},
  {Beuzekom}, {van den Brand}, {Van Den Broeck}, {Vander-Hyde}, {van der
  Schaaf}, {van Heijningen}, {Vardaro}, {Vargas}, {Varma}, {Vass},
  {Vas{\'u}th}, {Vecchio}, {Vedovato}, {Veitch}, {Veitch}, {Venkateswara},
  {Venneberg}, {Venugopalan}, {Verkindt}, {Verma}, {Veske}, {Vetrano},
  {Vicer{\'e}}, {Viets}, {Villa-Ortega}, {Vinet}, {Vitale}, {Vo}, {Vocca},
  {Vorvick}, {Vyatchanin}, {Wade}, {Wade}, {Wade}, {Walet}, {Walker},
  {Wallace}, {Wallace}, {Walsh}, {Wang}, {Wang}, {Wang}, {Wang}, {Ward},
  {Warner}, {Was}, {Washington}, {Watchi}, {Weaver}, {Wei}, {Weinert},
  {Weinstein}, {Weiss}, {Wellmann}, {Wen}, {We{\ss}els}, {Westhouse}, {Wette},
  {Whelan}, {White}, {White}, {Whiting}, {Whittle}, {Wilken}, {Williams},
  {Williams}, {Williamson}, {Willis}, {Willke}, {Wilson}, {Wimmer}, {Winkler},
  {Wipf}, {Woan}, {Woehler}, {Wofford}, {Wong}, {Wrangel}, {Wright}, {Wu},
  {Wysocki}, {Xiao}, {Yamamoto}, {Yang}, {Yang}, {Yang}, {Yap}, {Yeeles},
  {Yoon}, {Yu}, {Yu}, {Yuen}, {Zadro{\.z}ny}, {Zanolin}, {Zelenova}, {Zendri},
  {Zevin}, {Zhang}, {Zhang}, {Zhang}, {Zhang}, {Zhao}, {Zhao}, {Zhou}, {Zhou},
  {Zhu}, {Zimmerman}, {Zucker}, {Zweizig}, {LIGO Scientific Collaboration}, \&
  {Virgo Collaboration}}]{Abbott2021ApJ...913L...7A}
{Abbott}, R., {Abbott}, T.~D., {Abraham}, S., {et~al.} 2021, \apjl, 913, L7

\bibitem[{{Abohalima} \& {Frebel}(2018)}]{Abohalima_2018}
{Abohalima}, A. \& {Frebel}, A. 2018, \apjs, 238, 36

\bibitem[{{Aguado} {et~al.}(2021){Aguado}, {Myeong}, {Belokurov}, {Evans},
  {Koposov}, {Allende Prieto}, {Lanfranchi}, {Matteucci}, {Shetrone},
  {Sbordone}, {Navarrete}, {Gonz{\'a}lez Hern{\'a}ndez}, {Chanam{\'e}},
  {Peralta de Arriba}, \& {Yuan}}]{Aguado2021}
{Aguado}, D.~S., {Myeong}, G.~C., {Belokurov}, V., {et~al.} 2021, \mnras, 500,
  889

\bibitem[{{Allen} {et~al.}(2012){Allen}, {Ryan}, {Rossi}, {Beers}, \&
  {Tsangarides}}]{ALL12}
{Allen}, D.~M., {Ryan}, S.~G., {Rossi}, S., {Beers}, T.~C., \& {Tsangarides},
  S.~A. 2012, \aap, 548, A34

\bibitem[{{Aoki} {et~al.}(2002{\natexlab{a}}){Aoki}, {Ando}, {Honda}, {Iye},
  {Izumiura}, {Kajino}, {Kambe}, {Kawanomonoto}, {Noguchi}, {Okita},
  {Sadakane}, {Sato}, {Shelton}, {Takada-Hidai}, {Takeda}, {Watanabe}, \&
  {Yoshida}}]{AOK02B}
{Aoki}, W., {Ando}, H., {Honda}, S., {et~al.} 2002{\natexlab{a}}, \pasj, 54,
  427

\bibitem[{{Aoki} {et~al.}(2007{\natexlab{a}}){Aoki}, {Beers}, {Christlieb},
  {Norris}, {Ryan}, \& {Tsangarides}}]{AOK07A}
{Aoki}, W., {Beers}, T.~C., {Christlieb}, N., {et~al.} 2007{\natexlab{a}},
  \apj, 655, 492

\bibitem[{{Aoki} {et~al.}(2013){Aoki}, {Beers}, {Lee}, {Honda}, {Ito},
  {Takada-Hidai}, {Frebel}, {Suda}, {Fujimoto}, {Carollo}, \&
  {Sivarani}}]{AOK13}
{Aoki}, W., {Beers}, T.~C., {Lee}, Y.~S., {et~al.} 2013, \aj, 145, 13

\bibitem[{{Aoki} {et~al.}(2008){Aoki}, {Beers}, {Sivarani}, {Marsteller},
  {Lee}, {Honda}, {Norris}, {Ryan}, \& {Carollo}}]{AOK08}
{Aoki}, W., {Beers}, T.~C., {Sivarani}, T., {et~al.} 2008, \apj, 678, 1351

\bibitem[{{Aoki} {et~al.}(2005){Aoki}, {Honda}, {Beers}, {Kajino}, {Ando},
  {Norris}, {Ryan}, {Izumiura}, {Sadakane}, \& {Takada-Hidai}}]{AOK05}
{Aoki}, W., {Honda}, S., {Beers}, T.~C., {et~al.} 2005, \apj, 632, 611

\bibitem[{{Aoki} {et~al.}(2007{\natexlab{b}}){Aoki}, {Honda}, {Sadakane}, \&
  {Arimoto}}]{AOK07C}
{Aoki}, W., {Honda}, S., {Sadakane}, K., \& {Arimoto}, N. 2007{\natexlab{b}},
  \pasj, 59, L15

\bibitem[{{Aoki} {et~al.}(2012){Aoki}, {Ito}, \& {Tajitsu}}]{AOK12}
{Aoki}, W., {Ito}, H., \& {Tajitsu}, A. 2012, \apjl, 751, L6

\bibitem[{{Aoki} {et~al.}(2002{\natexlab{b}}){Aoki}, {Norris}, {Ryan}, {Beers},
  \& {Ando}}]{AOK02d}
{Aoki}, W., {Norris}, J.~E., {Ryan}, S.~G., {Beers}, T.~C., \& {Ando}, H.
  2002{\natexlab{b}}, \pasj, 54, 933

\bibitem[{{Aoki} {et~al.}(2014){Aoki}, {Tominaga}, {Beers}, {Honda}, \&
  {Lee}}]{AOK14}
{Aoki}, W., {Tominaga}, N., {Beers}, T.~C., {Honda}, S., \& {Lee}, Y.~S. 2014,
  Science, 345, 912

\bibitem[{{Arcones} {et~al.}(2007){Arcones}, {Janka}, \&
  {Scheck}}]{Arcones2007}
{Arcones}, A., {Janka}, H.~T., \& {Scheck}, L. 2007, \aap, 467, 1227

\bibitem[{{Arcones} \& {Thielemann}(2013)}]{Arcones_2012}
{Arcones}, A. \& {Thielemann}, F.~K. 2013, Journal of Physics G Nuclear
  Physics, 40, 013201

\bibitem[{{Argast} {et~al.}(2004){Argast}, {Samland}, {Thielemann}, \&
  {Qian}}]{argast2004neutron}
{Argast}, D., {Samland}, M., {Thielemann}, F.~K., \& {Qian}, Y.~Z. 2004, \aap,
  416, 997

\bibitem[{{Barbuy} {et~al.}(2005){Barbuy}, {Spite}, {Spite}, {Hill}, {Cayrel},
  {Plez}, \& {Petitjean}}]{BAB05}
{Barbuy}, B., {Spite}, M., {Spite}, F., {et~al.} 2005, \aap, 429, 1031

\bibitem[{{Barklem} {et~al.}(2005){Barklem}, {Christlieb}, {Beers}, {Hill},
  {Bessell}, {Holmberg}, {Marsteller}, {Rossi}, {Zickgraf}, \&
  {Reimers}}]{BAR05}
{Barklem}, P.~S., {Christlieb}, N., {Beers}, T.~C., {et~al.} 2005, \aap, 439,
  129

\bibitem[{{Bauswein} {et~al.}(2013){Bauswein}, {Goriely}, \&
  {Janka}}]{2013ApJ...773...78B}
{Bauswein}, A., {Goriely}, S., \& {Janka}, H.~T. 2013, \apj, 773, 78

\bibitem[{{Behara} {et~al.}(2010){Behara}, {Bonifacio}, {Ludwig}, {Sbordone},
  {Gonz{\'a}lez Hern{\'a}ndez}, \& {Caffau}}]{BEH10}
{Behara}, N.~T., {Bonifacio}, P., {Ludwig}, H.~G., {et~al.} 2010, \aap, 513,
  A72

\bibitem[{{Belczynski} {et~al.}(2020){Belczynski}, {Klencki}, {Fields},
  {Olejak}, {Berti}, {Meynet}, {Fryer}, {Holz}, {O'Shaughnessy}, {Brown},
  {Bulik}, {Leung}, {Nomoto}, {Madau}, {Hirschi}, {Kaiser}, {Jones}, {Mondal},
  {Chruslinska}, {Drozda}, {Gerosa}, {Doctor}, {Giersz}, {Ekstrom}, {Georgy},
  {Askar}, {Baibhav}, {Wysocki}, {Natan}, {Farr}, {Wiktorowicz}, {Coleman
  Miller}, {Farr}, \& {Lasota}}]{Belcynski2020}
{Belczynski}, K., {Klencki}, J., {Fields}, C.~E., {et~al.} 2020, \aap, 636,
  A104

\bibitem[{{Belokurov} {et~al.}(2018){Belokurov}, {Erkal}, {Evans}, {Koposov},
  \& {Deason}}]{Belokurov2018}
{Belokurov}, V., {Erkal}, D., {Evans}, N.~W., {Koposov}, S.~E., \& {Deason},
  A.~J. 2018, \mnras, 478, 611

\bibitem[{{Belokurov} {et~al.}(2006){Belokurov}, {Zucker}, {Evans}, {Gilmore},
  {Vidrih}, {Bramich}, {Newberg}, {Wyse}, {Irwin}, {Fellhauer}, {Hewett},
  {Walton}, {Wilkinson}, {Cole}, {Yanny}, {Rockosi}, {Beers}, {Bell},
  {Brinkmann}, {Ivezi{\'c}}, \& {Lupton}}]{Belokurov2006}
{Belokurov}, V., {Zucker}, D.~B., {Evans}, N.~W., {et~al.} 2006, \apjl, 642,
  L137

\bibitem[{{Bensby} {et~al.}(2011){Bensby}, {Ad{\'e}n}, {Mel{\'e}ndez}, {Gould},
  {Feltzing}, {Asplund}, {Johnson}, {Lucatello}, {Yee}, {Ram{\'\i}rez},
  {Cohen}, {Thompson}, {Bond}, {Gal-Yam}, {Han}, {Sumi}, {Suzuki}, {Wada},
  {Miyake}, {Furusawa}, {Ohmori}, {Saito}, {Tristram}, \& {Bennett}}]{BEN11}
{Bensby}, T., {Ad{\'e}n}, D., {Mel{\'e}ndez}, J., {et~al.} 2011, \aap, 533,
  A134

\bibitem[{{Berger}(2014)}]{berger2014short}
{Berger}, E. 2014, \araa, 52, 43

\bibitem[{{Bisterzo} {et~al.}(2015){Bisterzo}, {Gallino}, {K{\"a}ppeler},
  {Wiescher}, {Imbriani}, {Straniero}, {Cristallo}, {G{\"o}rres}, \&
  {deBoer}}]{2015MNRAS.449..506B}
{Bisterzo}, S., {Gallino}, R., {K{\"a}ppeler}, F., {et~al.} 2015, \mnras, 449,
  506

\bibitem[{{Bonetti} {et~al.}(2019){Bonetti}, {Perego}, {Dotti}, \&
  {Cescutti}}]{bonetti2019neutron}
{Bonetti}, M., {Perego}, A., {Dotti}, M., \& {Cescutti}, G. 2019, \mnras, 490,
  296

\bibitem[{{Bonifacio} {et~al.}(2012){Bonifacio}, {Sbordone}, {Caffau},
  {Ludwig}, {Spite}, {Gonz{\'a}lez Hern{\'a}ndez}, \&
  {Behara}}]{bonifacio2012chemical}
{Bonifacio}, P., {Sbordone}, L., {Caffau}, E., {et~al.} 2012, \aap, 542, A87

\bibitem[{{Bonifacio} {et~al.}(2009){Bonifacio}, {Spite}, {Cayrel}, {Hill},
  {Spite}, {Fran{\c{c}}ois}, {Plez}, {Ludwig}, {Caffau}, {Molaro}, {Depagne},
  {Andersen}, {Barbuy}, {Beers}, {Nordstr{\"o}m}, \& {Primas}}]{BON09}
{Bonifacio}, P., {Spite}, M., {Cayrel}, R., {et~al.} 2009, \aap, 501, 519

\bibitem[{{Brauer} {et~al.}(2019){Brauer}, {Ji}, {Frebel}, {Dooley},
  {G{\'o}mez}, \& {O'Shea}}]{Brauer2019ApJ...871..247B}
{Brauer}, K., {Ji}, A.~P., {Frebel}, A., {et~al.} 2019, \apj, 871, 247

\bibitem[{{Bullock} \& {Johnston}(2005)}]{Bullock2005}
{Bullock}, J.~S. \& {Johnston}, K.~V. 2005, \apj, 635, 931

\bibitem[{{Burris} {et~al.}(2000){Burris}, {Pilachowski}, {Armandroff},
  {Sneden}, {Cowan}, \& {Roe}}]{BUR00}
{Burris}, D.~L., {Pilachowski}, C.~A., {Armandroff}, T.~E., {et~al.} 2000,
  \apj, 544, 302

\bibitem[{{Caffau} {et~al.}(2011){Caffau}, {Bonifacio}, {Fran{\c{c}}ois},
  {Spite}, {Spite}, {Zaggia}, {Ludwig}, {Monaco}, {Sbordone}, {Cayrel},
  {Hammer}, {Randich}, {Hill}, \& {Molaro}}]{CAF11B}
{Caffau}, E., {Bonifacio}, P., {Fran{\c{c}}ois}, P., {et~al.} 2011, \aap, 534,
  A4

\bibitem[{{Calura} {et~al.}(2008){Calura}, {Lanfranchi}, \&
  {Matteucci}}]{Calura2008A&A...484..107C}
{Calura}, F., {Lanfranchi}, G.~A., \& {Matteucci}, F. 2008, \aap, 484, 107

\bibitem[{Cameron(1982)}]{cameron1982elemental}
Cameron, A. 1982, Essays in Nuclear Astrophysics, 23

\bibitem[{{Carretta} {et~al.}(2002){Carretta}, {Gratton}, {Cohen}, {Beers}, \&
  {Christlieb}}]{CAR02}
{Carretta}, E., {Gratton}, R., {Cohen}, J.~G., {Beers}, T.~C., \& {Christlieb},
  N. 2002, \aj, 124, 481

\bibitem[{{Cavallo} {et~al.}(2021){Cavallo}, {Cescutti}, \&
  {Matteucci}}]{cavallo2021}
{Cavallo}, L., {Cescutti}, G., \& {Matteucci}, F. 2021, \mnras, 503, 1

\bibitem[{{Cayrel} {et~al.}(2004{\natexlab{a}}){Cayrel}, {Depagne}, {Spite},
  {Hill}, {Spite}, {Fran{\c{c}}ois}, {Plez}, {Beers}, {Primas}, {Andersen},
  {Barbuy}, {Bonifacio}, {Molaro}, \& {Nordstr{\"o}m}}]{Cayrel2004}
{Cayrel}, R., {Depagne}, E., {Spite}, M., {et~al.} 2004{\natexlab{a}}, \aap,
  416, 1117

\bibitem[{{Cayrel} {et~al.}(2004{\natexlab{b}}){Cayrel}, {Depagne}, {Spite},
  {Hill}, {Spite}, {Fran{\c{c}}ois}, {Plez}, {Beers}, {Primas}, {Andersen},
  {Barbuy}, {Bonifacio}, {Molaro}, \& {Nordstr{\"o}m}}]{CAY04}
{Cayrel}, R., {Depagne}, E., {Spite}, M., {et~al.} 2004{\natexlab{b}}, \aap,
  416, 1117

\bibitem[{{Cescutti}(2008)}]{cescutti2008}
{Cescutti}, G. 2008, \aap, 481, 691

\bibitem[{{Cescutti} {et~al.}(2006){Cescutti}, {Fran{\c{c}}ois}, {Matteucci},
  {Cayrel}, \& {Spite}}]{cescutti2006chemical}
{Cescutti}, G., {Fran{\c{c}}ois}, P., {Matteucci}, F., {Cayrel}, R., \&
  {Spite}, M. 2006, \aap, 448, 557

\bibitem[{{Cescutti} {et~al.}(2015){Cescutti}, {Romano}, {Matteucci},
  {Chiappini}, \& {Hirschi}}]{cescutti2015role}
{Cescutti}, G., {Romano}, D., {Matteucci}, F., {Chiappini}, C., \& {Hirschi},
  R. 2015, \aap, 577, A139

\bibitem[{{Chiappini} {et~al.}(2008){Chiappini}, {Ekstr{\"o}m}, {Meynet},
  {Hirschi}, {Maeder}, \& {Charbonnel}}]{chiappini2008new}
{Chiappini}, C., {Ekstr{\"o}m}, S., {Meynet}, G., {et~al.} 2008, \aap, 479, L9

\bibitem[{{Christlieb} {et~al.}(2004){Christlieb}, {Gustafsson}, {Korn},
  {Barklem}, {Beers}, {Bessell}, {Karlsson}, \& {Mizuno-Wiedner}}]{CHR04}
{Christlieb}, N., {Gustafsson}, B., {Korn}, A.~J., {et~al.} 2004, \apj, 603,
  708

\bibitem[{{Chruslinska} {et~al.}(2018){Chruslinska}, {Belczynski}, {Klencki},
  \& {Benacquista}}]{chruslinska2018double}
{Chruslinska}, M., {Belczynski}, K., {Klencki}, J., \& {Benacquista}, M. 2018,
  \mnras, 474, 2937

\bibitem[{{Ciolfi}(2020)}]{Ciolfi2020}
{Ciolfi}, R. 2020, Frontiers in Astronomy and Space Sciences, 7, 27

\bibitem[{{Cohen} {et~al.}(2008){Cohen}, {Christlieb}, {McWilliam}, {Shectman},
  {Thompson}, {Melendez}, {Wisotzki}, \& {Reimers}}]{COH08}
{Cohen}, J.~G., {Christlieb}, N., {McWilliam}, A., {et~al.} 2008, \apj, 672,
  320

\bibitem[{{Cohen} {et~al.}(2004){Cohen}, {Christlieb}, {McWilliam}, {Shectman},
  {Thompson}, {Wasserburg}, {Ivans}, {Dehn}, {Karlsson}, \& {Melendez}}]{COH04}
{Cohen}, J.~G., {Christlieb}, N., {McWilliam}, A., {et~al.} 2004, \apj, 612,
  1107

\bibitem[{{Cohen} {et~al.}(2013){Cohen}, {Christlieb}, {Thompson}, {McWilliam},
  {Shectman}, {Reimers}, {Wisotzki}, \& {Kirby}}]{COH13}
{Cohen}, J.~G., {Christlieb}, N., {Thompson}, I., {et~al.} 2013, \apj, 778, 56

\bibitem[{{Cooper} {et~al.}(2010){Cooper}, {Cole}, {Frenk}, {White}, {Helly},
  {Benson}, {De Lucia}, {Helmi}, {Jenkins}, {Navarro}, {Springel}, \&
  {Wang}}]{Cooper2010}
{Cooper}, A.~P., {Cole}, S., {Frenk}, C.~S., {et~al.} 2010, \mnras, 406, 744

\bibitem[{{C{\^o}t{\'e}} {et~al.}(2019){C{\^o}t{\'e}}, {Eichler}, {Arcones},
  {Hansen}, {Simonetti}, {Frebel}, {Fryer}, {Pignatari}, {Reichert},
  {Belczynski}, \& {Matteucci}}]{cote2019neutron}
{C{\^o}t{\'e}}, B., {Eichler}, M., {Arcones}, A., {et~al.} 2019, \apj, 875, 106

\bibitem[{{Coulter} {et~al.}(2017){Coulter}, {Foley}, {Kilpatrick}, {Drout},
  {Piro}, {Shappee}, {Siebert}, {Simon}, {Ulloa}, {Kasen}, {Madore},
  {Murguia-Berthier}, {Pan}, {Prochaska}, {Ramirez-Ruiz}, {Rest}, \&
  {Rojas-Bravo}}]{coulter2017swope}
{Coulter}, D.~A., {Foley}, R.~J., {Kilpatrick}, C.~D., {et~al.} 2017, Science,
  358, 1556

\bibitem[{{Cowan} {et~al.}(2002){Cowan}, {Sneden}, {Burles}, {Ivans}, {Beers},
  {Truran}, {Lawler}, {Primas}, {Fuller}, {Pfeiffer}, \& {Kratz}}]{COW02}
{Cowan}, J.~J., {Sneden}, C., {Burles}, S., {et~al.} 2002, \apj, 572, 861

\bibitem[{{Cowan} {et~al.}(1991){Cowan}, {Thielemann}, \&
  {Truran}}]{cowan1991r}
{Cowan}, J.~J., {Thielemann}, F.-K., \& {Truran}, J.~W. 1991, \physrep, 208,
  267

\bibitem[{{Cowperthwaite} {et~al.}(2017){Cowperthwaite}, {Berger}, {Villar},
  {Metzger}, {Nicholl}, {Chornock}, {Blanchard}, {Fong}, {Margutti},
  {Soares-Santos}, {Alexander}, {Allam}, {Annis}, {Brout}, {Brown}, {Butler},
  {Chen}, {Diehl}, {Doctor}, {Drout}, {Eftekhari}, {Farr}, {Finley}, {Foley},
  {Frieman}, {Fryer}, {Garc{\'\i}a-Bellido}, {Gill}, {Guillochon}, {Herner},
  {Holz}, {Kasen}, {Kessler}, {Marriner}, {Matheson}, {Neilsen}, {Quataert},
  {Palmese}, {Rest}, {Sako}, {Scolnic}, {Smith}, {Tucker}, {Williams},
  {Balbinot}, {Carlin}, {Cook}, {Durret}, {Li}, {Lopes}, {Louren{\c{c}}o},
  {Marshall}, {Medina}, {Muir}, {Mu{\~n}oz}, {Sauseda}, {Schlegel}, {Secco},
  {Vivas}, {Wester}, {Zenteno}, {Zhang}, {Abbott}, {Banerji}, {Bechtol},
  {Benoit-L{\'e}vy}, {Bertin}, {Buckley-Geer}, {Burke}, {Capozzi}, {Carnero
  Rosell}, {Carrasco Kind}, {Castander}, {Crocce}, {Cunha}, {D'Andrea}, {da
  Costa}, {Davis}, {DePoy}, {Desai}, {Dietrich}, {Drlica-Wagner}, {Eifler},
  {Evrard}, {Fernandez}, {Flaugher}, {Fosalba}, {Gaztanaga}, {Gerdes},
  {Giannantonio}, {Goldstein}, {Gruen}, {Gruendl}, {Gutierrez}, {Honscheid},
  {Jain}, {James}, {Jeltema}, {Johnson}, {Johnson}, {Kent}, {Krause}, {Kron},
  {Kuehn}, {Nuropatkin}, {Lahav}, {Lima}, {Lin}, {Maia}, {March}, {Martini},
  {McMahon}, {Menanteau}, {Miller}, {Miquel}, {Mohr}, {Neilsen}, {Nichol},
  {Ogando}, {Plazas}, {Roe}, {Romer}, {Roodman}, {Rykoff}, {Sanchez},
  {Scarpine}, {Schindler}, {Schubnell}, {Sevilla-Noarbe}, {Smith}, {Smith},
  {Sobreira}, {Suchyta}, {Swanson}, {Tarle}, {Thomas}, {Thomas}, {Troxel},
  {Vikram}, {Walker}, {Wechsler}, {Weller}, {Yanny}, \&
  {Zuntz}}]{Cowperthwaite2017}
{Cowperthwaite}, P.~S., {Berger}, E., {Villar}, V.~A., {et~al.} 2017, \apjl,
  848, L17

\bibitem[{{Cui} {et~al.}(2013){Cui}, {Sivarani}, \& {Christlieb}}]{CUI13}
{Cui}, W.~Y., {Sivarani}, T., \& {Christlieb}, N. 2013, \aap, 558, A36

\bibitem[{{Curtis} {et~al.}(2019){Curtis}, {Ebinger}, {Fr{\"o}hlich}, {Hempel},
  {Perego}, {Liebend{\"o}rfer}, \& {Thielemann}}]{Curtis_2018}
{Curtis}, S., {Ebinger}, K., {Fr{\"o}hlich}, C., {et~al.} 2019, \apj, 870, 2

\bibitem[{{Dalton} {et~al.}(2012){Dalton}, {Trager}, {Abrams}, {Carter},
  {Bonifacio}, {Aguerri}, {MacIntosh}, {Evans}, {Lewis}, {Navarro}, {Agocs},
  {Dee}, {Rousset}, {Tosh}, {Middleton}, {Pragt}, {Terrett}, {Brock}, {Benn},
  {Verheijen}, {Cano Infantes}, {Bevil}, {Steele}, {Mottram}, {Bates},
  {Gribbin}, {Rey}, {Rodriguez}, {Delgado}, {Guinouard}, {Walton}, {Irwin},
  {Jagourel}, {Stuik}, {Gerlofsma}, {Roelfsma}, {Skillen}, {Ridings},
  {Balcells}, {Daban}, {Gouvret}, {Venema}, \& {Girard}}]{WEAVE}
{Dalton}, G., {Trager}, S.~C., {Abrams}, D.~C., {et~al.} 2012, in Society of
  Photo-Optical Instrumentation Engineers (SPIE) Conference Series, Vol. 8446,
  Ground-based and Airborne Instrumentation for Astronomy IV, 84460P

\bibitem[{{D'Avanzo}(2015)}]{d2015short}
{D'Avanzo}, P. 2015, Journal of High Energy Astrophysics, 7, 73

\bibitem[{{D'Avanzo} {et~al.}(2014){D'Avanzo}, {Salvaterra}, {Bernardini},
  {Nava}, {Campana}, {Covino}, {D'Elia}, {Ghirlanda}, {Ghisellini}, {Melandri},
  {Sbarufatti}, {Vergani}, \& {Tagliaferri}}]{Davanzo2014}
{D'Avanzo}, P., {Salvaterra}, R., {Bernardini}, M.~G., {et~al.} 2014, \mnras,
  442, 2342

\bibitem[{{de Jong} {et~al.}(2014){de Jong}, {Barden}, {Bellido-Tirado},
  {Brynnel}, {Chiappini}, {Depagne}, {Haynes}, {Johl}, {Phillips}, {Schnurr},
  {Schwope}, {Walcher}, {Bauer}, {Cescutti}, {Cioni}, {Dionies}, {Enke},
  {Haynes}, {Kelz}, {Kitaura}, {Lamer}, {Minchev}, {M{\"u}ller}, {Nuza},
  {Olaya}, {Piffl}, {Popow}, {Saviauk}, {Steinmetz}, {Ural}, {Valentini},
  {Winkler}, {Wisotzki}, {Ansorge}, {Banerji}, {Gonzalez Solares}, {Irwin},
  {Kennicutt}, {King}, {McMahon}, {Koposov}, {Parry}, {Sun}, {Walton},
  {Finger}, {Iwert}, {Krumpe}, {Lizon}, {Mainieri}, {Amans}, {Bonifacio},
  {Cohen}, {Fran{\c{c}}ois}, {Jagourel}, {Mignot}, {Royer}, {Sartoretti},
  {Bender}, {Hess}, {Lang-Bardl}, {Muschielok}, {Schlichter}, {B{\"o}hringer},
  {Boller}, {Bongiorno}, {Brusa}, {Dwelly}, {Merloni}, {Nandra}, {Salvato},
  {Pragt}, {Navarro}, {Gerlofsma}, {Roelfsema}, {Dalton}, {Middleton}, {Tosh},
  {Boeche}, {Caffau}, {Christlieb}, {Grebel}, {Hansen}, {Koch}, {Ludwig},
  {Mandel}, {Quirrenbach}, {Sbordone}, {Seifert}, {Thimm}, {Helmi}, {trager},
  {Bensby}, {Feltzing}, {Ruchti}, {Edvardsson}, {Korn}, {Lind}, {Boland },
  {Colless}, {Frost}, {Gilbert}, {Gillingham}, {Lawrence}, {Legg}, {Saunders},
  {Sheinis}, {Driver}, {Robotham}, {Bacon}, {Caillier}, {Kosmalski}, {Laurent},
  \& {Richard}}]{4MOST}
{de Jong}, R.~S., {Barden}, S., {Bellido-Tirado}, O., {et~al.} 2014, in Society
  of Photo-Optical Instrumentation Engineers (SPIE) Conference Series, Vol.
  9147, Ground-based and Airborne Instrumentation for Astronomy V, 91470M

\bibitem[{{Dominik} {et~al.}(2012){Dominik}, {Belczynski}, {Fryer}, {Holz},
  {Berti}, {Bulik}, {Mandel}, \& {O'Shaughnessy}}]{Dominik_2012}
{Dominik}, M., {Belczynski}, K., {Fryer}, C., {et~al.} 2012, \apj, 759, 52

\bibitem[{{Eichler} {et~al.}(1989){Eichler}, {Livio}, {Piran}, \&
  {Schramm}}]{1989Natur.340..126E}
{Eichler}, D., {Livio}, M., {Piran}, T., \& {Schramm}, D.~N. 1989, \nat, 340,
  126

\bibitem[{{Fattahi} {et~al.}(2019){Fattahi}, {Belokurov}, {Deason}, {Frenk},
  {G{\'o}mez}, {Grand}, {Marinacci}, {Pakmor}, \& {Springel}}]{Fattahi2019}
{Fattahi}, A., {Belokurov}, V., {Deason}, A.~J., {et~al.} 2019, \mnras, 484,
  4471

\bibitem[{{Fattahi} {et~al.}(2020){Fattahi}, {Deason}, {Frenk}, {Simpson},
  {G{\'o}mez}, {Grand}, {Monachesi}, {Marinacci}, \& {Pakmor}}]{Fattahi2020}
{Fattahi}, A., {Deason}, A.~J., {Frenk}, C.~S., {et~al.} 2020, \mnras, 497,
  4459

\bibitem[{{Fong} {et~al.}(2017){Fong}, {Berger}, {Blanchard}, {Margutti},
  {Cowperthwaite}, {Chornock}, {Alexander}, {Metzger}, {Villar}, {Nicholl},
  {Eftekhari}, {Williams}, {Annis}, {Brout}, {Brown}, {Chen}, {Doctor},
  {Diehl}, {Holz}, {Rest}, {Sako}, \& {Soares-Santos}}]{Fong_2017}
{Fong}, W., {Berger}, E., {Blanchard}, P.~K., {et~al.} 2017, \apjl, 848, L23

\bibitem[{{Frebel} {et~al.}(2007){Frebel}, {Christlieb}, {Norris}, {Thom},
  {Beers}, \& {Rhee}}]{FRE07b}
{Frebel}, A., {Christlieb}, N., {Norris}, J.~E., {et~al.} 2007, \apjl, 660,
  L117

\bibitem[{{Frebel} {et~al.}(2010){Frebel}, {Kirby}, \& {Simon}}]{Frebel2010}
{Frebel}, A., {Kirby}, E.~N., \& {Simon}, J.~D. 2010, \nat, 464, 72

\bibitem[{{Freiburghaus} {et~al.}(1999){Freiburghaus}, {Rosswog}, \&
  {Thielemann}}]{freiburghaus1999r}
{Freiburghaus}, C., {Rosswog}, S., \& {Thielemann}, F.~K. 1999, \apjl, 525,
  L121

\bibitem[{{Frischknecht} {et~al.}(2016){Frischknecht}, {Hirschi}, {Pignatari},
  {Maeder}, {Meynet}, {Chiappini}, {Thielemann}, {Rauscher}, {Georgy}, \&
  {Ekstr{\"o}m}}]{Frisch16}
{Frischknecht}, U., {Hirschi}, R., {Pignatari}, M., {et~al.} 2016, \mnras, 456,
  1803

\bibitem[{{Fruchter} {et~al.}(2006){Fruchter}, {Levan}, {Strolger},
  {Vreeswijk}, {Thorsett}, {Bersier}, {Burud}, {Castro Cer{\'o}n},
  {Castro-Tirado}, {Conselice}, {Dahlen}, {Ferguson}, {Fynbo}, {Garnavich},
  {Gibbons}, {Gorosabel}, {Gull}, {Hjorth}, {Holland}, {Kouveliotou}, {Levay},
  {Livio}, {Metzger}, {Nugent}, {Petro}, {Pian}, {Rhoads}, {Riess}, {Sahu},
  {Smette}, {Tanvir}, {Wijers}, \& {Woosley}}]{Fruchter2006Natur.441..463F}
{Fruchter}, A.~S., {Levan}, A.~J., {Strolger}, L., {et~al.} 2006, \nat, 441,
  463

\bibitem[{{Fulbright}(2000{\natexlab{a}})}]{fulbright2000abundances}
{Fulbright}, J.~P. 2000{\natexlab{a}}, \aj, 120, 1841

\bibitem[{{Fulbright}(2000{\natexlab{b}})}]{FUL00}
{Fulbright}, J.~P. 2000{\natexlab{b}}, \aj, 120, 1841

\bibitem[{{Gaia Collaboration} {et~al.}(2018){Gaia Collaboration}, {Brown},
  {Vallenari}, {Prusti}, {de Bruijne}, {Babusiaux}, {Bailer-Jones}, {Biermann},
  {Evans}, {Eyer}, {Jansen}, {Jordi}, {Klioner}, {Lammers}, {Lindegren},
  {Luri}, {Mignard}, {Panem}, {Pourbaix}, {Randich}, {Sartoretti}, {Siddiqui},
  {Soubiran}, {van Leeuwen}, {Walton}, {Arenou}, {Bastian}, {Cropper},
  {Drimmel}, {Katz}, {Lattanzi}, {Bakker}, {Cacciari}, {Casta{\~n}eda},
  {Chaoul}, {Cheek}, {De Angeli}, {Fabricius}, {Guerra}, {Holl}, {Masana},
  {Messineo}, {Mowlavi}, {Nienartowicz}, {Panuzzo}, {Portell}, {Riello},
  {Seabroke}, {Tanga}, {Th{\'e}venin}, {Gracia-Abril}, {Comoretto},
  {Garcia-Reinaldos}, {Teyssier}, {Altmann}, {Andrae}, {Audard},
  {Bellas-Velidis}, {Benson}, {Berthier}, {Blomme}, {Burgess}, {Busso},
  {Carry}, {Cellino}, {Clementini}, {Clotet}, {Creevey}, {Davidson}, {De
  Ridder}, {Delchambre}, {Dell'Oro}, {Ducourant},
  {Fern{\'a}ndez-Hern{\'a}ndez}, {Fouesneau}, {Fr{\'e}mat}, {Galluccio},
  {Garc{\'\i}a-Torres}, {Gonz{\'a}lez-N{\'u}{\~n}ez}, {Gonz{\'a}lez-Vidal},
  {Gosset}, {Guy}, {Halbwachs}, {Hambly}, {Harrison}, {Hern{\'a}ndez},
  {Hestroffer}, {Hodgkin}, {Hutton}, {Jasniewicz}, {Jean-Antoine-Piccolo},
  {Jordan}, {Korn}, {Krone-Martins}, {Lanzafame}, {Lebzelter}, {L{\"o}ffler},
  {Manteiga}, {Marrese}, {Mart{\'\i}n-Fleitas}, {Moitinho}, {Mora}, {Muinonen},
  {Osinde}, {Pancino}, {Pauwels}, {Petit}, {Recio-Blanco}, {Richards},
  {Rimoldini}, {Robin}, {Sarro}, {Siopis}, {Smith}, {Sozzetti}, {S{\"u}veges},
  {Torra}, {van Reeven}, {Abbas}, {Abreu Aramburu}, {Accart}, {Aerts},
  {Altavilla}, {{\'A}lvarez}, {Alvarez}, {Alves}, {Anderson}, {Andrei},
  {Anglada Varela}, {Antiche}, {Antoja}, {Arcay}, {Astraatmadja}, {Bach},
  {Baker}, {Balaguer-N{\'u}{\~n}ez}, {Balm}, {Barache}, {Barata}, {Barbato},
  {Barblan}, {Barklem}, {Barrado}, {Barros}, {Barstow}, {Bartholom{\'e}
  Mu{\~n}oz}, {Bassilana}, {Becciani}, {Bellazzini}, {Berihuete}, {Bertone},
  {Bianchi}, {Bienaym{\'e}}, {Blanco-Cuaresma}, {Boch}, {Boeche}, {Bombrun},
  {Borrachero}, {Bossini}, {Bouquillon}, {Bourda}, {Bragaglia}, {Bramante},
  {Breddels}, {Bressan}, {Brouillet}, {Br{\"u}semeister}, {Brugaletta},
  {Bucciarelli}, {Burlacu}, {Busonero}, {Butkevich}, {Buzzi}, {Caffau},
  {Cancelliere}, {Cannizzaro}, {Cantat-Gaudin}, {Carballo}, {Carlucci},
  {Carrasco}, {Casamiquela}, {Castellani}, {Castro-Ginard}, {Charlot},
  {Chemin}, {Chiavassa}, {Cocozza}, {Costigan}, {Cowell}, {Crifo}, {Crosta},
  {Crowley}, {Cuypers}, {Dafonte}, {Damerdji}, {Dapergolas}, {David}, {David},
  {de Laverny}, {De Luise}, {De March}, {de Martino}, {de Souza}, {de Torres},
  {Debosscher}, {del Pozo}, {Delbo}, {Delgado}, {Delgado}, {Di Matteo},
  {Diakite}, {Diener}, {Distefano}, {Dolding}, {Drazinos}, {Dur{\'a}n},
  {Edvardsson}, {Enke}, {Eriksson}, {Esquej}, {Eynard Bontemps}, {Fabre},
  {Fabrizio}, {Faigler}, {Falc{\~a}o}, {Farr{\`a}s Casas}, {Federici},
  {Fedorets}, {Fernique}, {Figueras}, {Filippi}, {Findeisen}, {Fonti},
  {Fraile}, {Fraser}, {Fr{\'e}zouls}, {Gai}, {Galleti}, {Garabato},
  {Garc{\'\i}a-Sedano}, {Garofalo}, {Garralda}, {Gavel}, {Gavras}, {Gerssen},
  {Geyer}, {Giacobbe}, {Gilmore}, {Girona}, {Giuffrida}, {Glass}, {Gomes},
  {Granvik}, {Gueguen}, {Guerrier}, {Guiraud}, {Guti{\'e}rrez-S{\'a}nchez},
  {Haigron}, {Hatzidimitriou}, {Hauser}, {Haywood}, {Heiter}, {Helmi}, {Heu},
  {Hilger}, {Hobbs}, {Hofmann}, {Holland}, {Huckle}, {Hypki}, {Icardi},
  {Jan{\ss}en}, {Jevardat de Fombelle}, {Jonker}, {Juh{\'a}sz}, {Julbe},
  {Karampelas}, {Kewley}, {Klar}, {Kochoska}, {Kohley}, {Kolenberg},
  {Kontizas}, {Kontizas}, {Koposov}, {Kordopatis}, {Kostrzewa-Rutkowska},
  {Koubsky}, {Lambert}, {Lanza}, {Lasne}, {Lavigne}, {Le Fustec}, {Le
  Poncin-Lafitte}, {Lebreton}, {Leccia}, {Leclerc}, {Lecoeur-Taibi},
  {Lenhardt}, {Leroux}, {Liao}, {Licata}, {Lindstr{\o}m}, {Lister}, {Livanou},
  {Lobel}, {L{\'o}pez}, {Managau}, {Mann}, {Mantelet}, {Marchal}, {Marchant},
  {Marconi}, {Marinoni}, {Marschalk{\'o}}, {Marshall}, {Martino}, {Marton},
  {Mary}, {Massari}, {Matijevi{\v{c}}}, {Mazeh}, {McMillan}, {Messina},
  {Michalik}, {Millar}, {Molina}, {Molinaro}, {Moln{\'a}r}, {Montegriffo},
  {Mor}, {Morbidelli}, {Morel}, {Morris}, {Mulone}, {Muraveva}, {Musella},
  {Nelemans}, {Nicastro}, {Noval}, {O'Mullane}, {Ord{\'e}novic},
  {Ord{\'o}{\~n}ez-Blanco}, {Osborne}, {Pagani}, {Pagano}, {Pailler},
  {Palacin}, {Palaversa}, {Panahi}, {Pawlak}, {Piersimoni}, {Pineau}, {Plachy},
  {Plum}, {Poggio}, {Poujoulet}, {Pr{\v{s}}a}, {Pulone}, {Racero}, {Ragaini},
  {Rambaux}, {Ramos-Lerate}, {Regibo}, {Reyl{\'e}}, {Riclet}, {Ripepi}, {Riva},
  {Rivard}, {Rixon}, {Roegiers}, {Roelens}, {Romero-G{\'o}mez}, {Rowell},
  {Royer}, {Ruiz-Dern}, {Sadowski}, {Sagrist{\`a} Sell{\'e}s}, {Sahlmann},
  {Salgado}, {Salguero}, {Sanna}, {Santana-Ros}, {Sarasso}, {Savietto},
  {Schultheis}, {Sciacca}, {Segol}, {Segovia}, {S{\'e}gransan}, {Shih},
  {Siltala}, {Silva}, {Smart}, {Smith}, {Solano}, {Solitro}, {Sordo}, {Soria
  Nieto}, {Souchay}, {Spagna}, {Spoto}, {Stampa}, {Steele},
  {Steidelm{\"u}ller}, {Stephenson}, {Stoev}, {Suess}, {Surdej}, {Szabados},
  {Szegedi-Elek}, {Tapiador}, {Taris}, {Tauran}, {Taylor}, {Teixeira},
  {Terrett}, {Teyssandier}, {Thuillot}, {Titarenko}, {Torra Clotet}, {Turon},
  {Ulla}, {Utrilla}, {Uzzi}, {Vaillant}, {Valentini}, {Valette}, {van Elteren},
  {Van Hemelryck}, {van Leeuwen}, {Vaschetto}, {Vecchiato}, {Veljanoski},
  {Viala}, {Vicente}, {Vogt}, {von Essen}, {Voss}, {Votruba}, {Voutsinas},
  {Walmsley}, {Weiler}, {Wertz}, {Wevers}, {Wyrzykowski}, {Yoldas},
  {{\v{Z}}erjal}, {Ziaeepour}, {Zorec}, {Zschocke}, {Zucker}, {Zurbach}, \&
  {Zwitter}}]{Gaia2018}
{Gaia Collaboration}, {Brown}, A.~G.~A., {Vallenari}, A., {et~al.} 2018, \aap,
  616, A1

\bibitem[{{Gaia Collaboration} {et~al.}(2016){Gaia Collaboration}, {Brown},
  {Vallenari}, {Prusti}, {de Bruijne}, {Mignard}, {Drimmel}, {Babusiaux},
  {Bailer-Jones}, {Bastian}, {Biermann}, {Evans}, {Eyer}, {Jansen}, {Jordi},
  {Katz}, {Klioner}, {Lammers}, {Lindegren}, {Luri}, {O'Mullane}, {Panem},
  {Pourbaix}, {Randich}, {Sartoretti}, {Siddiqui}, {Soubiran}, {Valette}, {van
  Leeuwen}, {Walton}, {Aerts}, {Arenou}, {Cropper}, {H{\o}g}, {Lattanzi},
  {Grebel}, {Holland}, {Huc}, {Passot}, {Perryman}, {Bramante}, {Cacciari},
  {Casta{\~n}eda}, {Chaoul}, {Cheek}, {De Angeli}, {Fabricius}, {Guerra},
  {Hern{\'a}ndez}, {Jean-Antoine-Piccolo}, {Masana}, {Messineo}, {Mowlavi},
  {Nienartowicz}, {Ord{\'o}{\~n}ez-Blanco}, {Panuzzo}, {Portell}, {Richards},
  {Riello}, {Seabroke}, {Tanga}, {Th{\'e}venin}, {Torra}, {Els},
  {Gracia-Abril}, {Comoretto}, {Garcia-Reinaldos}, {Lock}, {Mercier},
  {Altmann}, {Andrae}, {Astraatmadja}, {Bellas-Velidis}, {Benson}, {Berthier},
  {Blomme}, {Busso}, {Carry}, {Cellino}, {Clementini}, {Cowell}, {Creevey},
  {Cuypers}, {Davidson}, {De Ridder}, {de Torres}, {Delchambre}, {Dell'Oro},
  {Ducourant}, {Fr{\'e}mat}, {Garc{\'\i}a-Torres}, {Gosset}, {Halbwachs},
  {Hambly}, {Harrison}, {Hauser}, {Hestroffer}, {Hodgkin}, {Huckle}, {Hutton},
  {Jasniewicz}, {Jordan}, {Kontizas}, {Korn}, {Lanzafame}, {Manteiga},
  {Moitinho}, {Muinonen}, {Osinde}, {Pancino}, {Pauwels}, {Petit},
  {Recio-Blanco}, {Robin}, {Sarro}, {Siopis}, {Smith}, {Smith}, {Sozzetti},
  {Thuillot}, {van Reeven}, {Viala}, {Abbas}, {Abreu Aramburu}, {Accart},
  {Aguado}, {Allan}, {Allasia}, {Altavilla}, {{\'A}lvarez}, {Alves},
  {Anderson}, {Andrei}, {Anglada Varela}, {Antiche}, {Antoja}, {Ant{\'o}n},
  {Arcay}, {Bach}, {Baker}, {Balaguer-N{\'u}{\~n}ez}, {Barache}, {Barata},
  {Barbier}, {Barblan}, {Barrado y Navascu{\'e}s}, {Barros}, {Barstow},
  {Becciani}, {Bellazzini}, {Bello Garc{\'\i}a}, {Belokurov}, {Bendjoya},
  {Berihuete}, {Bianchi}, {Bienaym{\'e}}, {Billebaud}, {Blagorodnova},
  {Blanco-Cuaresma}, {Boch}, {Bombrun}, {Borrachero}, {Bouquillon}, {Bourda},
  {Bouy}, {Bragaglia}, {Breddels}, {Brouillet}, {Br{\"u}semeister},
  {Bucciarelli}, {Burgess}, {Burgon}, {Burlacu}, {Busonero}, {Buzzi}, {Caffau},
  {Cambras}, {Campbell}, {Cancelliere}, {Cantat-Gaudin}, {Carlucci},
  {Carrasco}, {Castellani}, {Charlot}, {Charnas}, {Chiavassa}, {Clotet},
  {Cocozza}, {Collins}, {Costigan}, {Crifo}, {Cross}, {Crosta}, {Crowley},
  {Dafonte}, {Damerdji}, {Dapergolas}, {David}, {David}, {De Cat}, {de Felice},
  {de Laverny}, {De Luise}, {De March}, {de Martino}, {de Souza}, {Debosscher},
  {del Pozo}, {Delbo}, {Delgado}, {Delgado}, {Di Matteo}, {Diakite},
  {Distefano}, {Dolding}, {Dos Anjos}, {Drazinos}, {Duran}, {Dzigan},
  {Edvardsson}, {Enke}, {Evans}, {Eynard Bontemps}, {Fabre}, {Fabrizio},
  {Faigler}, {Falc{\~a}o}, {Farr{\`a}s Casas}, {Federici}, {Fedorets},
  {Fern{\'a}ndez-Hern{\'a}ndez}, {Fernique}, {Fienga}, {Figueras}, {Filippi},
  {Findeisen}, {Fonti}, {Fouesneau}, {Fraile}, {Fraser}, {Fuchs}, {Gai},
  {Galleti}, {Galluccio}, {Garabato}, {Garc{\'\i}a-Sedano}, {Garofalo},
  {Garralda}, {Gavras}, {Gerssen}, {Geyer}, {Gilmore}, {Girona}, {Giuffrida},
  {Gomes}, {Gonz{\'a}lez-Marcos}, {Gonz{\'a}lez-N{\'u}{\~n}ez},
  {Gonz{\'a}lez-Vidal}, {Granvik}, {Guerrier}, {Guillout}, {Guiraud},
  {G{\'u}rpide}, {Guti{\'e}rrez-S{\'a}nchez}, {Guy}, {Haigron},
  {Hatzidimitriou}, {Haywood}, {Heiter}, {Helmi}, {Hobbs}, {Hofmann}, {Holl},
  {Holland}, {Hunt}, {Hypki}, {Icardi}, {Irwin}, {Jevardat de Fombelle},
  {Jofr{\'e}}, {Jonker}, {Jorissen}, {Julbe}, {Karampelas}, {Kochoska},
  {Kohley}, {Kolenberg}, {Kontizas}, {Koposov}, {Kordopatis}, {Koubsky},
  {Krone-Martins}, {Kudryashova}, {Kull}, {Bachchan}, {Lacoste-Seris}, {Lanza},
  {Lavigne}, {Le Poncin-Lafitte}, {Lebreton}, {Lebzelter}, {Leccia}, {Leclerc},
  {Lecoeur-Taibi}, {Lemaitre}, {Lenhardt}, {Leroux}, {Liao}, {Licata},
  {Lindstr{\o}m}, {Lister}, {Livanou}, {Lobel}, {L{\"o}ffler}, {L{\'o}pez},
  {Lorenz}, {MacDonald}, {Magalh{\~a}es Fernandes}, {Managau}, {Mann},
  {Mantelet}, {Marchal}, {Marchant}, {Marconi}, {Marinoni}, {Marrese},
  {Marschalk{\'o}}, {Marshall}, {Mart{\'\i}n-Fleitas}, {Martino}, {Mary},
  {Matijevi{\v{c}}}, {Mazeh}, {McMillan}, {Messina}, {Michalik}, {Millar},
  {Miranda}, {Molina}, {Molinaro}, {Molinaro}, {Moln{\'a}r}, {Moniez},
  {Montegriffo}, {Mor}, {Mora}, {Morbidelli}, {Morel}, {Morgenthaler},
  {Morris}, {Mulone}, {Muraveva}, {Musella}, {Narbonne}, {Nelemans},
  {Nicastro}, {Noval}, {Ord{\'e}novic}, {Ordieres-Mer{\'e}}, {Osborne},
  {Pagani}, {Pagano}, {Pailler}, {Palacin}, {Palaversa}, {Parsons}, {Pecoraro},
  {Pedrosa}, {Pentik{\"a}inen}, {Pichon}, {Piersimoni}, {Pineau}, {Plachy},
  {Plum}, {Poujoulet}, {Pr{\v{s}}a}, {Pulone}, {Ragaini}, {Rago}, {Rambaux},
  {Ramos-Lerate}, {Ranalli}, {Rauw}, {Read}, {Regibo}, {Reyl{\'e}}, {Ribeiro},
  {Rimoldini}, {Ripepi}, {Riva}, {Rixon}, {Roelens}, {Romero-G{\'o}mez},
  {Rowell}, {Royer}, {Ruiz-Dern}, {Sadowski}, {Sagrist{\`a} Sell{\'e}s},
  {Sahlmann}, {Salgado}, {Salguero}, {Sarasso}, {Savietto}, {Schultheis},
  {Sciacca}, {Segol}, {Segovia}, {Segransan}, {Shih}, {Smareglia}, {Smart},
  {Solano}, {Solitro}, {Sordo}, {Soria Nieto}, {Souchay}, {Spagna}, {Spoto},
  {Stampa}, {Steele}, {Steidelm{\"u}ller}, {Stephenson}, {Stoev}, {Suess},
  {S{\"u}veges}, {Surdej}, {Szabados}, {Szegedi-Elek}, {Tapiador}, {Taris},
  {Tauran}, {Taylor}, {Teixeira}, {Terrett}, {Tingley}, {Trager}, {Turon},
  {Ulla}, {Utrilla}, {Valentini}, {van Elteren}, {Van Hemelryck}, {van
  Leeuwen}, {Varadi}, {Vecchiato}, {Veljanoski}, {Via}, {Vicente}, {Vogt},
  {Voss}, {Votruba}, {Voutsinas}, {Walmsley}, {Weiler}, {Weingrill}, {Wevers},
  {Wyrzykowski}, {Yoldas}, {{\v{Z}}erjal}, {Zucker}, {Zurbach}, {Zwitter},
  {Alecu}, {Allen}, {Allende Prieto}, {Amorim}, {Anglada-Escud{\'e}},
  {Arsenijevic}, {Azaz}, {Balm}, {Beck}, {Bernstein}, {Bigot}, {Bijaoui},
  {Blasco}, {Bonfigli}, {Bono}, {Boudreault}, {Bressan}, {Brown}, {Brunet},
  {Bunclark}, {Buonanno}, {Butkevich}, {Carret}, {Carrion}, {Chemin},
  {Ch{\'e}reau}, {Corcione}, {Darmigny}, {de Boer}, {de Teodoro}, {de Zeeuw},
  {Delle Luche}, {Domingues}, {Dubath}, {Fodor}, {Fr{\'e}zouls}, {Fries},
  {Fustes}, {Fyfe}, {Gallardo}, {Gallegos}, {Gardiol}, {Gebran}, {Gomboc},
  {G{\'o}mez}, {Grux}, {Gueguen}, {Heyrovsky}, {Hoar}, {Iannicola}, {Isasi
  Parache}, {Janotto}, {Joliet}, {Jonckheere}, {Keil}, {Kim}, {Klagyivik},
  {Klar}, {Knude}, {Kochukhov}, {Kolka}, {Kos}, {Kutka}, {Lainey}, {LeBouquin},
  {Liu}, {Loreggia}, {Makarov}, {Marseille}, {Martayan}, {Martinez-Rubi},
  {Massart}, {Meynadier}, {Mignot}, {Munari}, {Nguyen}, {Nordlander}, {Ocvirk},
  {O'Flaherty}, {Olias Sanz}, {Ortiz}, {Osorio}, {Oszkiewicz}, {Ouzounis},
  {Palmer}, {Park}, {Pasquato}, {Peltzer}, {Peralta}, {P{\'e}turaud},
  {Pieniluoma}, {Pigozzi}, {Poels}, {Prat}, {Prod'homme}, {Raison}, {Rebordao},
  {Risquez}, {Rocca-Volmerange}, {Rosen}, {Ruiz-Fuertes}, {Russo}, {Sembay},
  {Serraller Vizcaino}, {Short}, {Siebert}, {Silva}, {Sinachopoulos}, {Slezak},
  {Soffel}, {Sosnowska}, {Strai{\v{z}}ys}, {ter Linden}, {Terrell}, {Theil},
  {Tiede}, {Troisi}, {Tsalmantza}, {Tur}, {Vaccari}, {Vachier}, {Valles}, {Van
  Hamme}, {Veltz}, {Virtanen}, {Wallut}, {Wichmann}, {Wilkinson}, {Ziaeepour},
  \& {Zschocke}}]{Gaia2016}
{Gaia Collaboration}, {Brown}, A.~G.~A., {Vallenari}, A., {et~al.} 2016, \aap,
  595, A2

\bibitem[{{Ghirlanda} {et~al.}(2016){Ghirlanda}, {Salafia}, {Pescalli},
  {Ghisellini}, {Salvaterra}, {Chassande-Mottin}, {Colpi}, {Nappo}, {D'Avanzo},
  {Melandri}, {Bernardini}, {Branchesi}, {Campana}, {Ciolfi}, {Covino},
  {G{\"o}tz}, {Vergani}, {Zennaro}, \& {Tagliaferri}}]{Ghirlanda2016}
{Ghirlanda}, G., {Salafia}, O.~S., {Pescalli}, A., {et~al.} 2016, \aap, 594,
  A84

\bibitem[{{Giacobbo} \& {Mapelli}(2018)}]{giacobbo2018progenitors}
{Giacobbo}, N. \& {Mapelli}, M. 2018, \mnras, 480, 2011

\bibitem[{{Giacobbo} \& {Mapelli}(2019)}]{GiacobboMapeli2018b}
{Giacobbo}, N. \& {Mapelli}, M. 2019, \mnras, 482, 2234

\bibitem[{{Goldstein} {et~al.}(2017){Goldstein}, {Veres}, {Burns}, {Briggs},
  {Hamburg}, {Kocevski}, {Wilson-Hodge}, {Preece}, {Poolakkil}, {Roberts},
  {Hui}, {Connaughton}, {Racusin}, {von Kienlin}, {Dal Canton}, {Christensen},
  {Littenberg}, {Siellez}, {Blackburn}, {Broida}, {Bissaldi}, {Cleveland},
  {Gibby}, {Giles}, {Kippen}, {McBreen}, {McEnery}, {Meegan}, {Paciesas}, \&
  {Stanbro}}]{Goldstein2017}
{Goldstein}, A., {Veres}, P., {Burns}, E., {et~al.} 2017, \apjl, 848, L14

\bibitem[{{Graur} {et~al.}(2011){Graur}, {Poznanski}, {Maoz}, {Yasuda},
  {Totani}, {Fukugita}, {Filippenko}, {Foley}, {Silverman}, {Gal-Yam},
  {Horesh}, \& {Jannuzi}}]{graur2011supernovae}
{Graur}, O., {Poznanski}, D., {Maoz}, D., {et~al.} 2011, \mnras, 417, 916

\bibitem[{{Greggio}(2005)}]{Greggio2005}
{Greggio}, L. 2005, \aap, 441, 1055

\bibitem[{{Greggio} {et~al.}(2021){Greggio}, {Simonetti}, \&
  {Matteucci}}]{Greggio2021}
{Greggio}, L., {Simonetti}, P., \& {Matteucci}, F. 2021, \mnras, 500, 1755

\bibitem[{{Guetta} \& {Piran}(2006)}]{Guetta2006}
{Guetta}, D. \& {Piran}, T. 2006, \aap, 453, 823

\bibitem[{{Hansen} {et~al.}(2011){Hansen}, {Nordstr{\"o}m}, {Bonifacio},
  {Spite}, {Andersen}, {Beers}, {Cayrel}, {Spite}, {Molaro}, {Barbuy},
  {Depagne}, {Fran{\c{c}}ois}, {Hill}, {Plez}, \& {Sivarani}}]{HAN11}
{Hansen}, C.~J., {Nordstr{\"o}m}, B., {Bonifacio}, P., {et~al.} 2011, \aap,
  527, A65

\bibitem[{{Hansen} {et~al.}(2012){Hansen}, {Primas}, {Hartman}, {Kratz},
  {Wanajo}, {Leibundgut}, {Farouqi}, {Hallmann}, {Christlieb}, \&
  {Nilsson}}]{HAN12}
{Hansen}, C.~J., {Primas}, F., {Hartman}, H., {et~al.} 2012, \aap, 545, A31

\bibitem[{{Hansen} {et~al.}(2015){Hansen}, {Hansen}, {Christlieb}, {Beers},
  {Yong}, {Bessell}, {Frebel}, {Garc{\'\i}a P{\'e}rez}, {Placco}, {Norris}, \&
  {Asplund}}]{HAN15}
{Hansen}, T., {Hansen}, C.~J., {Christlieb}, N., {et~al.} 2015, \apj, 807, 173

\bibitem[{{Hattori} {et~al.}(2022){Hattori}, {Okuno}, \&
  {Roederer}}]{Hattori2022arXiv220704110H}
{Hattori}, K., {Okuno}, A., \& {Roederer}, I.~U. 2022, arXiv e-prints,
  arXiv:2207.04110

\bibitem[{{Hayek} {et~al.}(2009){Hayek}, {Wiesendahl}, {Christlieb},
  {Eriksson}, {Korn}, {Barklem}, {Hill}, {Beers}, {Farouqi}, {Pfeiffer}, \&
  {Kratz}}]{HAY09}
{Hayek}, W., {Wiesendahl}, U., {Christlieb}, N., {et~al.} 2009, \aap, 504, 511

\bibitem[{{Helmi} {et~al.}(2018){Helmi}, {Babusiaux}, {Koppelman}, {Massari},
  {Veljanoski}, \& {Brown}}]{Helmi2018}
{Helmi}, A., {Babusiaux}, C., {Koppelman}, H.~H., {et~al.} 2018, \nat, 563, 85

\bibitem[{{Heringer} {et~al.}(2017){Heringer}, {Pritchet}, {Kezwer}, {Graham},
  {Sand}, \& {Bildfell}}]{heringer2016type}
{Heringer}, E., {Pritchet}, C., {Kezwer}, J., {et~al.} 2017, \apj, 834, 15

\bibitem[{{Hirai} {et~al.}(2022){Hirai}, {Beers}, {Chiba}, {Aoki}, {Shank},
  {Saitoh}, {Okamoto}, \& {Makino}}]{Hirai2022arXiv220604060H}
{Hirai}, Y., {Beers}, T.~C., {Chiba}, M., {et~al.} 2022, arXiv e-prints,
  arXiv:2206.04060

\bibitem[{{Hirai} {et~al.}(2015){Hirai}, {Ishimaru}, {Saitoh}, {Fujii},
  {Hidaka}, \& {Kajino}}]{Hirai_2015}
{Hirai}, Y., {Ishimaru}, Y., {Saitoh}, T.~R., {et~al.} 2015, \apj, 814, 41

\bibitem[{{Hollek} {et~al.}(2011){Hollek}, {Frebel}, {Roederer}, {Sneden},
  {Shetrone}, {Beers}, {Kang}, \& {Thom}}]{HOL11}
{Hollek}, J.~K., {Frebel}, A., {Roederer}, I.~U., {et~al.} 2011, \apj, 742, 54

\bibitem[{{Honda} {et~al.}(2011){Honda}, {Aoki}, {Beers}, \&
  {Takada-Hidai}}]{HON11}
{Honda}, S., {Aoki}, W., {Beers}, T.~C., \& {Takada-Hidai}, M. 2011, \apj, 730,
  77

\bibitem[{{Honda} {et~al.}(2004{\natexlab{a}}){Honda}, {Aoki}, {Kajino},
  {Ando}, {Beers}, {Izumiura}, {Sadakane}, \&
  {Takada-Hidai}}]{honda2004spectroscopic}
{Honda}, S., {Aoki}, W., {Kajino}, T., {et~al.} 2004{\natexlab{a}}, \apj, 607,
  474

\bibitem[{{Honda} {et~al.}(2004{\natexlab{b}}){Honda}, {Aoki}, {Kajino},
  {Ando}, {Beers}, {Izumiura}, {Sadakane}, \& {Takada-Hidai}}]{HON04}
{Honda}, S., {Aoki}, W., {Kajino}, T., {et~al.} 2004{\natexlab{b}}, \apj, 607,
  474

\bibitem[{{Hotokezaka} {et~al.}(2013){Hotokezaka}, {Kiuchi}, {Kyutoku},
  {Muranushi}, {Sekiguchi}, {Shibata}, \& {Taniguchi}}]{2013PhRvD..88d4026H}
{Hotokezaka}, K., {Kiuchi}, K., {Kyutoku}, K., {et~al.} 2013, \prd, 88, 044026

\bibitem[{{Howard} {et~al.}(1972){Howard}, {Arnett}, {Clayton}, \&
  {Woosley}}]{Howard1972ApJ}
{Howard}, W.~M., {Arnett}, W.~D., {Clayton}, D.~D., \& {Woosley}, S.~E. 1972,
  \apj, 175, 201

\bibitem[{{Howard} {et~al.}(1986){Howard}, {Mathews}, {Takahashi}, \&
  {Ward}}]{howard1986parametric}
{Howard}, W.~M., {Mathews}, G.~J., {Takahashi}, K., \& {Ward}, R.~A. 1986,
  \apj, 309, 633

\bibitem[{{Ishigaki} {et~al.}(2010){Ishigaki}, {Chiba}, \& {Aoki}}]{ISH10}
{Ishigaki}, M., {Chiba}, M., \& {Aoki}, W. 2010, \pasj, 62, 143

\bibitem[{{Ishigaki} {et~al.}(2013){Ishigaki}, {Aoki}, \& {Chiba}}]{ISH13}
{Ishigaki}, M.~N., {Aoki}, W., \& {Chiba}, M. 2013, \apj, 771, 67

\bibitem[{{Ishimaru} {et~al.}(2015){Ishimaru}, {Wanajo}, \&
  {Prantzos}}]{Ishimaru2015}
{Ishimaru}, Y., {Wanajo}, S., \& {Prantzos}, N. 2015, \apjl, 804, L35

\bibitem[{{Ivans} {et~al.}(2006){Ivans}, {Simmerer}, {Sneden}, {Lawler},
  {Cowan}, {Gallino}, \& {Bisterzo}}]{IVA06}
{Ivans}, I.~I., {Simmerer}, J., {Sneden}, C., {et~al.} 2006, \apj, 645, 613

\bibitem[{{Ivans} {et~al.}(2003){Ivans}, {Sneden}, {James}, {Preston},
  {Fulbright}, {H{\"o}flich}, {Carney}, \& {Wheeler}}]{IVA03}
{Ivans}, I.~I., {Sneden}, C., {James}, C.~R., {et~al.} 2003, \apj, 592, 906

\bibitem[{{Ivezi{\'c}} {et~al.}(2012){Ivezi{\'c}}, {Beers}, \&
  {Juri{\'c}}}]{Ivezic2012}
{Ivezi{\'c}}, {\v{Z}}., {Beers}, T.~C., \& {Juri{\'c}}, M. 2012, \araa, 50, 251

\bibitem[{{Jacobson} {et~al.}(2015){Jacobson}, {Keller}, {Frebel}, {Casey},
  {Asplund}, {Bessell}, {Da Costa}, {Lind}, {Marino}, {Norris}, {Pe{\~n}a},
  {Schmidt}, {Tisserand}, {Walsh}, {Yong}, \& {Yu}}]{JAC15}
{Jacobson}, H.~R., {Keller}, S., {Frebel}, A., {et~al.} 2015, \apj, 807, 171

\bibitem[{{Ji} {et~al.}(2016){Ji}, {Frebel}, {Chiti}, \&
  {Simon}}]{Ji2016Natur.531..610J}
{Ji}, A.~P., {Frebel}, A., {Chiti}, A., \& {Simon}, J.~D. 2016, \nat, 531, 610

\bibitem[{{Johnson}(2002)}]{JOH02a}
{Johnson}, J.~A. 2002, \apjs, 139, 219

\bibitem[{{Johnson} \& {Bolte}(2004)}]{JOH04}
{Johnson}, J.~A. \& {Bolte}, M. 2004, \apj, 605, 462

\bibitem[{{Johnston} {et~al.}(2008){Johnston}, {Bullock}, {Sharma}, {Font},
  {Robertson}, \& {Leitner}}]{Johnston2008}
{Johnston}, K.~V., {Bullock}, J.~S., {Sharma}, S., {et~al.} 2008, \apj, 689,
  936

\bibitem[{{Jonsell} {et~al.}(2006){Jonsell}, {Barklem}, {Gustafsson},
  {Christlieb}, {Hill}, {Beers}, \& {Holmberg}}]{JON06}
{Jonsell}, K., {Barklem}, P.~S., {Gustafsson}, B., {et~al.} 2006, \aap, 451,
  651

\bibitem[{{Jonsell} {et~al.}(2005){Jonsell}, {Edvardsson}, {Gustafsson},
  {Magain}, {Nissen}, \& {Asplund}}]{JON05}
{Jonsell}, K., {Edvardsson}, B., {Gustafsson}, B., {et~al.} 2005, \aap, 440,
  321

\bibitem[{{Kirby} {et~al.}(2013){Kirby}, {Cohen}, {Guhathakurta}, {Cheng},
  {Bullock}, \& {Gallazzi}}]{Kirby2013}
{Kirby}, E.~N., {Cohen}, J.~G., {Guhathakurta}, P., {et~al.} 2013, \apj, 779,
  102

\bibitem[{{Koch} \& {Edvardsson}(2002)}]{koch2002europium}
{Koch}, A. \& {Edvardsson}, B. 2002, \aap, 381, 500

\bibitem[{{Komiya} \& {Shigeyama}(2016)}]{Komiya2016}
{Komiya}, Y. \& {Shigeyama}, T. 2016, \apj, 830, 76

\bibitem[{{Korobkin} {et~al.}(2012){Korobkin}, {Rosswog}, {Arcones}, \&
  {Winteler}}]{korobkin2012astrophysical}
{Korobkin}, O., {Rosswog}, S., {Arcones}, A., \& {Winteler}, C. 2012, \mnras,
  426, 1940

\bibitem[{{Krisciunas} {et~al.}(2017){Krisciunas}, {Contreras}, {Burns},
  {Phillips}, {Stritzinger}, {Morrell}, {Hamuy}, {Anais}, {Boldt}, {Busta},
  {Campillay}, {Castell{\'o}n}, {Folatelli}, {Freedman}, {Gonz{\'a}lez},
  {Hsiao}, {Krzeminski}, {Persson}, {Roth}, {Salgado}, {Ser{\'o}n}, {Suntzeff},
  {Torres}, {Filippenko}, {Li}, {Madore}, {DePoy}, {Marshall}, {Rheault}, \&
  {Villanueva}}]{krisciunas2017carnegie}
{Krisciunas}, K., {Contreras}, C., {Burns}, C.~R., {et~al.} 2017, \aj, 154, 211

\bibitem[{{Lai} {et~al.}(2008){Lai}, {Bolte}, {Johnson}, {Lucatello}, {Heger},
  \& {Woosley}}]{LAI08}
{Lai}, D.~K., {Bolte}, M., {Johnson}, J.~A., {et~al.} 2008, \apj, 681, 1524

\bibitem[{{Lai} {et~al.}(2007){Lai}, {Johnson}, {Bolte}, \&
  {Lucatello}}]{LAI07}
{Lai}, D.~K., {Johnson}, J.~A., {Bolte}, M., \& {Lucatello}, S. 2007, \apj,
  667, 1185

\bibitem[{{Lanfranchi} \&
  {Matteucci}(2003)}]{LanfranchiMatteucci2003MNRAS.345...71L}
{Lanfranchi}, G.~A. \& {Matteucci}, F. 2003, \mnras, 345, 71

\bibitem[{{Leaman} {et~al.}(2011){Leaman}, {Li}, {Chornock}, \&
  {Filippenko}}]{Leaman2011}
{Leaman}, J., {Li}, W., {Chornock}, R., \& {Filippenko}, A.~V. 2011, \mnras,
  412, 1419

\bibitem[{{Li} {et~al.}(2015{\natexlab{a}}){Li}, {Aoki}, {Zhao}, {Honda},
  {Christlieb}, \& {Suda}}]{LI15b}
{Li}, H., {Aoki}, W., {Zhao}, G., {et~al.} 2015{\natexlab{a}}, \pasj, 67, 84

\bibitem[{{Li} {et~al.}(2013){Li}, {Ludwig}, {Caffau}, {Christlieb}, \&
  {Zhao}}]{LI13}
{Li}, H.~N., {Ludwig}, H.~G., {Caffau}, E., {Christlieb}, N., \& {Zhao}, G.
  2013, \apj, 765, 51

\bibitem[{{Li} {et~al.}(2015{\natexlab{b}}){Li}, {Zhao}, {Christlieb}, {Wang},
  {Wang}, {Zhang}, {Hou}, \& {Yuan}}]{LI15a}
{Li}, H.-N., {Zhao}, G., {Christlieb}, N., {et~al.} 2015{\natexlab{b}}, \apj,
  798, 110

\bibitem[{{Li} {et~al.}(2011){Li}, {Chornock}, {Leaman}, {Filippenko},
  {Poznanski}, {Wang}, {Ganeshalingam}, \& {Mannucci}}]{Li2011}
{Li}, W., {Chornock}, R., {Leaman}, J., {et~al.} 2011, \mnras, 412, 1473

\bibitem[{{Limongi} \& {Chieffi}(2018)}]{LC18}
{Limongi}, M. \& {Chieffi}, A. 2018, \apjs, 237, 13

\bibitem[{{Lucatello} {et~al.}(2003){Lucatello}, {Gratton}, {Cohen}, {Beers},
  {Christlieb}, {Carretta}, \& {Ram{\'\i}rez}}]{LUC03}
{Lucatello}, S., {Gratton}, R., {Cohen}, J.~G., {et~al.} 2003, \aj, 125, 875

\bibitem[{{Mackereth} \& {Bovy}(2020)}]{Mackereth2020}
{Mackereth}, J.~T. \& {Bovy}, J. 2020, \mnras, 492, 3631

\bibitem[{{Madau} \& {Dickinson}(2014)}]{MD2014}
{Madau}, P. \& {Dickinson}, M. 2014, \araa, 52, 415

\bibitem[{{Maeder} \& {Meynet}(1989)}]{maeder1989grids}
{Maeder}, A. \& {Meynet}, G. 1989, \aap, 210, 155

\bibitem[{{Majewski} {et~al.}(2003){Majewski}, {Skrutskie}, {Weinberg}, \&
  {Ostheimer}}]{Majewski2003}
{Majewski}, S.~R., {Skrutskie}, M.~F., {Weinberg}, M.~D., \& {Ostheimer}, J.~C.
  2003, \apj, 599, 1082

\bibitem[{{Maoz} \& {Badenes}(2010)}]{maoz2010supernova}
{Maoz}, D. \& {Badenes}, C. 2010, \mnras, 407, 1314

\bibitem[{{Maoz} \& {Mannucci}(2012)}]{maoz2012type}
{Maoz}, D. \& {Mannucci}, F. 2012, \pasa, 29, 447

\bibitem[{{Mashonkina} {et~al.}(2010){Mashonkina}, {Christlieb}, {Barklem},
  {Hill}, {Beers}, \& {Velichko}}]{MAS10}
{Mashonkina}, L., {Christlieb}, N., {Barklem}, P.~S., {et~al.} 2010, \aap, 516,
  A46

\bibitem[{{Mashonkina} {et~al.}(2014){Mashonkina}, {Christlieb}, \&
  {Eriksson}}]{MAS14}
{Mashonkina}, L., {Christlieb}, N., \& {Eriksson}, K. 2014, \aap, 569, A43

\bibitem[{{Masseron} {et~al.}(2012){Masseron}, {Johnson}, {Lucatello},
  {Karakas}, {Plez}, {Beers}, \& {Christlieb}}]{MAS12}
{Masseron}, T., {Johnson}, J.~A., {Lucatello}, S., {et~al.} 2012, \apj, 751, 14

\bibitem[{{Masseron} {et~al.}(2006){Masseron}, {van Eck}, {Famaey}, {Goriely},
  {Plez}, {Siess}, {Beers}, {Primas}, \& {Jorissen}}]{MAS06}
{Masseron}, T., {van Eck}, S., {Famaey}, B., {et~al.} 2006, \aap, 455, 1059

\bibitem[{{Matteucci} \& {Greggio}(1986)}]{1986A&A...154..279M}
{Matteucci}, F. \& {Greggio}, L. 1986, \aap, 154, 279

\bibitem[{{Matteucci} {et~al.}(2006){Matteucci}, {Panagia}, {Pipino},
  {Mannucci}, {Recchi}, \& {Della Valle}}]{matteucci2006}
{Matteucci}, F., {Panagia}, N., {Pipino}, A., {et~al.} 2006, \mnras, 372, 265

\bibitem[{{Matteucci} {et~al.}(2014){Matteucci}, {Romano}, {Arcones},
  {Korobkin}, \& {Rosswog}}]{matteucci2014europium}
{Matteucci}, F., {Romano}, D., {Arcones}, A., {Korobkin}, O., \& {Rosswog}, S.
  2014, \mnras, 438, 2177

\bibitem[{{McWilliam}(1998)}]{mcwilliam98}
{McWilliam}, A. 1998, \aj, 115, 1640

\bibitem[{{McWilliam} {et~al.}(1995){McWilliam}, {Preston}, {Sneden}, \&
  {Searle}}]{MCW95}
{McWilliam}, A., {Preston}, G.~W., {Sneden}, C., \& {Searle}, L. 1995, \aj,
  109, 2757

\bibitem[{{Mennekens} \& {Vanbeveren}(2014)}]{mennekens2014massive}
{Mennekens}, N. \& {Vanbeveren}, D. 2014, \aap, 564, A134

\bibitem[{{Molero} {et~al.}(2021){Molero}, {Simonetti}, {Matteucci}, \& {della
  Valle}}]{molero2021predicted}
{Molero}, M., {Simonetti}, P., {Matteucci}, F., \& {della Valle}, M. 2021,
  \mnras, 500, 1071

\bibitem[{{M{\"o}sta} {et~al.}(2015){M{\"o}sta}, {Ott}, {Radice}, {Roberts},
  {Schnetter}, \& {Haas}}]{2015Natur.528..376M}
{M{\"o}sta}, P., {Ott}, C.~D., {Radice}, D., {et~al.} 2015, \nat, 528, 376

\bibitem[{{Naiman} {et~al.}(2018){Naiman}, {Pillepich}, {Springel},
  {Ramirez-Ruiz}, {Torrey}, {Vogelsberger}, {Pakmor}, {Nelson}, {Marinacci},
  {Hernquist}, {Weinberger}, \& {Genel}}]{Naiman2018}
{Naiman}, J.~P., {Pillepich}, A., {Springel}, V., {et~al.} 2018, \mnras, 477,
  1206

\bibitem[{{Newberg} {et~al.}(2002){Newberg}, {Yanny}, {Rockosi}, {Grebel},
  {Rix}, {Brinkmann}, {Csabai}, {Hennessy}, {Hindsley}, {Ibata}, {Ivezi{\'c}},
  {Lamb}, {Nash}, {Odenkirchen}, {Rave}, {Schneider}, {Smith}, {Stolte}, \&
  {York}}]{Newberg2002}
{Newberg}, H.~J., {Yanny}, B., {Rockosi}, C., {et~al.} 2002, \apj, 569, 245

\bibitem[{{Nishimura} {et~al.}(2015){Nishimura}, {Takiwaki}, \&
  {Thielemann}}]{Nishimura2015}
{Nishimura}, N., {Takiwaki}, T., \& {Thielemann}, F.-K. 2015, \apj, 810, 109

\bibitem[{{Norris} {et~al.}(1997){Norris}, {Ryan}, {Beers}, \&
  {Deliyannis}}]{NOR97A}
{Norris}, J.~E., {Ryan}, S.~G., {Beers}, T.~C., \& {Deliyannis}, C.~P. 1997,
  \apj, 485, 370

\bibitem[{{Oechslin} {et~al.}(2007){Oechslin}, {Janka}, \&
  {Marek}}]{2007A&A...467..395O}
{Oechslin}, R., {Janka}, H.~T., \& {Marek}, A. 2007, \aap, 467, 395

\bibitem[{{Panov} {et~al.}(2008){Panov}, {Korneev}, \&
  {Thielemann}}]{2008AstL...34..189P}
{Panov}, I.~V., {Korneev}, I.~Y., \& {Thielemann}, F.~K. 2008, Astronomy
  Letters, 34, 189

\bibitem[{{Perego} {et~al.}(2014){Perego}, {Rosswog}, {Cabez{\'o}n},
  {Korobkin}, {K{\"a}ppeli}, {Arcones}, \&
  {Liebend{\"o}rfer}}]{2014MNRAS.443.3134P}
{Perego}, A., {Rosswog}, S., {Cabez{\'o}n}, R.~M., {et~al.} 2014, \mnras, 443,
  3134

\bibitem[{{Perego} {et~al.}(2021){Perego}, {Thielemann}, \&
  {Cescutti}}]{Perego2021hgwa.bookE..13P}
{Perego}, A., {Thielemann}, F.~K., \& {Cescutti}, G. 2021, in Handbook of
  Gravitational Wave Astronomy (Springer Singapore), 13

\bibitem[{{Placco} {et~al.}(2014){Placco}, {Frebel}, {Beers}, {Christlieb},
  {Lee}, {Kennedy}, {Rossi}, \& {Santucci}}]{PAL14a}
{Placco}, V.~M., {Frebel}, A., {Beers}, T.~C., {et~al.} 2014, \apj, 781, 40

\bibitem[{{Placco} {et~al.}(2013){Placco}, {Frebel}, {Beers}, {Karakas},
  {Kennedy}, {Rossi}, {Christlieb}, \& {Stancliffe}}]{PLA13}
{Placco}, V.~M., {Frebel}, A., {Beers}, T.~C., {et~al.} 2013, \apj, 770, 104

\bibitem[{{Prantzos} {et~al.}(2020){Prantzos}, {Abia}, {Cristallo}, {Limongi},
  \& {Chieffi}}]{Prantzos2020}
{Prantzos}, N., {Abia}, C., {Cristallo}, S., {Limongi}, M., \& {Chieffi}, A.
  2020, \mnras, 491, 1832

\bibitem[{{Preston} {et~al.}(2001){Preston}, {Martins}, \& {Rundle}}]{PRE01}
{Preston}, E.~F., {Martins}, J. S.~S., \& {Rundle}, J.~B. 2001, arXiv e-prints,
  cond

\bibitem[{{Preston} \& {Sneden}(2000)}]{PRE00}
{Preston}, G.~W. \& {Sneden}, C. 2000, \aj, 120, 1014

\bibitem[{{Preston} {et~al.}(2006){Preston}, {Thompson}, {Sneden},
  {Stachowski}, \& {Shectman}}]{PRE06}
{Preston}, G.~W., {Thompson}, I.~B., {Sneden}, C., {Stachowski}, G., \&
  {Shectman}, S.~A. 2006, \aj, 132, 1714

\bibitem[{{Radice} {et~al.}(2018){Radice}, {Perego}, {Hotokezaka}, {Fromm},
  {Bernuzzi}, \& {Roberts}}]{Radice2018}
{Radice}, D., {Perego}, A., {Hotokezaka}, K., {et~al.} 2018, \apj, 869, 130

\bibitem[{{Reichert} {et~al.}(2020){Reichert}, {Hansen}, {Hanke},
  {Sk{\'u}lad{\'o}ttir}, {Arcones}, \& {Grebel}}]{Reichert2020}
{Reichert}, M., {Hansen}, C.~J., {Hanke}, M., {et~al.} 2020, \aap, 641, A127

\bibitem[{{Rodney} {et~al.}(2014){Rodney}, {Riess}, {Strolger}, {Dahlen},
  {Graur}, {Casertano}, {Dickinson}, {Ferguson}, {Garnavich}, {Hayden}, {Jha},
  {Jones}, {Kirshner}, {Koekemoer}, {McCully}, {Mobasher}, {Patel}, {Weiner},
  {Cenko}, {Clubb}, {Cooper}, {Filippenko}, {Frederiksen}, {Hjorth},
  {Leibundgut}, {Matheson}, {Nayyeri}, {Penner}, {Trump}, {Silverman}, {U},
  {Azalee Bostroem}, {Challis}, {Rajan}, {Wolff}, {Faber}, {Grogin}, \&
  {Kocevski}}]{rodney2014type}
{Rodney}, S.~A., {Riess}, A.~G., {Strolger}, L.-G., {et~al.} 2014, \aj, 148, 13

\bibitem[{{Roederer} {et~al.}(2008){Roederer}, {Frebel}, {Shetrone}, {Allende
  Prieto}, {Rhee}, {Gallino}, {Bisterzo}, {Sneden}, {Beers}, \&
  {Cowan}}]{ROE08}
{Roederer}, I.~U., {Frebel}, A., {Shetrone}, M.~D., {et~al.} 2008, \apj, 679,
  1549

\bibitem[{{Roederer} {et~al.}(2016){Roederer}, {Mateo}, {Bailey}, {Song},
  {Bell}, {Crane}, {Loebman}, {Nidever}, {Olszewski}, {Shectman}, {Thompson},
  {Valluri}, \& {Walker}}]{Roederer2016AJ....151...82R}
{Roederer}, I.~U., {Mateo}, M., {Bailey}, John~I., I., {et~al.} 2016, \aj, 151,
  82

\bibitem[{{Roederer} {et~al.}(2014){Roederer}, {Preston}, {Thompson},
  {Shectman}, {Sneden}, {Burley}, \& {Kelson}}]{roederer2014search}
{Roederer}, I.~U., {Preston}, G.~W., {Thompson}, I.~B., {et~al.} 2014, \aj,
  147, 136

\bibitem[{{Roederer} {et~al.}(2010){Roederer}, {Sneden}, {Thompson}, {Preston},
  \& {Shectman}}]{ROE10}
{Roederer}, I.~U., {Sneden}, C., {Thompson}, I.~B., {Preston}, G.~W., \&
  {Shectman}, S.~A. 2010, \apj, 711, 573

\bibitem[{{Roy}(2021)}]{Roy2021Galax...9...79R}
{Roy}, A. 2021, Galaxies, 9, 79

\bibitem[{{Ryan} {et~al.}(1996){Ryan}, {Norris}, \& {Beers}}]{RYA96}
{Ryan}, S.~G., {Norris}, J.~E., \& {Beers}, T.~C. 1996, \apj, 471, 254

\bibitem[{{Ryan} {et~al.}(1991){Ryan}, {Norris}, \& {Bessell}}]{RYA91}
{Ryan}, S.~G., {Norris}, J.~E., \& {Bessell}, M.~S. 1991, \aj, 102, 303

\bibitem[{{Salpeter}(1955)}]{Salpeter1955ApJ...121..161S}
{Salpeter}, E.~E. 1955, \apj, 121, 161

\bibitem[{{Savchenko} {et~al.}(2017){Savchenko}, {Ferrigno}, {Kuulkers},
  {Bazzano}, {Bozzo}, {Brandt}, {Chenevez}, {Courvoisier}, {Diehl}, {Domingo},
  {Hanlon}, {Jourdain}, {von Kienlin}, {Laurent}, {Lebrun}, {Lutovinov},
  {Martin-Carrillo}, {Mereghetti}, {Natalucci}, {Rodi}, {Roques}, {Sunyaev}, \&
  {Ubertini}}]{Savchenko2017}
{Savchenko}, V., {Ferrigno}, C., {Kuulkers}, E., {et~al.} 2017, \apjl, 848, L15

\bibitem[{{Scalo}(1986)}]{scalo1986stellar}
{Scalo}, J.~M. 1986, \fcp, 11, 1

\bibitem[{{Sch{\"o}nrich} \& {Weinberg}(2019)}]{schonrich2019chemical}
{Sch{\"o}nrich}, R.~A. \& {Weinberg}, D.~H. 2019, \mnras, 487, 580

\bibitem[{{Siegel} {et~al.}(2019){Siegel}, {Barnes}, \&
  {Metzger}}]{Siegel_2019}
{Siegel}, D.~M., {Barnes}, J., \& {Metzger}, B.~D. 2019, \nat, 569, 241

\bibitem[{{Simonetti} {et~al.}(2019){Simonetti}, {Matteucci}, {Greggio}, \&
  {Cescutti}}]{simonetti2019new}
{Simonetti}, P., {Matteucci}, F., {Greggio}, L., \& {Cescutti}, G. 2019,
  \mnras, 486, 2896

\bibitem[{{Siqueira Mello} {et~al.}(2014){Siqueira Mello}, {Hill}, {Barbuy},
  {Spite}, {Spite}, {Beers}, {Caffau}, {Bonifacio}, {Cayrel}, {Fran{\c{c}}ois},
  {Schatz}, \& {Wanajo}}]{SIQ14}
{Siqueira Mello}, C., {Hill}, V., {Barbuy}, B., {et~al.} 2014, \aap, 565, A93

\bibitem[{{Sivarani} {et~al.}(2006){Sivarani}, {Beers}, {Bonifacio}, {Molaro},
  {Cayrel}, {Herwig}, {Spite}, {Spite}, {Plez}, {Andersen}, {Barbuy},
  {Depagne}, {Hill}, {Fran{\c{c}}ois}, {Nordstr{\"o}m}, \& {Primas}}]{SIV06}
{Sivarani}, T., {Beers}, T.~C., {Bonifacio}, P., {et~al.} 2006, \aap, 459, 125

\bibitem[{{Sneden} {et~al.}(2003){Sneden}, {Cowan}, {Lawler}, {Ivans},
  {Burles}, {Beers}, {Primas}, {Hill}, {Truran}, {Fuller}, {Pfeiffer}, \&
  {Kratz}}]{SNE03}
{Sneden}, C., {Cowan}, J.~J., {Lawler}, J.~E., {et~al.} 2003, \apj, 591, 936

\bibitem[{{Spite} {et~al.}(2014){Spite}, {Spite}, {Bonifacio}, {Caffau},
  {Fran{\c{c}}ois}, \& {Sbordone}}]{SPI14}
{Spite}, M., {Spite}, F., {Bonifacio}, P., {et~al.} 2014, \aap, 571, A40

\bibitem[{{Symbalisty} \& {Schramm}(1982)}]{1982ApL....22..143S}
{Symbalisty}, E. \& {Schramm}, D.~N. 1982, \aplett, 22, 143

\bibitem[{{Tanaka} {et~al.}(2017){Tanaka}, {Utsumi}, {Mazzali}, {Tominaga},
  {Yoshida}, {Sekiguchi}, {Morokuma}, {Motohara}, {Ohta}, {Kawabata}, {Abe},
  {Aoki}, {Asakura}, {Baar}, {Barway}, {Bond}, {Doi}, {Fujiyoshi}, {Furusawa},
  {Honda}, {Itoh}, {Kawabata}, {Kawai}, {Kim}, {Lee}, {Miyazaki}, {Morihana},
  {Nagashima}, {Nagayama}, {Nakaoka}, {Nakata}, {Ohsawa}, {Ohshima}, {Okita},
  {Saito}, {Sumi}, {Tajitsu}, {Takahashi}, {Takayama}, {Tamura}, {Tanaka},
  {Terai}, {Tristram}, {Yasuda}, \& {Zenko}}]{Tanaka2017}
{Tanaka}, M., {Utsumi}, Y., {Mazzali}, P.~A., {et~al.} 2017, \pasj, 69, 102

\bibitem[{{Tang} {et~al.}(2020){Tang}, {Eldridge}, {Stanway}, \&
  {Bray}}]{Tang2020}
{Tang}, P.~N., {Eldridge}, J.~J., {Stanway}, E.~R., \& {Bray}, J.~C. 2020,
  \mnras, 493, L6

\bibitem[{{Tanvir} {et~al.}(2013){Tanvir}, {Levan}, {Fruchter}, {Hjorth},
  {Hounsell}, {Wiersema}, \& {Tunnicliffe}}]{2013Natur.500..547T}
{Tanvir}, N.~R., {Levan}, A.~J., {Fruchter}, A.~S., {et~al.} 2013, \nat, 500,
  547

\bibitem[{{Tauris} {et~al.}(2017){Tauris}, {Kramer}, {Freire}, {Wex}, {Janka},
  {Langer}, {Podsiadlowski}, {Bozzo}, {Chaty}, {Kruckow}, {van den Heuvel},
  {Antoniadis}, {Breton}, \& {Champion}}]{tauris2017formation}
{Tauris}, T.~M., {Kramer}, M., {Freire}, P.~C.~C., {et~al.} 2017, \apj, 846,
  170

\bibitem[{Thornton {et~al.}(1998)Thornton, Gaudlitz, Janka, \&
  Steinmetz}]{thornton1998energy}
Thornton, K., Gaudlitz, M., Janka, H.-T., \& Steinmetz, M. 1998, The
  Astrophysical Journal, 500, 95

\bibitem[{{Totani} {et~al.}(2008){Totani}, {Morokuma}, {Oda}, {Doi}, \&
  {Yasuda}}]{totani2008delay}
{Totani}, T., {Morokuma}, T., {Oda}, T., {Doi}, M., \& {Yasuda}, N. 2008,
  \pasj, 60, 1327

\bibitem[{{Tremonti} {et~al.}(2004){Tremonti}, {Heckman}, {Kauffmann},
  {Brinchmann}, {Charlot}, {White}, {Seibert}, {Peng}, {Schlegel}, {Uomoto},
  {Fukugita}, \& {Brinkmann}}]{Tremonti2004ApJ...613..898T}
{Tremonti}, C.~A., {Heckman}, T.~M., {Kauffmann}, G., {et~al.} 2004, \apj, 613,
  898

\bibitem[{{Troja} {et~al.}(2019){Troja}, {van Eerten}, {Ryan}, {Ricci},
  {Burgess}, {Wieringa}, {Piro}, {Cenko}, \& {Sakamoto}}]{Troja2019}
{Troja}, E., {van Eerten}, H., {Ryan}, G., {et~al.} 2019, \mnras, 489, 1919

\bibitem[{{Tutukov} \& {Yungelson}(1993)}]{Tutukov1993}
{Tutukov}, A.~V. \& {Yungelson}, L.~R. 1993, \mnras, 260, 675

\bibitem[{{Villar} {et~al.}(2017){Villar}, {Guillochon}, {Berger}, {Metzger},
  {Cowperthwaite}, {Nicholl}, {Alexander}, {Blanchard}, {Chornock},
  {Eftekhari}, {Fong}, {Margutti}, \& {Williams}}]{Villar2017}
{Villar}, V.~A., {Guillochon}, J., {Berger}, E., {et~al.} 2017, \apjl, 851, L21

\bibitem[{{Vincenzo} {et~al.}(2019){Vincenzo}, {Spitoni}, {Calura},
  {Matteucci}, {Silva Aguirre}, {Miglio}, \&
  {Cescutti}}]{Vincenzo2019MNRAS.487L..47V}
{Vincenzo}, F., {Spitoni}, E., {Calura}, F., {et~al.} 2019, \mnras, 487, L47

\bibitem[{{Virgili} {et~al.}(2011){Virgili}, {Zhang}, {O'Brien}, \&
  {Troja}}]{Virgili2011}
{Virgili}, F.~J., {Zhang}, B., {O'Brien}, P., \& {Troja}, E. 2011, \apj, 727,
  109

\bibitem[{{Wainwright} {et~al.}(2007){Wainwright}, {Berger}, \&
  {Penprase}}]{Wainwright2007ApJ...657..367W}
{Wainwright}, C., {Berger}, E., \& {Penprase}, B.~E. 2007, \apj, 657, 367

\bibitem[{{Wanajo} {et~al.}(2001){Wanajo}, {Kajino}, {Mathews}, \&
  {Otsuki}}]{wanajo2001r}
{Wanajo}, S., {Kajino}, T., {Mathews}, G.~J., \& {Otsuki}, K. 2001, \apj, 554,
  578

\bibitem[{{Wanajo} {et~al.}(2018){Wanajo}, {M{\"u}ller}, {Janka}, \&
  {Heger}}]{Wanajo_2018}
{Wanajo}, S., {M{\"u}ller}, B., {Janka}, H.-T., \& {Heger}, A. 2018, \apj, 852,
  40

\bibitem[{{Wanajo} {et~al.}(2014){Wanajo}, {Sekiguchi}, {Nishimura}, {Kiuchi},
  {Kyutoku}, \& {Shibata}}]{Wanajo_2014}
{Wanajo}, S., {Sekiguchi}, Y., {Nishimura}, N., {et~al.} 2014, \apjl, 789, L39

\bibitem[{Wanderman \& Piran(2015)}]{Wanderman}
Wanderman, D. \& Piran, T. 2015, Monthly Notices of the Royal Astronomical
  Society, 448, 3026

\bibitem[{{Watson} {et~al.}(2019){Watson}, {Hansen}, {Selsing}, {Koch},
  {Malesani}, {Andersen}, {Fynbo}, {Arcones}, {Bauswein}, {Covino}, {Grado},
  {Heintz}, {Hunt}, {Kouveliotou}, {Leloudas}, {Levan}, {Mazzali}, \&
  {Pian}}]{Watson2019Natur.574..497W}
{Watson}, D., {Hansen}, C.~J., {Selsing}, J., {et~al.} 2019, \nat, 574, 497

\bibitem[{{Wehmeyer} {et~al.}(2015){Wehmeyer}, {Pignatari}, \&
  {Thielemann}}]{Wehmeyer2015}
{Wehmeyer}, B., {Pignatari}, M., \& {Thielemann}, F.~K. 2015, \mnras, 452, 1970

\bibitem[{{Westin} {et~al.}(2000){Westin}, {Sneden}, {Gustafsson}, \&
  {Cowan}}]{WES00}
{Westin}, J., {Sneden}, C., {Gustafsson}, B., \& {Cowan}, J.~J. 2000, \apj,
  530, 783

\bibitem[{{Winteler} {et~al.}(2012){Winteler}, {K{\"a}ppeli}, {Perego},
  {Arcones}, {Vasset}, {Nishimura}, {Liebend{\"o}rfer}, \&
  {Thielemann}}]{winteler2012magnetorotationally}
{Winteler}, C., {K{\"a}ppeli}, R., {Perego}, A., {et~al.} 2012, \apjl, 750, L22

\bibitem[{{Woolf} {et~al.}(1995){Woolf}, {Tomkin}, \& {Lambert}}]{woolf1995r}
{Woolf}, V.~M., {Tomkin}, J., \& {Lambert}, D.~L. 1995, \apj, 453, 660

\bibitem[{{Woosley} \& {Bloom}(2006)}]{Woosley2006ARA&A..44..507W}
{Woosley}, S.~E. \& {Bloom}, J.~S. 2006, \araa, 44, 507

\bibitem[{{Woosley} \& {Heger}(2007)}]{WOOSLEY2007269}
{Woosley}, S.~E. \& {Heger}, A. 2007, \physrep, 442, 269

\bibitem[{{Woosley} {et~al.}(1994){Woosley}, {Wilson}, {Mathews}, {Hoffman}, \&
  {Meyer}}]{woosley1994r}
{Woosley}, S.~E., {Wilson}, J.~R., {Mathews}, G.~J., {Hoffman}, R.~D., \&
  {Meyer}, B.~S. 1994, \apj, 433, 229

\bibitem[{{Yong} {et~al.}(2013){Yong}, {Norris}, {Bessell}, {Christlieb},
  {Asplund}, {Beers}, {Barklem}, {Frebel}, \& {Ryan}}]{YON13}
{Yong}, D., {Norris}, J.~E., {Bessell}, M.~S., {et~al.} 2013, \apj, 762, 26

\bibitem[{{Yoon} {et~al.}(2006){Yoon}, {Langer}, \&
  {Norman}}]{2006A&A...460..199Y}
{Yoon}, S.~C., {Langer}, N., \& {Norman}, C. 2006, \aap, 460, 199

\bibitem[{{Zhang} {et~al.}(2009){Zhang}, {Ishigaki}, {Aoki}, {Zhao}, \&
  {Chiba}}]{ZHA09}
{Zhang}, L., {Ishigaki}, M., {Aoki}, W., {Zhao}, G., \& {Chiba}, M. 2009, \apj,
  706, 1095

\end{thebibliography}

\end{document}